\documentclass[twocolumn,traditabstract,longauth]{aa}

\usepackage{dcolumn}
\usepackage{epsf}
\usepackage{txfonts}
\usepackage{graphicx}
\usepackage{booktabs}
\usepackage{rotating}
\usepackage{natbib}
\usepackage{url}
\usepackage{amssymb}
\usepackage{fixltx2e}
\bibpunct{(}{)}{;}{a}{}{,} 

\newcommand{\HLP}{\tt HEALPix}
\newcommand{\hi}{\ion{H}{i}}

\newcommand{\kms}{km\,s$^{-1}$}

\def\IRAS{{IRAS\/}}
\def\WMAP{{WMAP\/}}
\def\COBE{{COBE\/}}
\def\Herschel{\textit{Herschel}}
\newcommand{\IRIS}{{IRIS\/}}
\newcommand{\DIRBE}{{DIRBE\/}}
\newcommand{\FIRAS}{{FIRAS\/}}
\newcommand{\Ebv}{$E(B-V)$}
\newcommand{\radiance}{$\cal{R}$}

\newcommand{\Ldust}{$L_{\rm H}$}

\newcommand{\opacity}{$\sigma_{\mathrm{e}\,353}$}
\newcommand{\XCOunits}{H$_2$\,cm$^{-2}$\,K$^{-1}$\,km$^{-1}$\,s}

\usepackage{color}

\def\setsymbol#1#2{\expandafter\def\csname #1\endcsname{#2}}
\def\getsymbol#1{\csname #1\endcsname}

\def\Planck{\textit{Planck}}

\def\allresultspapers{\nocite{planck2013-p01, planck2013-p02, planck2013-p02a, planck2013-p02d, planck2013-p02b, planck2013-p03, planck2013-p03c, planck2013-p03f, planck2013-p03d, planck2013-p03e, planck2013-p01a, planck2013-p06, planck2013-p03a, planck2013-pip88, planck2013-p08, planck2013-p11, planck2013-p12, planck2013-p13, planck2013-p14, planck2013-p15, planck2013-p05b, planck2013-p17, planck2013-p09, planck2013-p09a, planck2013-p20, planck2013-p19, planck2013-pipaberration, planck2013-p05, planck2013-p05a, planck2013-pip56, planck2013-p06b}}

\newbox\tablebox    \newdimen\tablewidth
\def\leaderfil{\leaders\hbox to 5pt{\hss.\hss}\hfil}
\def\tablenote#1 #2\par{\begingroup \parindent=0.8em
    \abovedisplayshortskip=0pt\belowdisplayshortskip=0pt
    \noindent
    $$\hss\vbox{\hsize\tablewidth \hangindent=\parindent \hangafter=1 \noindent
    \hbox to \parindent{$^#1$\hss}\strut#2\strut\par}\hss$$
    \endgroup}

\def\L2{\ifmmode L_2\else $L_2$\fi}

\def\DeltaT{\ifmmode \Delta T\else $\Delta T$\fi}
\def\deltat{\ifmmode \Delta t\else $\Delta t$\fi}
\def\fknee{\ifmmode f_{\rm knee}\else $f_{\rm knee}$\fi}
\def\Fmax{\ifmmode F_{\rm max}\else $F_{\rm max}$\fi}
\def\solar{\ifmmode{\rm M}_{\mathord\odot}\else${\rm M}_{\mathord\odot}$\fi}
\def\Msolar{\ifmmode{\rm M}_{\mathord\odot}\else${\rm M}_{\mathord\odot}$\fi}
\def\Lsolar{\ifmmode{\rm L}_{\mathord\odot}\else${\rm L}_{\mathord\odot}$\fi}

\def\inv{\ifmmode^{-1}\else$^{-1}$\fi}
\def\mo{\ifmmode^{-1}\else$^{-1}$\fi}
\def\sup#1{\ifmmode ^{\rm #1}\else $^{\rm #1}$\fi}
\def\expo#1{\ifmmode \times 10^{#1}\else $\times 10^{#1}$\fi}
\def\,{\thinspace}
\def\lsim{\mathrel{\raise .4ex\hbox{\rlap{$<$}\lower 1.2ex\hbox{$\sim$}}}}
\def\gsim{\mathrel{\raise .4ex\hbox{\rlap{$>$}\lower 1.2ex\hbox{$\sim$}}}}

\def\simprop{\mathrel{\raise .4ex\hbox{\rlap{$\propto$}\lower 1.2ex\hbox{$\sim$}}}}
\def\deg{\ifmmode^\circ\else$^\circ$\fi}
\def\pdeg{\ifmmode $\setbox0=\hbox{$^{\circ}$}\rlap{\hskip.11\wd0 .}$^{\circ}
          \else \setbox0=\hbox{$^{\circ}$}\rlap{\hskip.11\wd0 .}$^{\circ}$\fi}
\def\arcs{\ifmmode {^{\scriptstyle\prime\prime}}
          \else $^{\scriptstyle\prime\prime}$\fi}
\def\arcm{\ifmmode {^{\scriptstyle\prime}}
          \else $^{\scriptstyle\prime}$\fi}
\newdimen\sa  \newdimen\sb
\def\parcs{\sa=.07em \sb=.03em
     \ifmmode \hbox{\rlap{.}}^{\scriptstyle\prime\kern -\sb\prime}\hbox{\kern -\sa}
     \else \rlap{.}$^{\scriptstyle\prime\kern -\sb\prime}$\kern -\sa\fi}
\def\parcm{\sa=.08em \sb=.03em
     \ifmmode \hbox{\rlap{.}\kern\sa}^{\scriptstyle\prime}\hbox{\kern-\sb}
     \else \rlap{.}\kern\sa$^{\scriptstyle\prime}$\kern-\sb\fi}
\def\ra[#1 #2 #3.#4]{#1\sup{h}#2\sup{m}#3\sup{s}\llap.#4}
\def\dec[#1 #2 #3.#4]{#1\deg#2\arcm#3\arcs\llap.#4}
\def\deco[#1 #2 #3]{#1\deg#2\arcm#3\arcs}
\def\rra[#1 #2]{#1\sup{h}#2\sup{m}}

\def\dots{\relax\ifmmode \ldots\else $\ldots$\fi}
\def\WHzsr{\ifmmode $W\,Hz\mo\,sr\mo$\else W\,Hz\mo\,sr\mo\fi}
\def\mHz{\ifmmode $\,mHz$\else \,mHz\fi}
\def\GHz{\ifmmode $\,GHz$\else \,GHz\fi}
\def\mKs{\ifmmode $\,mK\,s$^{1/2}\else \,mK\,s$^{1/2}$\fi}
\def\muKs{\ifmmode \,\mu$K\,s$^{1/2}\else \,$\mu$K\,s$^{1/2}$\fi}
\def\muKRJs{\ifmmode \,\mu$K$_{\rm RJ}$\,s$^{1/2}\else \,$\mu$K$_{\rm RJ}$\,s$^{1/2}$\fi}
\def\muKHz{\ifmmode \,\mu$K\,Hz$^{-1/2}\else \,$\mu$K\,Hz$^{-1/2}$\fi}
\def\MJysr{\ifmmode \,$MJy\,sr\mo$\else \,MJy\,sr\mo\fi}
\def\MJysrmK{\ifmmode \,$MJy\,sr\mo$\,mK$_{\rm CMB}\mo\else \,MJy\,sr\mo\,mK$_{\rm CMB}\mo$\fi}
\def\microns{\ifmmode \,\mu$m$\else \,$\mu$m\fi}

\def\muK{\ifmmode \,\mu$K$\else \,$\mu$\hbox{K}\fi}
\def\microK{\ifmmode \,\mu$K$\else \,$\mu$\hbox{K}\fi}
\def\muW{\ifmmode \,\mu$W$\else \,$\mu$\hbox{W}\fi}
\def\kms{\ifmmode $\,km\,s$^{-1}\else \,km\,s$^{-1}$\fi}
\def\kmsMpc{\ifmmode $\,\kms\,Mpc\mo$\else \,\kms\,Mpc\mo\fi}
\setsymbol{LFI:center:frequency:70GHz:units}{70.3\,GHz}
\setsymbol{LFI:center:frequency:44GHz:units}{44.1\,GHz}
\setsymbol{LFI:center:frequency:30GHz:units}{28.5\,GHz}

\setsymbol{LFI:center:frequency:70GHz}{70.3}
\setsymbol{LFI:center:frequency:44GHz}{44.1}
\setsymbol{LFI:center:frequency:30GHz}{28.5}

\setsymbol{LFI:center:frequency:LFI18:Rad:M:units}{71.7\GHz}
\setsymbol{LFI:center:frequency:LFI19:Rad:M:units}{67.5\GHz}
\setsymbol{LFI:center:frequency:LFI20:Rad:M:units}{69.2\GHz}
\setsymbol{LFI:center:frequency:LFI21:Rad:M:units}{70.4\GHz}
\setsymbol{LFI:center:frequency:LFI22:Rad:M:units}{71.5\GHz}
\setsymbol{LFI:center:frequency:LFI23:Rad:M:units}{70.8\GHz}
\setsymbol{LFI:center:frequency:LFI24:Rad:M:units}{44.4\GHz}
\setsymbol{LFI:center:frequency:LFI25:Rad:M:units}{44.0\GHz}
\setsymbol{LFI:center:frequency:LFI26:Rad:M:units}{43.9\GHz}
\setsymbol{LFI:center:frequency:LFI27:Rad:M:units}{28.3\GHz}
\setsymbol{LFI:center:frequency:LFI28:Rad:M:units}{28.8\GHz}
\setsymbol{LFI:center:frequency:LFI18:Rad:S:units}{70.1\GHz}
\setsymbol{LFI:center:frequency:LFI19:Rad:S:units}{69.6\GHz}
\setsymbol{LFI:center:frequency:LFI20:Rad:S:units}{69.5\GHz}
\setsymbol{LFI:center:frequency:LFI21:Rad:S:units}{69.5\GHz}
\setsymbol{LFI:center:frequency:LFI22:Rad:S:units}{72.8\GHz}
\setsymbol{LFI:center:frequency:LFI23:Rad:S:units}{71.3\GHz}
\setsymbol{LFI:center:frequency:LFI24:Rad:S:units}{44.1\GHz}
\setsymbol{LFI:center:frequency:LFI25:Rad:S:units}{44.1\GHz}
\setsymbol{LFI:center:frequency:LFI26:Rad:S:units}{44.1\GHz}
\setsymbol{LFI:center:frequency:LFI27:Rad:S:units}{28.5\GHz}
\setsymbol{LFI:center:frequency:LFI28:Rad:S:units}{28.2\GHz}

\setsymbol{LFI:center:frequency:LFI18:Rad:M}{71.7}
\setsymbol{LFI:center:frequency:LFI19:Rad:M}{67.5}
\setsymbol{LFI:center:frequency:LFI20:Rad:M}{69.2}
\setsymbol{LFI:center:frequency:LFI21:Rad:M}{70.4}
\setsymbol{LFI:center:frequency:LFI22:Rad:M}{71.5}
\setsymbol{LFI:center:frequency:LFI23:Rad:M}{70.8}
\setsymbol{LFI:center:frequency:LFI24:Rad:M}{44.4}
\setsymbol{LFI:center:frequency:LFI25:Rad:M}{44.0}
\setsymbol{LFI:center:frequency:LFI26:Rad:M}{43.9}
\setsymbol{LFI:center:frequency:LFI27:Rad:M}{28.3}
\setsymbol{LFI:center:frequency:LFI28:Rad:M}{28.8}
\setsymbol{LFI:center:frequency:LFI18:Rad:S}{70.1}
\setsymbol{LFI:center:frequency:LFI19:Rad:S}{69.6}
\setsymbol{LFI:center:frequency:LFI20:Rad:S}{69.5}
\setsymbol{LFI:center:frequency:LFI21:Rad:S}{69.5}
\setsymbol{LFI:center:frequency:LFI22:Rad:S}{72.8}
\setsymbol{LFI:center:frequency:LFI23:Rad:S}{71.3}
\setsymbol{LFI:center:frequency:LFI24:Rad:S}{44.1}
\setsymbol{LFI:center:frequency:LFI25:Rad:S}{44.1}
\setsymbol{LFI:center:frequency:LFI26:Rad:S}{44.1}
\setsymbol{LFI:center:frequency:LFI27:Rad:S}{28.5}
\setsymbol{LFI:center:frequency:LFI28:Rad:S}{28.2}

\setsymbol{LFI:white:noise:sensitivity:70GHz:units}{134.7\muKs}
\setsymbol{LFI:white:noise:sensitivity:44GHz:units}{164.7\muKs}
\setsymbol{LFI:white:noise:sensitivity:30GHz:units}{143.4\muKs}

\setsymbol{LFI:white:noise:sensitivity:70GHz}{134.7}
\setsymbol{LFI:white:noise:sensitivity:44GHz}{164.7}
\setsymbol{LFI:white:noise:sensitivity:30GHz}{143.4}

\setsymbol{LFI:white:noise:sensitivity:LFI18:Rad:M:units}{512.0\muKs}
\setsymbol{LFI:white:noise:sensitivity:LFI19:Rad:M:units}{581.4\muKs}
\setsymbol{LFI:white:noise:sensitivity:LFI20:Rad:M:units}{590.8\muKs}
\setsymbol{LFI:white:noise:sensitivity:LFI21:Rad:M:units}{455.2\muKs}
\setsymbol{LFI:white:noise:sensitivity:LFI22:Rad:M:units}{492.0\muKs}
\setsymbol{LFI:white:noise:sensitivity:LFI23:Rad:M:units}{507.7\muKs}
\setsymbol{LFI:white:noise:sensitivity:LFI24:Rad:M:units}{462.2\muKs}
\setsymbol{LFI:white:noise:sensitivity:LFI25:Rad:M:units}{413.6\muKs}
\setsymbol{LFI:white:noise:sensitivity:LFI26:Rad:M:units}{478.6\muKs}
\setsymbol{LFI:white:noise:sensitivity:LFI27:Rad:M:units}{277.7\muKs}
\setsymbol{LFI:white:noise:sensitivity:LFI28:Rad:M:units}{312.3\muKs}
\setsymbol{LFI:white:noise:sensitivity:LFI18:Rad:S:units}{465.7\muKs}
\setsymbol{LFI:white:noise:sensitivity:LFI19:Rad:S:units}{555.6\muKs}
\setsymbol{LFI:white:noise:sensitivity:LFI20:Rad:S:units}{623.2\muKs}
\setsymbol{LFI:white:noise:sensitivity:LFI21:Rad:S:units}{564.1\muKs}
\setsymbol{LFI:white:noise:sensitivity:LFI22:Rad:S:units}{534.4\muKs}
\setsymbol{LFI:white:noise:sensitivity:LFI23:Rad:S:units}{542.4\muKs}
\setsymbol{LFI:white:noise:sensitivity:LFI24:Rad:S:units}{399.2\muKs}
\setsymbol{LFI:white:noise:sensitivity:LFI25:Rad:S:units}{392.6\muKs}
\setsymbol{LFI:white:noise:sensitivity:LFI26:Rad:S:units}{418.6\muKs}
\setsymbol{LFI:white:noise:sensitivity:LFI27:Rad:S:units}{302.9\muKs}
\setsymbol{LFI:white:noise:sensitivity:LFI28:Rad:S:units}{285.3\muKs}

\setsymbol{LFI:white:noise:sensitivity:LFI18:Rad:M}{512.0}
\setsymbol{LFI:white:noise:sensitivity:LFI19:Rad:M}{581.4}
\setsymbol{LFI:white:noise:sensitivity:LFI20:Rad:M}{590.8}
\setsymbol{LFI:white:noise:sensitivity:LFI21:Rad:M}{455.2}
\setsymbol{LFI:white:noise:sensitivity:LFI22:Rad:M}{492.0}
\setsymbol{LFI:white:noise:sensitivity:LFI23:Rad:M}{507.7}
\setsymbol{LFI:white:noise:sensitivity:LFI24:Rad:M}{462.2}
\setsymbol{LFI:white:noise:sensitivity:LFI25:Rad:M}{413.6}
\setsymbol{LFI:white:noise:sensitivity:LFI26:Rad:M}{478.6}
\setsymbol{LFI:white:noise:sensitivity:LFI27:Rad:M}{277.7}
\setsymbol{LFI:white:noise:sensitivity:LFI28:Rad:M}{312.3}
\setsymbol{LFI:white:noise:sensitivity:LFI18:Rad:S}{465.7}
\setsymbol{LFI:white:noise:sensitivity:LFI19:Rad:S}{555.6}
\setsymbol{LFI:white:noise:sensitivity:LFI20:Rad:S}{623.2}
\setsymbol{LFI:white:noise:sensitivity:LFI21:Rad:S}{564.1}
\setsymbol{LFI:white:noise:sensitivity:LFI22:Rad:S}{534.4}
\setsymbol{LFI:white:noise:sensitivity:LFI23:Rad:S}{542.4}
\setsymbol{LFI:white:noise:sensitivity:LFI24:Rad:S}{399.2}
\setsymbol{LFI:white:noise:sensitivity:LFI25:Rad:S}{392.6}
\setsymbol{LFI:white:noise:sensitivity:LFI26:Rad:S}{418.6}
\setsymbol{LFI:white:noise:sensitivity:LFI27:Rad:S}{302.9}
\setsymbol{LFI:white:noise:sensitivity:LFI28:Rad:S}{285.3}

\setsymbol{LFI:knee:frequency:70GHz:units}{29.5\mHz}
\setsymbol{LFI:knee:frequency:44GHz:units}{56.2\mHz}
\setsymbol{LFI:knee:frequency:30GHz:units}{113.7\mHz}

\setsymbol{LFI:knee:frequency:70GHz}{29.5}
\setsymbol{LFI:knee:frequency:44GHz}{56.2}
\setsymbol{LFI:knee:frequency:30GHz}{113.7}

\setsymbol{LFI:knee:frequency:LFI18:Rad:M:units}{16.3\mHz}
\setsymbol{LFI:knee:frequency:LFI19:Rad:M:units}{15.1\mHz}
\setsymbol{LFI:knee:frequency:LFI20:Rad:M:units}{18.7\mHz}
\setsymbol{LFI:knee:frequency:LFI21:Rad:M:units}{37.2\mHz}
\setsymbol{LFI:knee:frequency:LFI22:Rad:M:units}{12.7\mHz}
\setsymbol{LFI:knee:frequency:LFI23:Rad:M:units}{34.6\mHz}
\setsymbol{LFI:knee:frequency:LFI24:Rad:M:units}{46.2\mHz}
\setsymbol{LFI:knee:frequency:LFI25:Rad:M:units}{24.9\mHz}
\setsymbol{LFI:knee:frequency:LFI26:Rad:M:units}{67.6\mHz}
\setsymbol{LFI:knee:frequency:LFI27:Rad:M:units}{187.4\mHz}
\setsymbol{LFI:knee:frequency:LFI28:Rad:M:units}{122.2\mHz}
\setsymbol{LFI:knee:frequency:LFI18:Rad:S:units}{17.7\mHz}
\setsymbol{LFI:knee:frequency:LFI19:Rad:S:units}{22.0\mHz}
\setsymbol{LFI:knee:frequency:LFI20:Rad:S:units}{8.7\mHz}
\setsymbol{LFI:knee:frequency:LFI21:Rad:S:units}{25.9\mHz}
\setsymbol{LFI:knee:frequency:LFI22:Rad:S:units}{15.8\mHz}
\setsymbol{LFI:knee:frequency:LFI23:Rad:S:units}{129.8\mHz}
\setsymbol{LFI:knee:frequency:LFI24:Rad:S:units}{100.9\mHz}
\setsymbol{LFI:knee:frequency:LFI25:Rad:S:units}{38.9\mHz}
\setsymbol{LFI:knee:frequency:LFI26:Rad:S:units}{58.9\mHz}
\setsymbol{LFI:knee:frequency:LFI27:Rad:S:units}{104.4\mHz}
\setsymbol{LFI:knee:frequency:LFI28:Rad:S:units}{40.7\mHz}

\setsymbol{LFI:knee:frequency:LFI18:Rad:M}{16.3}
\setsymbol{LFI:knee:frequency:LFI19:Rad:M}{15.1}
\setsymbol{LFI:knee:frequency:LFI20:Rad:M}{18.7}
\setsymbol{LFI:knee:frequency:LFI21:Rad:M}{37.2}
\setsymbol{LFI:knee:frequency:LFI22:Rad:M}{12.7}
\setsymbol{LFI:knee:frequency:LFI23:Rad:M}{34.6}
\setsymbol{LFI:knee:frequency:LFI24:Rad:M}{46.2}
\setsymbol{LFI:knee:frequency:LFI25:Rad:M}{24.9}
\setsymbol{LFI:knee:frequency:LFI26:Rad:M}{67.6}
\setsymbol{LFI:knee:frequency:LFI27:Rad:M}{187.4}
\setsymbol{LFI:knee:frequency:LFI28:Rad:M}{122.2}
\setsymbol{LFI:knee:frequency:LFI18:Rad:S}{17.7}
\setsymbol{LFI:knee:frequency:LFI19:Rad:S}{22.0}
\setsymbol{LFI:knee:frequency:LFI20:Rad:S}{8.7}
\setsymbol{LFI:knee:frequency:LFI21:Rad:S}{25.9}
\setsymbol{LFI:knee:frequency:LFI22:Rad:S}{15.8}
\setsymbol{LFI:knee:frequency:LFI23:Rad:S}{129.8}
\setsymbol{LFI:knee:frequency:LFI24:Rad:S}{100.9}
\setsymbol{LFI:knee:frequency:LFI25:Rad:S}{38.9}
\setsymbol{LFI:knee:frequency:LFI26:Rad:S}{58.9}
\setsymbol{LFI:knee:frequency:LFI27:Rad:S}{104.4}
\setsymbol{LFI:knee:frequency:LFI28:Rad:S}{40.7}

\setsymbol{LFI:slope:70GHz:units}{$-1.03$\mHz}
\setsymbol{LFI:slope:44GHz:units}{$-0.89$\mHz}
\setsymbol{LFI:slope:30GHz:units}{$-0.87$\mHz}

\setsymbol{LFI:slope:70GHz}{$-1.03$}
\setsymbol{LFI:slope:44GHz}{$-0.89$}
\setsymbol{LFI:slope:30GHz}{$-0.87$}

\setsymbol{LFI:slope:LFI18:Rad:M:units}{$-1.04$\mHz}
\setsymbol{LFI:slope:LFI19:Rad:M:units}{$-1.09$\mHz}
\setsymbol{LFI:slope:LFI20:Rad:M:units}{$-0.69$\mHz}
\setsymbol{LFI:slope:LFI21:Rad:M:units}{$-1.56$\mHz}
\setsymbol{LFI:slope:LFI22:Rad:M:units}{$-1.01$\mHz}
\setsymbol{LFI:slope:LFI23:Rad:M:units}{$-0.96$\mHz}
\setsymbol{LFI:slope:LFI24:Rad:M:units}{$-0.83$\mHz}
\setsymbol{LFI:slope:LFI25:Rad:M:units}{$-0.91$\mHz}
\setsymbol{LFI:slope:LFI26:Rad:M:units}{$-0.95$\mHz}
\setsymbol{LFI:slope:LFI27:Rad:M:units}{$-0.87$\mHz}
\setsymbol{LFI:slope:LFI28:Rad:M:units}{$-0.88$\mHz}
\setsymbol{LFI:slope:LFI18:Rad:S:units}{$-1.15$\mHz}
\setsymbol{LFI:slope:LFI19:Rad:S:units}{$-1.00$\mHz}
\setsymbol{LFI:slope:LFI20:Rad:S:units}{$-0.95$\mHz}
\setsymbol{LFI:slope:LFI21:Rad:S:units}{$-0.92$\mHz}
\setsymbol{LFI:slope:LFI22:Rad:S:units}{$-1.01$\mHz}
\setsymbol{LFI:slope:LFI23:Rad:S:units}{$-0.95$\mHz}
\setsymbol{LFI:slope:LFI24:Rad:S:units}{$-0.73$\mHz}
\setsymbol{LFI:slope:LFI25:Rad:S:units}{$-1.16$\mHz}
\setsymbol{LFI:slope:LFI26:Rad:S:units}{$-0.79$\mHz}
\setsymbol{LFI:slope:LFI27:Rad:S:units}{$-0.82$\mHz}
\setsymbol{LFI:slope:LFI28:Rad:S:units}{$-0.91$\mHz}

\setsymbol{LFI:slope:LFI18:Rad:M}{$-1.04$}
\setsymbol{LFI:slope:LFI19:Rad:M}{$-1.09$}
\setsymbol{LFI:slope:LFI20:Rad:M}{$-0.69$}
\setsymbol{LFI:slope:LFI21:Rad:M}{$-1.56$}
\setsymbol{LFI:slope:LFI22:Rad:M}{$-1.01$}
\setsymbol{LFI:slope:LFI23:Rad:M}{$-0.96$}
\setsymbol{LFI:slope:LFI24:Rad:M}{$-0.83$}
\setsymbol{LFI:slope:LFI25:Rad:M}{$-0.91$}
\setsymbol{LFI:slope:LFI26:Rad:M}{$-0.95$}
\setsymbol{LFI:slope:LFI27:Rad:M}{$-0.87$}
\setsymbol{LFI:slope:LFI28:Rad:M}{$-0.88$}
\setsymbol{LFI:slope:LFI18:Rad:S}{$-1.15$}
\setsymbol{LFI:slope:LFI19:Rad:S}{$-1.00$}
\setsymbol{LFI:slope:LFI20:Rad:S}{$-0.95$}
\setsymbol{LFI:slope:LFI21:Rad:S}{$-0.92$}
\setsymbol{LFI:slope:LFI22:Rad:S}{$-1.01$}
\setsymbol{LFI:slope:LFI23:Rad:S}{$-0.95$}
\setsymbol{LFI:slope:LFI24:Rad:S}{$-0.73$}
\setsymbol{LFI:slope:LFI25:Rad:S}{$-1.16$}
\setsymbol{LFI:slope:LFI26:Rad:S}{$-0.79$}
\setsymbol{LFI:slope:LFI27:Rad:S}{$-0.82$}
\setsymbol{LFI:slope:LFI28:Rad:S}{$-0.91$}

\setsymbol{LFI:FWHM:70GHz:units}{13\parcm01}
\setsymbol{LFI:FWHM:44GHz:units}{27\parcm92}
\setsymbol{LFI:FWHM:30GHz:units}{32\parcm65}

\setsymbol{LFI:FWHM:70GHz}{13.01}
\setsymbol{LFI:FWHM:44GHz}{27.92}
\setsymbol{LFI:FWHM:30GHz}{32.65}

\setsymbol{LFI:FWHM:LFI18:units}{13\parcm39}
\setsymbol{LFI:FWHM:LFI19:units}{13\parcm01}
\setsymbol{LFI:FWHM:LFI20:units}{12\parcm75}
\setsymbol{LFI:FWHM:LFI21:units}{12\parcm74}
\setsymbol{LFI:FWHM:LFI22:units}{12\parcm87}
\setsymbol{LFI:FWHM:LFI23:units}{13\parcm27}
\setsymbol{LFI:FWHM:LFI24:units}{22\parcm98}
\setsymbol{LFI:FWHM:LFI25:units}{30\parcm46}
\setsymbol{LFI:FWHM:LFI26:units}{30\parcm31}
\setsymbol{LFI:FWHM:LFI27:units}{32\parcm65}
\setsymbol{LFI:FWHM:LFI28:units}{32\parcm66}

\setsymbol{LFI:FWHM:LFI18}{13.39}
\setsymbol{LFI:FWHM:LFI19}{13.01}
\setsymbol{LFI:FWHM:LFI20}{12.75}
\setsymbol{LFI:FWHM:LFI21}{12.74}
\setsymbol{LFI:FWHM:LFI22}{12.87}
\setsymbol{LFI:FWHM:LFI23}{13.27}
\setsymbol{LFI:FWHM:LFI24}{22.98}
\setsymbol{LFI:FWHM:LFI25}{30.46}
\setsymbol{LFI:FWHM:LFI26}{30.31}
\setsymbol{LFI:FWHM:LFI27}{32.65}
\setsymbol{LFI:FWHM:LFI28}{32.66}

\setsymbol{LFI:FWHM:uncertainty:LFI18:units}{0.170\arcm}
\setsymbol{LFI:FWHM:uncertainty:LFI19:units}{0.174\arcm}
\setsymbol{LFI:FWHM:uncertainty:LFI20:units}{0.170\arcm}
\setsymbol{LFI:FWHM:uncertainty:LFI21:units}{0.156\arcm}
\setsymbol{LFI:FWHM:uncertainty:LFI22:units}{0.164\arcm}
\setsymbol{LFI:FWHM:uncertainty:LFI23:units}{0.171\arcm}
\setsymbol{LFI:FWHM:uncertainty:LFI24:units}{0.652\arcm}
\setsymbol{LFI:FWHM:uncertainty:LFI25:units}{1.075\arcm}
\setsymbol{LFI:FWHM:uncertainty:LFI26:units}{1.131\arcm}
\setsymbol{LFI:FWHM:uncertainty:LFI27:units}{1.266\arcm}
\setsymbol{LFI:FWHM:uncertainty:LFI28:units}{1.287\arcm}

\setsymbol{LFI:FWHM:uncertainty:LFI18}{0.170}
\setsymbol{LFI:FWHM:uncertainty:LFI19}{0.174}
\setsymbol{LFI:FWHM:uncertainty:LFI20}{0.170}
\setsymbol{LFI:FWHM:uncertainty:LFI21}{0.156}
\setsymbol{LFI:FWHM:uncertainty:LFI22}{0.164}
\setsymbol{LFI:FWHM:uncertainty:LFI23}{0.171}
\setsymbol{LFI:FWHM:uncertainty:LFI24}{0.652}
\setsymbol{LFI:FWHM:uncertainty:LFI25}{1.075}
\setsymbol{LFI:FWHM:uncertainty:LFI26}{1.131}
\setsymbol{LFI:FWHM:uncertainty:LFI27}{1.266}
\setsymbol{LFI:FWHM:uncertainty:LFI28}{1.287}

\setsymbol{HFI:center:frequency:100GHz:units}{100\,GHz}
\setsymbol{HFI:center:frequency:143GHz:units}{143\,GHz}
\setsymbol{HFI:center:frequency:217GHz:units}{217\,GHz}
\setsymbol{HFI:center:frequency:353GHz:units}{353\,GHz}
\setsymbol{HFI:center:frequency:545GHz:units}{545\,GHz}
\setsymbol{HFI:center:frequency:857GHz:units}{857\,GHz}

\setsymbol{HFI:center:frequency:100GHz}{100}
\setsymbol{HFI:center:frequency:143GHz}{143}
\setsymbol{HFI:center:frequency:217GHz}{217}
\setsymbol{HFI:center:frequency:353GHz}{353}
\setsymbol{HFI:center:frequency:545GHz}{545}
\setsymbol{HFI:center:frequency:857GHz}{857}

\setsymbol{HFI:Ndetectors:100GHz}{8}
\setsymbol{HFI:Ndetectors:143GHz}{11}
\setsymbol{HFI:Ndetectors:217GHz}{12}
\setsymbol{HFI:Ndetectors:353GHz}{12}
\setsymbol{HFI:Ndetectors:545GHz}{3}
\setsymbol{HFI:Ndetectors:857GHz}{4}

\setsymbol{HFI:FWHM:Maps:100GHz:units}{9\parcm88}
\setsymbol{HFI:FWHM:Maps:143GHz:units}{7\parcm18}
\setsymbol{HFI:FWHM:Maps:217GHz:units}{4\parcm87}
\setsymbol{HFI:FWHM:Maps:353GHz:units}{4\parcm65}
\setsymbol{HFI:FWHM:Maps:545GHz:units}{4\parcm72}
\setsymbol{HFI:FWHM:Maps:857GHz:units}{4\parcm39}
\setsymbol{HFI:FWHM:Maps:100GHz}{9.88}
\setsymbol{HFI:FWHM:Maps:143GHz}{7.18}
\setsymbol{HFI:FWHM:Maps:217GHz}{4.87}
\setsymbol{HFI:FWHM:Maps:353GHz}{4.65}
\setsymbol{HFI:FWHM:Maps:545GHz}{4.72}
\setsymbol{HFI:FWHM:Maps:857GHz}{4.39}

\setsymbol{HFI:beam:ellipticity:Maps:100GHz}{1.15}
\setsymbol{HFI:beam:ellipticity:Maps:143GHz}{1.01}
\setsymbol{HFI:beam:ellipticity:Maps:217GHz}{1.06}
\setsymbol{HFI:beam:ellipticity:Maps:353GHz}{1.05}
\setsymbol{HFI:beam:ellipticity:Maps:545GHz}{1.14}
\setsymbol{HFI:beam:ellipticity:Maps:857GHz}{1.19}

\setsymbol{HFI:FWHM:Mars:100GHz:units}{9\parcm37}
\setsymbol{HFI:FWHM:Mars:143GHz:units}{7\parcm04}
\setsymbol{HFI:FWHM:Mars:217GHz:units}{4\parcm68}
\setsymbol{HFI:FWHM:Mars:353GHz:units}{4\parcm43}
\setsymbol{HFI:FWHM:Mars:545GHz:units}{3\parcm80}
\setsymbol{HFI:FWHM:Mars:857GHz:units}{3\parcm67}

\setsymbol{HFI:FWHM:Mars:100GHz}{9.37}
\setsymbol{HFI:FWHM:Mars:143GHz}{7.04}
\setsymbol{HFI:FWHM:Mars:217GHz}{4.68}
\setsymbol{HFI:FWHM:Mars:353GHz}{4.43}
\setsymbol{HFI:FWHM:Mars:545GHz}{3.80}
\setsymbol{HFI:FWHM:Mars:857GHz}{3.67}

\setsymbol{HFI:beam:ellipticity:Mars:100GHz}{1.18}
\setsymbol{HFI:beam:ellipticity:Mars:143GHz}{1.03}
\setsymbol{HFI:beam:ellipticity:Mars:217GHz}{1.14}
\setsymbol{HFI:beam:ellipticity:Mars:353GHz}{1.09}
\setsymbol{HFI:beam:ellipticity:Mars:545GHz}{1.25}
\setsymbol{HFI:beam:ellipticity:Mars:857GHz}{1.03}

\setsymbol{HFI:CMB:relative:calibration:100GHz}{$\lsim 1\%$}
\setsymbol{HFI:CMB:relative:calibration:143GHz}{$\lsim 1\%$}
\setsymbol{HFI:CMB:relative:calibration:217GHz}{$\lsim 1\%$}
\setsymbol{HFI:CMB:relative:calibration:353GHz}{$\lsim 1\%$}
\setsymbol{HFI:CMB:relative:calibration:545GHz}{}
\setsymbol{HFI:CMB:relative:calibration:857GHz}{}

\setsymbol{HFI:CMB:absolute:calibration:100GHz}{$\lsim 2\%$}
\setsymbol{HFI:CMB:absolute:calibration:143GHz}{$\lsim 2\%$}
\setsymbol{HFI:CMB:absolute:calibration:217GHz}{$\lsim 2\%$}
\setsymbol{HFI:CMB:absolute:calibration:353GHz}{$\lsim 2\%$}
\setsymbol{HFI:CMB:absolute:calibration:545GHz}{}
\setsymbol{HFI:CMB:absolute:calibration:857GHz}{}

\setsymbol{HFI:FIRAS:gain:calibration:accuracy:statistical:100GHz}{}
\setsymbol{HFI:FIRAS:gain:calibration:accuracy:statistical:143GHz}{}
\setsymbol{HFI:FIRAS:gain:calibration:accuracy:statistical:217GHz}{}
\setsymbol{HFI:FIRAS:gain:calibration:accuracy:statistical:353GHz}{2.5\%}
\setsymbol{HFI:FIRAS:gain:calibration:accuracy:statistical:545GHz}{1\%}
\setsymbol{HFI:FIRAS:gain:calibration:accuracy:statistical:857GHz}{0.5\%}

\setsymbol{HFI:FIRAS:gain:calibration:accuracy:systematic:100GHz}{}
\setsymbol{HFI:FIRAS:gain:calibration:accuracy:systematic:143GHz}{}
\setsymbol{HFI:FIRAS:gain:calibration:accuracy:systematic:217GHz}{}
\setsymbol{HFI:FIRAS:gain:calibration:accuracy:systematic:353GHz}{}
\setsymbol{HFI:FIRAS:gain:calibration:accuracy:systematic:545GHz}{7\%}
\setsymbol{HFI:FIRAS:gain:calibration:accuracy:systematic:857GHz}{7\%}

\setsymbol{HFI:FIRAS:zero:point:accuracy:100GHz:units}{0.8\MJysr}
\setsymbol{HFI:FIRAS:zero:point:accuracy:143GHz:units}{}
\setsymbol{HFI:FIRAS:zero:point:accuracy:217GHz:units}{}
\setsymbol{HFI:FIRAS:zero:point:accuracy:353GHz:units}{1.4\MJysr}
\setsymbol{HFI:FIRAS:zero:point:accuracy:545GHz:units}{2.2\MJysr}
\setsymbol{HFI:FIRAS:zero:point:accuracy:857GHz:units}{1.7\MJysr}

\setsymbol{HFI:FIRAS:zero:point:accuracy:100GHz}{0.8}
\setsymbol{HFI:FIRAS:zero:point:accuracy:143GHz}{}
\setsymbol{HFI:FIRAS:zero:point:accuracy:217GHz}{}
\setsymbol{HFI:FIRAS:zero:point:accuracy:353GHz}{1.4}
\setsymbol{HFI:FIRAS:zero:point:accuracy:545GHz}{2.2}
\setsymbol{HFI:FIRAS:zero:point:accuracy:857GHz}{1.7}

\setsymbol{HFI:unit:conversion:100GHz:units}{0.2415\MJysrmK}
\setsymbol{HFI:unit:conversion:143GHz:units}{0.3694\MJysrmK}
\setsymbol{HFI:unit:conversion:217GHz:units}{0.4811\MJysrmK}
\setsymbol{HFI:unit:conversion:353GHz:units}{0.2883\MJysrmK}
\setsymbol{HFI:unit:conversion:545GHz:units}{0.05826\MJysrmK}
\setsymbol{HFI:unit:conversion:857GHz:units}{0.002238\MJysrmK}

\setsymbol{HFI:unit:conversion:100GHz}{0.2415}
\setsymbol{HFI:unit:conversion:143GHz}{0.3694}
\setsymbol{HFI:unit:conversion:217GHz}{0.4811}
\setsymbol{HFI:unit:conversion:353GHz}{0.2883}
\setsymbol{HFI:unit:conversion:545GHz}{0.05826}
\setsymbol{HFI:unit:conversion:857GHz}{0.002238}

\setsymbol{HFI:colour:correction:alpha=-2:V1.01:100GHz}{0.9893}
\setsymbol{HFI:colour:correction:alpha=-2:V1.01:143GHz}{0.9759}
\setsymbol{HFI:colour:correction:alpha=-2:V1.01:217GHz}{1.0007}
\setsymbol{HFI:colour:correction:alpha=-2:V1.01:353GHz}{1.0028}
\setsymbol{HFI:colour:correction:alpha=-2:V1.01:545GHz}{1.0019}
\setsymbol{HFI:colour:correction:alpha=-2:V1.01:857GHz}{0.9889}

\setsymbol{HFI:colour:correction:alpha=0:V1.01:100GHz}{1.0008}
\setsymbol{HFI:colour:correction:alpha=0:V1.01:143GHz}{1.0148}
\setsymbol{HFI:colour:correction:alpha=0:V1.01:217GHz}{0.9909}
\setsymbol{HFI:colour:correction:alpha=0:V1.01:353GHz}{0.9888}
\setsymbol{HFI:colour:correction:alpha=0:V1.01:545GHz}{0.9878}
\setsymbol{HFI:colour:correction:alpha=0:V1.01:857GHz}{1.0014}

\providecommand{\sorthelp}[1]{}

\begin{document}

\title{\Planck\ 2013 results. XI. All-sky model of thermal dust emission\thanks{Corresponding author: Marc-Antoine Miville-Desch\^enes, \newline e-mail: {\tt mamd@ias.u-psud.fr}}}

\author{\small
Planck Collaboration:
A.~Abergel\inst{62}
\and
P.~A.~R.~Ade\inst{90}
\and
N.~Aghanim\inst{62}
\and
M.~I.~R.~Alves\inst{62}
\and
G.~Aniano\inst{62}
\and
C.~Armitage-Caplan\inst{95}
\and
M.~Arnaud\inst{75}
\and
M.~Ashdown\inst{72, 6}
\and
F.~Atrio-Barandela\inst{19}
\and
J.~Aumont\inst{62}
% ...
}
\institute{\small
APC, AstroParticule et Cosmologie, Universit\'{e} Paris Diderot, CNRS/IN2P3, CEA/lrfu, Observatoire de Paris, Sorbonne Paris Cit\'{e}, 10, rue Alice Domon et L\'{e}onie Duquet, 75205 Paris Cedex 13, France\\
\and
Aalto University Mets\"{a}hovi Radio Observatory and Dept of Radio Science and Engineering, P.O. Box 13000, FI-00076 AALTO, Finland\\
\and
African Institute for Mathematical Sciences, 6-8 Melrose Road, Muizenberg, Cape Town, South Africa\\
\and
Agenzia Spaziale Italiana Science Data Center, Via del Politecnico snc, 00133, Roma, Italy\\
\and
Agenzia Spaziale Italiana, Viale Liegi 26, Roma, Italy\\
\and
Astrophysics Group, Cavendish Laboratory, University of Cambridge, J J Thomson Avenue, Cambridge CB3 0HE, U.K.\\
\and
Astrophysics \& Cosmology Research Unit, School of Mathematics, Statistics \& Computer Science, University of KwaZulu-Natal, Westville Campus, Private Bag X54001, Durban 4000, South Africa\\
\and
CITA, University of Toronto, 60 St. George St., Toronto, ON M5S 3H8, Canada\\
\and
CNRS, IRAP, 9 Av. colonel Roche, BP 44346, F-31028 Toulouse cedex 4, France\\
\and
California Institute of Technology, Pasadena, California, U.S.A.\\
\and
Centre for Theoretical Cosmology, DAMTP, University of Cambridge, Wilberforce Road, Cambridge CB3 0WA, U.K.\\
\and
Centro de Estudios de F\'{i}sica del Cosmos de Arag\'{o}n (CEFCA), Plaza San Juan, 1, planta 2, E-44001, Teruel, Spain\\
\and
Computational Cosmology Center, Lawrence Berkeley National Laboratory, Berkeley, California, U.S.A.\\
\and
Consejo Superior de Investigaciones Cient\'{\i}ficas (CSIC), Madrid, Spain\\
\and
DSM/Irfu/SPP, CEA-Saclay, F-91191 Gif-sur-Yvette Cedex, France\\
\and
DTU Space, National Space Institute, Technical University of Denmark, Elektrovej 327, DK-2800 Kgs. Lyngby, Denmark\\
\and
D\'{e}partement de Physique Th\'{e}orique, Universit\'{e} de Gen\`{e}ve, 24, Quai E. Ansermet,1211 Gen\`{e}ve 4, Switzerland\\
\and
D\'{e}partement de physique, de g\'{e}nie physique et d'optique, Universit\'{e} Laval, Qu\'{e}bec, Canada\\
\and
Departamento de F\'{\i}sica Fundamental, Facultad de Ciencias, Universidad de Salamanca, 37008 Salamanca, Spain\\
\and
Departamento de F\'{\i}sica, Universidad de Oviedo, Avda. Calvo Sotelo s/n, Oviedo, Spain\\
\and
Department of Astronomy and Astrophysics, University of Toronto, 50 Saint George Street, Toronto, Ontario, Canada\\
\and
Department of Astrophysics/IMAPP, Radboud University Nijmegen, P.O. Box 9010, 6500 GL Nijmegen, The Netherlands\\
\and
Department of Electrical Engineering and Computer Sciences, University of California, Berkeley, California, U.S.A.\\
\and
Department of Physics \& Astronomy, University of British Columbia, 6224 Agricultural Road, Vancouver, British Columbia, Canada\\
\and
Department of Physics and Astronomy, Dana and David Dornsife College of Letter, Arts and Sciences, University of Southern California, Los Angeles, CA 90089, U.S.A.\\
\and
Department of Physics and Astronomy, University College London, London WC1E 6BT, U.K.\\
\and
Department of Physics, Florida State University, Keen Physics Building, 77 Chieftan Way, Tallahassee, Florida, U.S.A.\\
\and
Department of Physics, Gustaf H\"{a}llstr\"{o}min katu 2a, University of Helsinki, Helsinki, Finland\\
\and
Department of Physics, Princeton University, Princeton, New Jersey, U.S.A.\\
\and
Department of Physics, University of California, One Shields Avenue, Davis, California, U.S.A.\\
\and
Department of Physics, University of California, Santa Barbara, California, U.S.A.\\
\and
Department of Physics, University of Illinois at Urbana-Champaign, 1110 West Green Street, Urbana, Illinois, U.S.A.\\
\and
Dipartimento di Fisica e Astronomia G. Galilei, Universit\`{a} degli Studi di Padova, via Marzolo 8, 35131 Padova, Italy\\
\and
Dipartimento di Fisica e Scienze della Terra, Universit\`{a} di Ferrara, Via Saragat 1, 44122 Ferrara, Italy\\
\and
Dipartimento di Fisica, Universit\`{a} La Sapienza, P. le A. Moro 2, Roma, Italy\\
\and
Dipartimento di Fisica, Universit\`{a} degli Studi di Milano, Via Celoria, 16, Milano, Italy\\
\and
Dipartimento di Fisica, Universit\`{a} degli Studi di Trieste, via A. Valerio 2, Trieste, Italy\\
\and
Dipartimento di Fisica, Universit\`{a} di Roma Tor Vergata, Via della Ricerca Scientifica, 1, Roma, Italy\\
\and
Discovery Center, Niels Bohr Institute, Blegdamsvej 17, Copenhagen, Denmark\\
\and
Dpto. Astrof\'{i}sica, Universidad de La Laguna (ULL), E-38206 La Laguna, Tenerife, Spain\\
\and
European Space Agency, ESAC, Planck Science Office, Camino bajo del Castillo, s/n, Urbanizaci\'{o}n Villafranca del Castillo, Villanueva de la Ca\~{n}ada, Madrid, Spain\\
\and
European Space Agency, ESTEC, Keplerlaan 1, 2201 AZ Noordwijk, The Netherlands\\
\and
Finnish Centre for Astronomy with ESO (FINCA), University of Turku, V\"{a}is\"{a}l\"{a}ntie 20, FIN-21500, Piikki\"{o}, Finland\\
\and
Helsinki Institute of Physics, Gustaf H\"{a}llstr\"{o}min katu 2, University of Helsinki, Helsinki, Finland\\
\and
INAF - Osservatorio Astrofisico di Catania, Via S. Sofia 78, Catania, Italy\\
\and
INAF - Osservatorio Astronomico di Padova, Vicolo dell'Osservatorio 5, Padova, Italy\\
\and
INAF - Osservatorio Astronomico di Roma, via di Frascati 33, Monte Porzio Catone, Italy\\
\and
INAF - Osservatorio Astronomico di Trieste, Via G.B. Tiepolo 11, Trieste, Italy\\
\and
INAF Istituto di Radioastronomia, Via P. Gobetti 101, 40129 Bologna, Italy\\
\and
INAF/IASF Bologna, Via Gobetti 101, Bologna, Italy\\
\and
INAF/IASF Milano, Via E. Bassini 15, Milano, Italy\\
\and
INFN, Sezione di Bologna, Via Irnerio 46, I-40126, Bologna, Italy\\
\and
INFN, Sezione di Roma 1, Universit\`{a} di Roma Sapienza, Piazzale Aldo Moro 2, 00185, Roma, Italy\\
\and
INFN/National Institute for Nuclear Physics, Via Valerio 2, I-34127 Trieste, Italy\\
\and
IPAG: Institut de Plan\'{e}tologie et d'Astrophysique de Grenoble, Universit\'{e} Joseph Fourier, Grenoble 1 / CNRS-INSU, UMR 5274, Grenoble, F-38041, France\\
\and
ISDC Data Centre for Astrophysics, University of Geneva, ch. d'Ecogia 16, Versoix, Switzerland\\
\and
IUCAA, Post Bag 4, Ganeshkhind, Pune University Campus, Pune 411 007, India\\
\and
Imperial College London, Astrophysics group, Blackett Laboratory, Prince Consort Road, London, SW7 2AZ, U.K.\\
\and
Infrared Processing and Analysis Center, California Institute of Technology, Pasadena, CA 91125, U.S.A.\\
\and
Institut N\'{e}el, CNRS, Universit\'{e} Joseph Fourier Grenoble I, 25 rue des Martyrs, Grenoble, France\\
\and
Institut Universitaire de France, 103, bd Saint-Michel, 75005, Paris, France\\
\and
Institut d'Astrophysique Spatiale, CNRS (UMR8617) Universit\'{e} Paris-Sud 11, B\^{a}timent 121, Orsay, France\\
\and
Institut d'Astrophysique de Paris, CNRS (UMR7095), 98 bis Boulevard Arago, F-75014, Paris, France\\
\and
Institute for Space Sciences, Bucharest-Magurale, Romania\\
\and
Institute of Astronomy and Astrophysics, Academia Sinica, Taipei, Taiwan\\
\and
Institute of Astronomy, University of Cambridge, Madingley Road, Cambridge CB3 0HA, U.K.\\
\and
Institute of Theoretical Astrophysics, University of Oslo, Blindern, Oslo, Norway\\
\and
Instituto de Astrof\'{\i}sica de Canarias, C/V\'{\i}a L\'{a}ctea s/n, La Laguna, Tenerife, Spain\\
\and
Instituto de F\'{\i}sica de Cantabria (CSIC-Universidad de Cantabria), Avda. de los Castros s/n, Santander, Spain\\
\and
Jet Propulsion Laboratory, California Institute of Technology, 4800 Oak Grove Drive, Pasadena, California, U.S.A.\\
\and
Jodrell Bank Centre for Astrophysics, Alan Turing Building, School of Physics and Astronomy, The University of Manchester, Oxford Road, Manchester, M13 9PL, U.K.\\
\and
Kavli Institute for Cosmology Cambridge, Madingley Road, Cambridge, CB3 0HA, U.K.\\
\and
LAL, Universit\'{e} Paris-Sud, CNRS/IN2P3, Orsay, France\\
\and
LERMA, CNRS, Observatoire de Paris, 61 Avenue de l'Observatoire, Paris, France\\
\and
Laboratoire AIM, IRFU/Service d'Astrophysique - CEA/DSM - CNRS - Universit\'{e} Paris Diderot, B\^{a}t. 709, CEA-Saclay, F-91191 Gif-sur-Yvette Cedex, France\\
\and
Laboratoire Traitement et Communication de l'Information, CNRS (UMR 5141) and T\'{e}l\'{e}com ParisTech, 46 rue Barrault F-75634 Paris Cedex 13, France\\
\and
Laboratoire de Physique Subatomique et de Cosmologie, Universit\'{e} Joseph Fourier Grenoble I, CNRS/IN2P3, Institut National Polytechnique de Grenoble, 53 rue des Martyrs, 38026 Grenoble cedex, France\\
\and
Laboratoire de Physique Th\'{e}orique, Universit\'{e} Paris-Sud 11 \& CNRS, B\^{a}timent 210, 91405 Orsay, France\\
\and
Lawrence Berkeley National Laboratory, Berkeley, California, U.S.A.\\
\and
Max-Planck-Institut f\"{u}r Astrophysik, Karl-Schwarzschild-Str. 1, 85741 Garching, Germany\\
\and
McGill Physics, Ernest Rutherford Physics Building, McGill University, 3600 rue University, Montr\'{e}al, QC, H3A 2T8, Canada\\
\and
MilliLab, VTT Technical Research Centre of Finland, Tietotie 3, Espoo, Finland\\
\and
National University of Ireland, Department of Experimental Physics, Maynooth, Co. Kildare, Ireland\\
\and
Niels Bohr Institute, Blegdamsvej 17, Copenhagen, Denmark\\
\and
Observational Cosmology, Mail Stop 367-17, California Institute of Technology, Pasadena, CA, 91125, U.S.A.\\
\and
Optical Science Laboratory, University College London, Gower Street, London, U.K.\\
\and
Princeton University Observatory, Peyton Hall, Princeton, NJ 08544-1001, U.S.A.\\
\and
SB-ITP-LPPC, EPFL, CH-1015, Lausanne, Switzerland\\
\and
SISSA, Astrophysics Sector, via Bonomea 265, 34136, Trieste, Italy\\
\and
School of Physics and Astronomy, Cardiff University, Queens Buildings, The Parade, Cardiff, CF24 3AA, U.K.\\
\and
Space Research Institute (IKI), Russian Academy of Sciences, Profsoyuznaya Str, 84/32, Moscow, 117997, Russia\\
\and
Space Sciences Laboratory, University of California, Berkeley, California, U.S.A.\\
\and
Special Astrophysical Observatory, Russian Academy of Sciences, Nizhnij Arkhyz, Zelenchukskiy region, Karachai-Cherkessian Republic, 369167, Russia\\
\and
Stanford University, Dept of Physics, Varian Physics Bldg, 382 Via Pueblo Mall, Stanford, California, U.S.A.\\
\and
Sub-Department of Astrophysics, University of Oxford, Keble Road, Oxford OX1 3RH, U.K.\\
% ...
}

\abstract{ This paper presents an all-sky model of dust emission from
the \Planck\ 353, 545, and 857\,GHz, and \IRAS\ 100\,$\mu$m data. Using
a modified blackbody fit to the data we present all-sky maps of the
dust optical depth, temperature, and spectral index over the
353--3000\,GHz range.  This model is a good representation of the
\IRAS\ and \Planck\ data at 5\arcm\ between 353 and 3000\,GHz
(850 and 100\,$\mu$m).
It shows variations of the order of 30\,\% compared with the widely-used
model of Finkbeiner, Davis, and Schlegel.  The \Planck\ data allow us
to estimate the dust temperature uniformly over the whole sky, down to
an angular resolution of 5\arcm, providing an improved estimate of the
dust optical depth compared to previous all-sky dust model, especially
in high-contrast molecular regions where the dust temperature varies
strongly at small scales in response to dust evolution, extinction,
and/or local production of heating photons.
An increase of the dust opacity at 353\,GHz, $\tau_{353}/N_{\rm H}$,
from the diffuse to the denser interstellar medium (ISM) is reported.
It is associated with a decrease in the observed dust temperature,
$T_{\rm obs}$, that could be due at least in part to the increased
dust opacity.  We also report an excess of dust emission at \hi\
column densities lower than $10^{20}$\,cm$^{-2}$ that could be the
signature of dust in the warm ionized medium.  In the diffuse ISM at
high Galactic latitude, we report an anticorrelation between
$\tau_{353}/N_{\rm H}$ and $T_{\rm obs}$ while the dust specific
luminosity, i.e., the total dust emission integrated over frequency
(the radiance) per hydrogen atom, stays about constant, confirming
one of the \Planck\ Early Results obtained on selected fields.  This
effect is compatible with the view that, in the diffuse ISM, $T_{\rm obs}$ 
responds to spatial variations of the dust opacity,
due to variations of dust properties,
in addition to (small) variations of the radiation field strength.
The implication is that in the diffuse high-latitude ISM $\tau_{353}$
is not as reliable a tracer of dust column density as we conclude it
is in molecular clouds where the correlation of $\tau_{353}$ with dust
extinction estimated using colour excess measurements on stars is strong. 
To estimate Galactic \Ebv\ in extragalactic fields at high latitude we
develop a new method based on the thermal dust radiance, instead of
the dust optical depth, calibrated to \Ebv\ using reddening
measurements of quasars deduced from Sloan Digital Sky Survey data. }

\keywords{methods: data analysis -- ISM: general -- ISM: dust, extinction -- submillimetre: ISM -- infrared: ISM -- opacity}

\titlerunning{All-sky model of thermal dust emission}

\authorrunning{Planck collaboration}

\maketitle

\section{Introduction}\label{sec:introduction}

This paper, one of a set associated with the 2013 release of data from
the \Planck\footnote{\Planck\ (\url{http://www.esa.int/Planck}) is a
  project of the European Space Agency (ESA) with instruments provided
  by two scientific consortia funded by ESA member states (in
  particular the lead countries France and Italy), with contributions
  from NASA (USA) and telescope reflectors provided by a collaboration
  between ESA and a scientific consortium led and funded by Denmark.}
mission \citep{planck2013-p01}, presents a new parametrization of dust
emission that covers the whole sky, at 5\arcm\ resolution, based on
data from 353 to 3000\,GHz (100 to 850\,$\mu$m).

Because it is well mixed with the gas and because of its direct
reaction to UV photons from stars, dust is a great tracer of
the interstellar medium (ISM) and of star formation activity.  On the other hand,
for many studies in extragalactic astrophysics and cosmology, Galactic
interstellar dust is a nuisance, a source of extinction and reddening
for UV to near-infrared observations and a contaminating emission in
the infrared to millimetre wavelengths.
Thanks to the sensitivity, spectral coverage, and angular resolution
of \Planck, this model of dust emission brings new constraints on the
dust spectral energy distribution (SED), on its variations across the sky,
and on the relationships between dust emission, dust extinction, and
gas column density. In particular, this model of dust emission
provides a new map of dust extinction at 5\arcm\ resolution, aimed at
helping extragalactic studies.

The emission in the submillimetre  range arises from the
bigger dust grains that are in thermal equilibrium with the ambient
radiation field. Thermal dust emission is influenced by a combination
of the dust column density, radiation field strength, and dust
properties (size distribution, chemical composition, and the grain
structure). When the effect of the radiation field can be estimated
(using the dust temperature as a probe) and the dust properties
assumed, the dust optical depth is possibly the most reliable tracer
of interstellar column density, and therefore of mass for objects at
known distances. Dust optical depth is used to estimate the mass of
interstellar clumps and cores \citep{ossenkopf1994} in particular with
the higher resolution \Herschel\ data \citep{launhardt2013}, 
to study the statistical properties of the ISM structure and its link
with gravity, interstellar turbulence, and stellar feedback
\citep{peretto2012,kainulainen2013}, and
as a way to sample the mass of the ISM in general
\citep{planck2011-7.0}. The accuracy of these
determinations depends on, among other things, the frequency range
over which the dust spectrum is observed. The combination of \Planck\
and \IRAS\ data offers a new view on interstellar dust by allowing us
to sample the dust spectrum from the Wien to the Rayleigh-Jeans sides,
at 5\arcm\ resolution over the whole sky.

Dust emission, with extinction and polarization, is a key element to
constrain the properties of interstellar dust
\citep{draine2007a,compiegne2011}. The dust emissivity (i.e., the
amount of emission per unit of gas column density) and the shape of
the dust SED provide information on
the nature of the dust particles, in particular their structure,
composition, and abundance, related to the dust-to-gas ratio. 

Changes in the dust emissivity and the shape of the dust SED can be
related to dust evolutionary processes. Interstellar dust grains are
thought to be the seeds from which larger particles form in the ISM,
up to planetesimals in circumstellar environments
\citep{brauer2007,beckwith2000,birnstiel2012}. This growth of solids
can be followed in earlier phases of the star-formation process, at
the protostellar phase and even before, in molecular clouds and in
the diffuse ISM. Many studies have revealed increases of the dust
emissivity with (column) density in molecular clouds accompanied by a
decrease in dust temperature
\citep{stepnik2003,schnee2008,planck2011-7.13,arab2012,martin2012,roy2013}.
One explanation is that grain structure is changing through
aggregation of smaller particles, enhancing the opacity
\citep{kohler2011}. \Planck's spectral coverage allows
us to model the big grain thermal emission, in particular its spectral
index that is related to the grain composition and structure
\citep{ormel2011,meny2007,kohler2012}. Because of its full-sky
coverage, \Planck\ can also reveal variations of the dust SED with
environment, enabling us to better understand the evolutionary track
of dust grains through the ISM phases.

Dust emission is one of the major foregrounds hampering the study of
the cosmic microwave background (CMB). The thermal dust emission
peaks at a frequency close to 2000\,GHz but its emission is still a
fair fraction of the CMB anisotropies in the 20--200\,GHz range where
they are measured. This is even more the case in polarization
\citep{miville-deschenes2011}. The model of dust emission proposed by
\citet{finkbeiner1999} based on data from previous satellite missions
(\IRAS\ and \COBE) made an important contribution to the field, in
guiding the design of CMB experiments and in helping the data analysis
by providing a spatial template and a spectral dependence of the dust
emission at CMB frequencies. It is still the basis of recent models of
Galactic foreground emission \citep{delabrouille2013}. With its
frequency coverage that bridges the gap between \IRAS\ and the CMB
range, its high sensitivity, and its better angular resolution,
\Planck\ offers the opportunity to develop a new model of thermal dust
emission.

Estimating reddening and extinction by foreground interstellar dust is
a major issue for observations of extragalactic objects in the UV to
near-infrared range. Major efforts have been made toward producing sky
maps that provide a way to correct for the chromatic extinction of
light by Galactic interstellar dust on any line of sight. First
\citet{burstein1978} used \hi\ as a proxy for dust extinction by
correlating integrated 21\,cm line emission with extinction estimated
from galaxy counts. It was subsequently discovered that \hi\ is not a
reliable tracer of total column density $N_{\rm H}$ for $N_\ion{H}{i}$
greater than a few $10^{20}$\,cm$^{-2}$ due to molecular gas
contributions
\citep{lebrun1982,boulanger1988,desert1988,heiles1988a,blitz1990,reach1994,boulanger1996}.
It was then proposed to use dust emission as a more direct way to
estimate dust extinction. By combining 100, 140, and 240\,$\mu$m data
(\DIRBE\ and \IRAS) \citet{schlegel1998} produced an all-sky map of
dust optical depth at 100\,$\mu$m that was then calibrated into dust
reddening by correlating with colour excesses measured for galaxies. The
work presented here is the direct continuation of these studies. Like
\citet{schlegel1998} we also propose a map of \Ebv\ based on a model
of dust emission calibrated using colour excess measurements of
extragalactic objects, here quasars.

The paper is organized as follows. 
The data used and the preprocessing steps are presented in
Sect.~\ref{sec:data}. 
The model of the dust emission, SED fit methodology, the exploration of potential biases, and the
all-sky maps of dust parameters are described in
Sect.~\ref{sec:mbbfit}. 
The results of the Galactic dust model are analysed in
Sect.~\ref{sec:discussion}.
Sections~\ref{sec:column_density} and \ref{sec:extinction} describe specifically how the dust
emission model compares to other tracers of column density. 
The \Planck\ dust products, the dust model maps, and the \Ebv\ map
aimed at helping extragalactic studies to estimate Galactic
extinction are detailed in Sect.~\ref{sec:products} and compared with
similar products in the literature.
Concluding remarks are given in Sect.~\ref{sec:conclusion}.


\begin{table*}[!ht]
\caption{\label{tab:data} Properties of the \IRAS\ and \Planck\ 
maps from which ZE has been removed. }
\tabskip=0pt
\begin{center}
\begin{tabular}{cc D{.}{.}{1.2} c c D{.}{.}{2.1}}\specialrule{\lightrulewidth}{0pt}{0pt} \specialrule{\lightrulewidth}{1.5pt}{\belowrulesep}
$\nu$ & $\lambda$ & \multicolumn{1}{c}{FWHM} & Offset & Dipole & \multicolumn{1}{c}{$c_\nu$} \\
\,[GHz] & [$\mu$m] & \multicolumn{1}{c}{[arcmin]} & [MJy\,sr$^{-1}$] & [MJy\,sr$^{-1}$] & \multicolumn{1}{c}{[\%]}\\ \midrule
3000 & 100 & 4.3 & $-0.174 \pm 0.005$ & $\:\:\,-$ & 13.6\\
$\:\:\:857$ & 350 & 4.63 & $\:\:\,0.093 \pm 0.009$ & $\:\:\,-$ & 10.0\\
$\:\:\:545$ & 550 & 4.84 & $\:\:\,0.095 \pm 0.014$ & $\:\:\,0.0148 \pm 0.0001$ & 10.0\\
$\:\:\:353$ & 850 & 4.86 & $\:\:\,0.085 \pm 0.006$ & $-0.0089 \pm 0.0001$ & 1.2\\ \bottomrule[\lightrulewidth]
\end{tabular}
\end{center}
{\bf Note:} Column 1: frequency. Column
  2: wavelength. Column 3: angular resolution \citep[see][]{planck2013-p03c}. Column 4: offset (and
  its uncertainty $\delta O_\nu$) removed from the maps to adjust them
  to a coherent Galactic zero level \citep[see][]{planck2013-p03f}.
  Column 5: amplitude of the residual dipole removed. The residual
  dipole removed at 353 and 545\,GHz is oriented toward
 $l = 263\fdg99$, $b = 48\fdg26$, the direction of the solar dipole
  estimated using \WMAP\ data \citep{hinshaw2009}. Column 6: 
  calibration uncertainty.
\end{table*}

\section{Data and preprocessing}

\label{sec:data}

The analysis presented here relies on the combination of the
\Planck\ data from the HFI instrument at 857, 545, and 353\,GHz (respectively 350, 550, and
850\,$\mu$m) with the \IRAS\ 100\,$\mu$m (3000\,GHz) data.

\subsection{\Planck\ data}

\label{sec:planckdata}

For \Planck\ we used the HFI 2013 delivery maps \citep{planck2013-p03}, corrected for zodiacal
emission \citep[ZE -- see][]{planck2013-pip88}. Each map was
smoothed to a common resolution of 5\arcm, assuming a Gaussian
beam.\footnote{Each map was smoothed using a Gaussian beam of FWHM,
  $f_s$, that complements the native FWHM, $f_i$
  (Table~\ref{tab:data}), of the map to bring it to 5\arcm: $f_s =
  \sqrt{5^2 - f_i^2}$.}  
The 353\,GHz map, natively built in units of
$K_{\rm CMB}$, was transformed to MJy~sr$^{-1}$ using the conversion
factor given by \citet{planck2013-p03d}. The CMB anisotropies map
provided by the {\tt SMICA} algorithm \citep{planck2013-p06}, which
has an angular resolution of 5\arcm, was removed from each \Planck\ HFI
map.

As shown in \citet{planck2013-p03a}, $^{12}$CO and $^{13}$CO
rotational lines fall in each of the \Planck\ HFI filters, except at
143\,GHz. At 857 and 545\,GHz the CO lines ($J$=5$\rightarrow$4 and
$J$=4$\rightarrow$3, respectively) are very faint compared to the
dust emission and they are not considered here. On the other hand,
emission from the $^{12}$CO $J$=3$\rightarrow$2 line was detected in
the 353\,GHz band \citep{planck2013-p03a}. Nevertheless, this emission
is still faint compared to the dust emission whereas the noise on the
\Planck\ CO emission estimate in the 353\,GHz band is quite high (see
\citet{planck2013-p03a} for details). The detection of the $^{12}$CO
$J$=3$\rightarrow$2 line emission by \Planck\ is above $3\sigma$ for
only 2.6\,\% of the sky. When detected above $3\sigma$, this emission
is on average 2\,\% of the 353\,GHz specific intensity. It contributes
5\,\% or more of the 353\,GHz specific intensity for only 0.3\,\% of
the sky. Given such a relatively small contribution we did not
subtract CO emission from the data so as not to compromise the
353\,GHz map through the adverse impact of the noise of the $^{12}$CO
$J$=3$\rightarrow$2 product.

\begin{figure*}
\centering
\includegraphics[draft=false, angle=0]{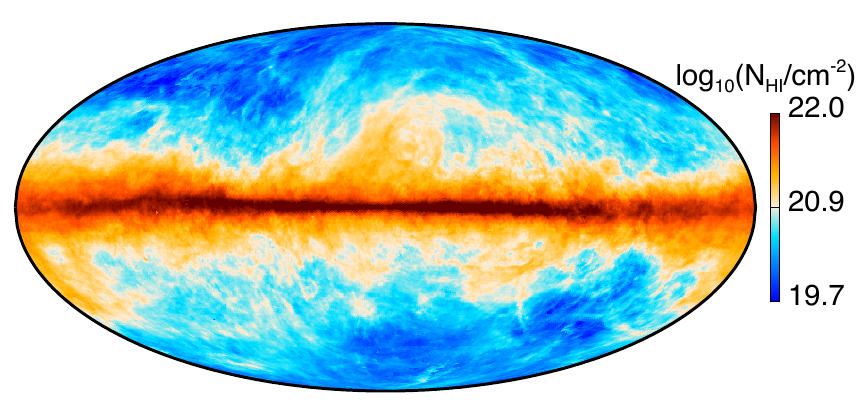}
\includegraphics[draft=false, angle=0]{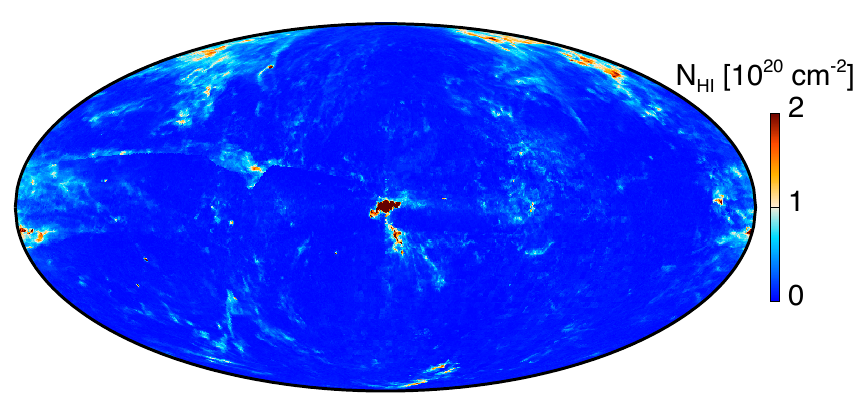}
\caption{\label{fig:hi_maskhi} All-sky Mollweide projections of
  \hi\ maps used in the determination of the offsets.  The centre of
  the map is toward the Galactic centre.  \emph{Left:} \hi\
  column density of low velocity clouds (LVC). \emph{Right:}
  intermediate velocity clouds (IVC). See text.}
\end{figure*}

\begin{figure*}
\centering
\includegraphics[draft=false, angle=0]{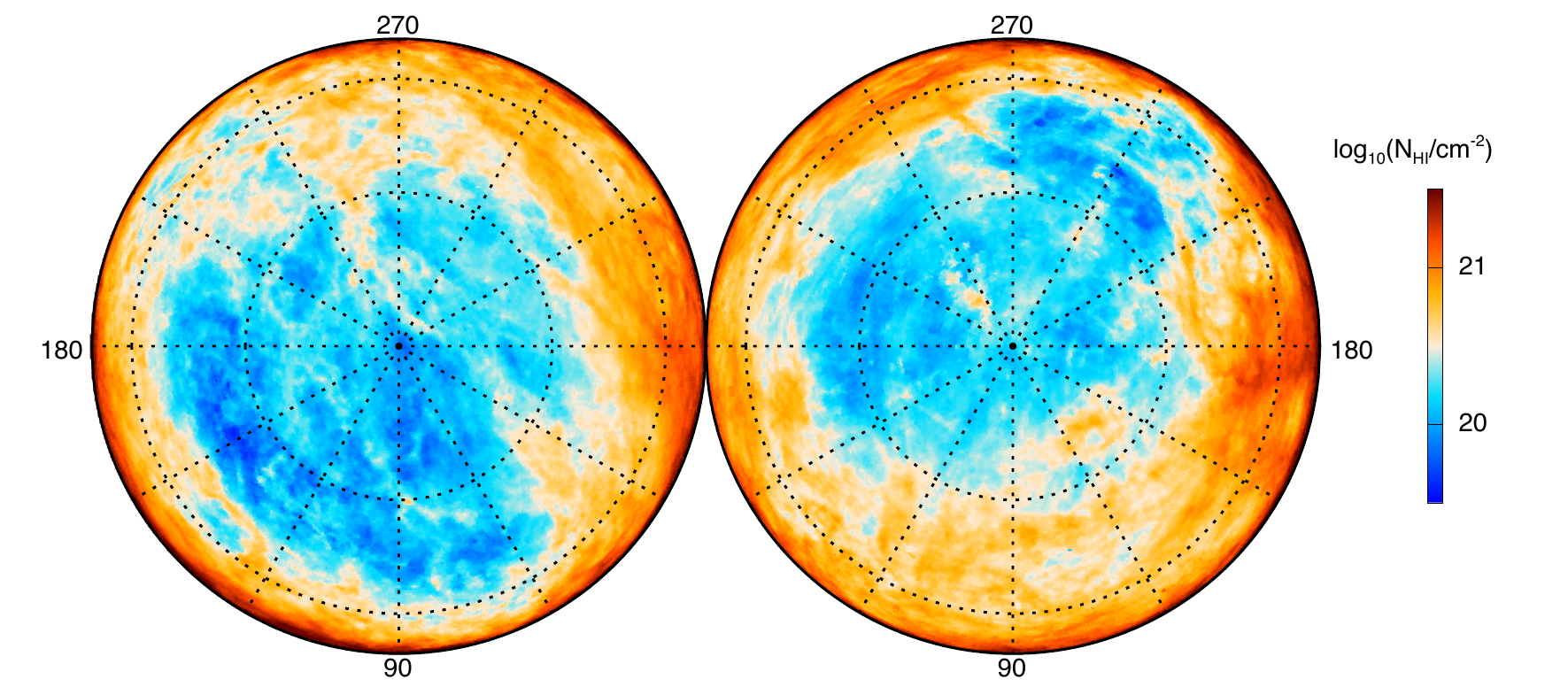}
\includegraphics[draft=false, angle=0]{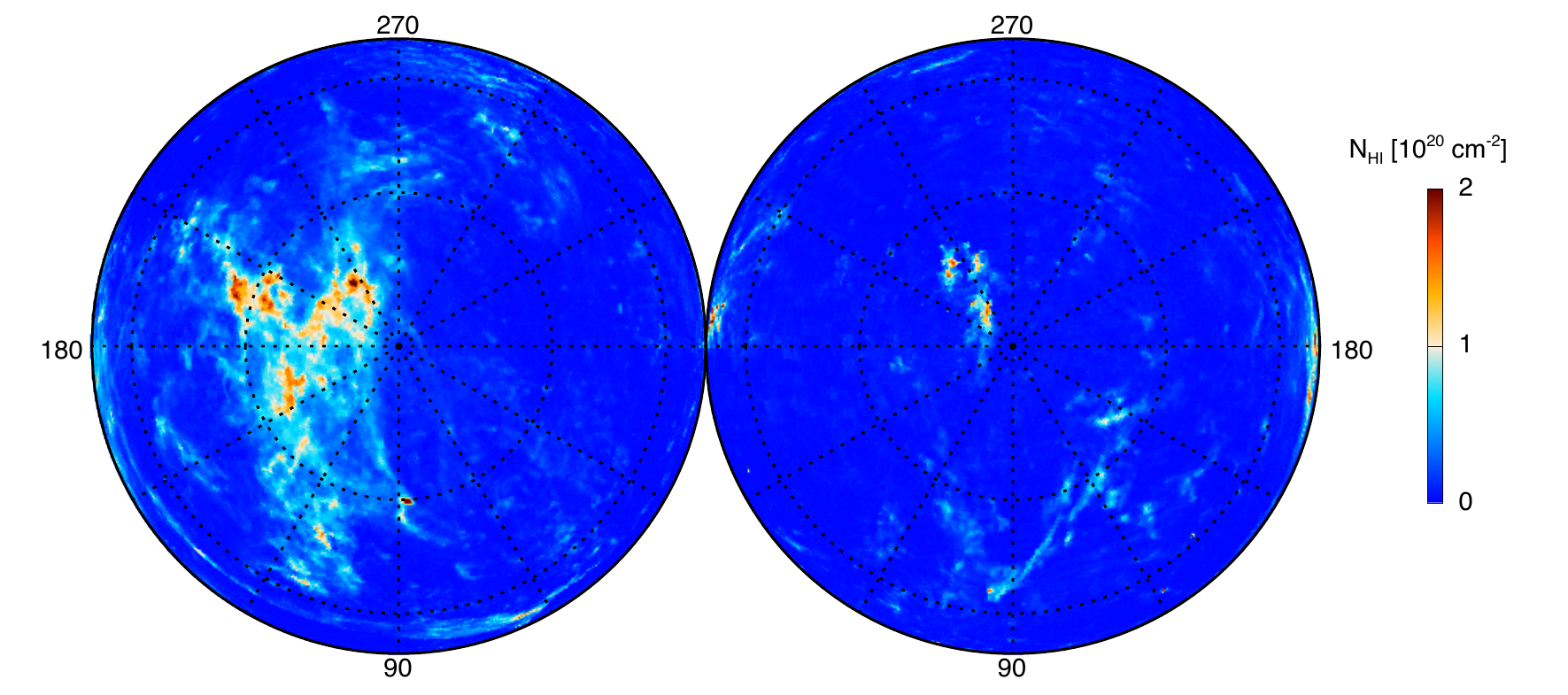}
\caption{\label{fig:fullres_polar_mapsm1} Polar orthographic
  projections of the LVC map (\textit{upper}) and the IVC map (\textit{lower}) shown in
  Fig.~\ref{fig:hi_maskhi}.  The left (right) panel is centred on the
  north (south) Galactic pole. Longitude increases clockwise
  (anticlockwise), with the two panels joining at $l=0$\deg\ and $b =
  0$\deg. Dotted lines representing constant longitude and latitude are
  spaced by 30\deg.  The radius from
  the pole is $\propto \cos(b)$, so this projection emphasizes features
  at high latitude.}
\end{figure*}

\begin{figure*}
\centering
\includegraphics[draft=false, angle=0]{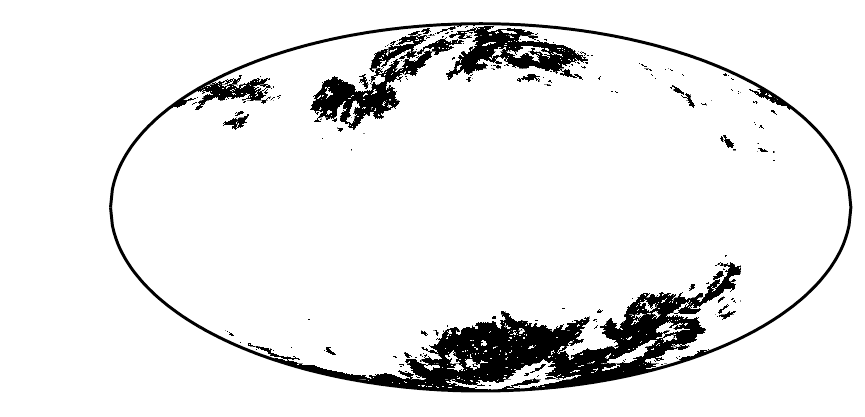}
\includegraphics[draft=false, angle=0]{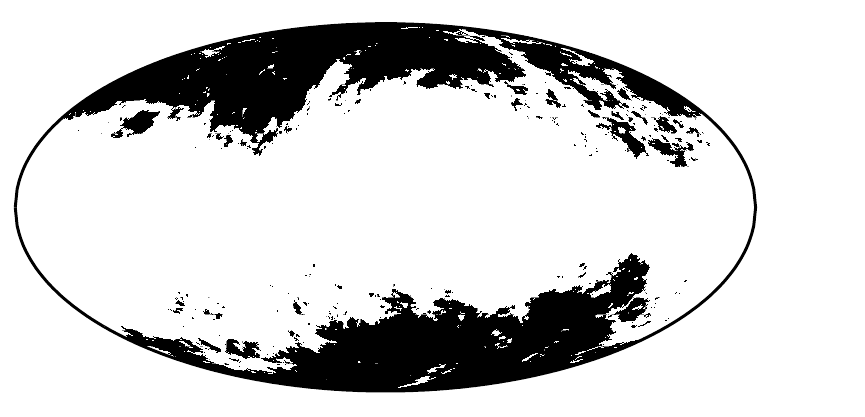}
\caption{\label{fig:hi_mask} Masks used to estimate the zero levels of
  the maps. \emph{Left:} ``low $N_\ion{H}{i}$ mask'' including pixels of
  the sky where the LVC column density is $<2\times10^{20}$\,cm$^{-2}$
  and the IVC column density is below
  $<0.1\times10^{20}$\,cm$^{-2}$. \emph{Right:} mask where the total
  \hi\ column density (LVC plus IVC) is lower than
  $3\times10^{20}$\,cm$^{-2}$.}
\end{figure*}

\begin{figure*}
\centering
\includegraphics[draft=false, angle=0]{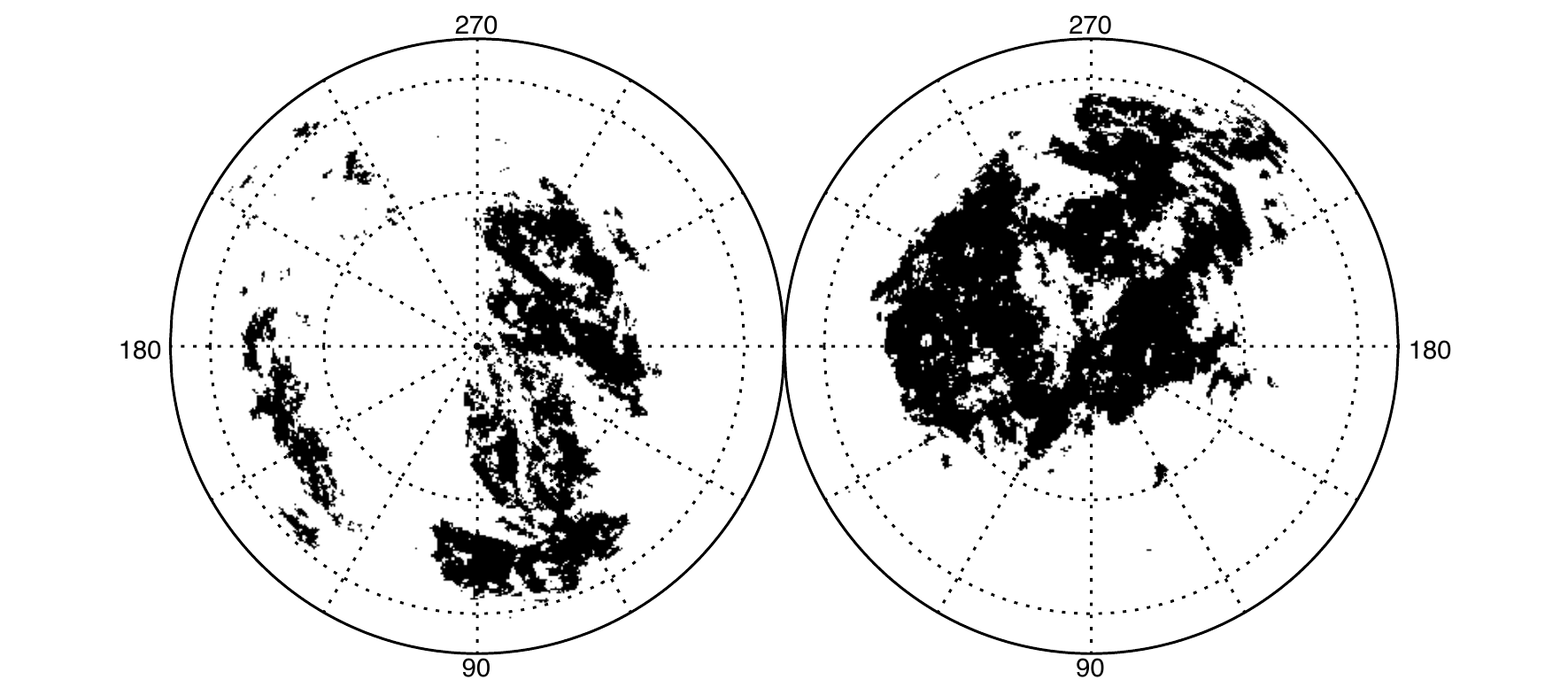}
\caption{\label{fig:fullres_polar_maps0} Polar orthographic
  projection of the low $N_\ion{H}{i}$ mask shown in Fig.~\ref{fig:hi_mask}
  (left).  }
\end{figure*}

\subsection{The 100\,$\mu$m map}
\label{sec:100um}

The 100\,$\mu$m map used in this analysis is a combination of the
\IRIS\ map \citep{miville-deschenes2005a} and the map of
\citet[hereafter SFD]{schlegel1998}, both projected on the
{\HLP}\footnote{\url{http://healpix.sourceforge.net}} grid
\citep{gorski2005} at $N_{\rm side}=2048$. Both \IRIS\ and SFD maps
were built by combining \IRAS\ and \DIRBE\ 100\,$\mu$m data.
Nevertheless these two maps show differences at large scales due to
the different assumptions used for the ZE removal. \citet{miville-deschenes2005a} used
the \DIRBE\ 100\,$\mu$m map from which ZE was removed by the \DIRBE\
team, using the model of \citet{kelsall1998}. On the other hand,
\citet{schlegel1998} used their own empirical approach to remove ZE
based on a scaling of the \DIRBE\ 25\,$\mu$m data. Because it is based
on data and not on a model, the SFD correction is closer to the
complex structure of the ZE and provides a better result. This can be
assessed by looking at the correlation of the \IRIS\ and SFD maps with
\hi\ in the diffuse areas of the sky ($1<N_\ion{H}{i}<2 \times
10^{20}$\,cm$^{-2}$),
as detailed in Appendix~\ref{sec:rmzody}.
The uncertainty of the slope of the correlation
with $N_\ion{H}{i}$ and the standard deviation of the residual is about
30\,\% lower for the SFD map compared to the \IRIS\ map. For that
reason (and others described in Appendix~\ref{sec:rmzody}) we favour the
use of the SFD map at large scales.

At scales smaller than 30\arcm, the \IRIS\ map has several advantages
over the SFD map.\footnote{It is at 30\arcm\ that both maps match in
  power -- see Fig.~15 of \citet{miville-deschenes2005a}. This scale
  is close to the resolution of the \DIRBE\ data (42\arcm) that were
  used in both products to set the large-scale emission.}
The \IRIS\
map is at the original angular resolution (4\farcm3) of the \IRAS\
data while SFD smoothed the map to 6\farcm1. \IRIS\ also benefits from
a non-linear gain correction that is coherent for point sources and
diffuse emission. Finally point sources were kept in the \IRIS\ map
whereas SFD removed some of them (mostly galaxies but also ISM
clumps). To combine the advantages of the two maps, we built a
100\,$\mu$m map, $I_{100}$, that is compatible with SFD at scales
larger than 30\arcm\ and compatible with \IRIS\ at smaller scales:
\begin{equation}
\label{eq:100micron}
I_{100} = I_{\rm IRIS} - I_{\rm IRIS} \otimes f_{\rm IRIS}^{30} + I_{\rm SFD} \otimes f_{\rm SFD}^{30}\,,
\end{equation}
where $I_{\rm IRIS}$ and $I_{\rm SFD}$ are, respectively, the \IRIS\ and
the SFD maps, and $f_{i}^{30}$ is the complementary Gaussian kernel needed
to bring the maps to 30\arcm\ resolution.

\begin{figure*}
\centering
\includegraphics[draft=false, angle=0]{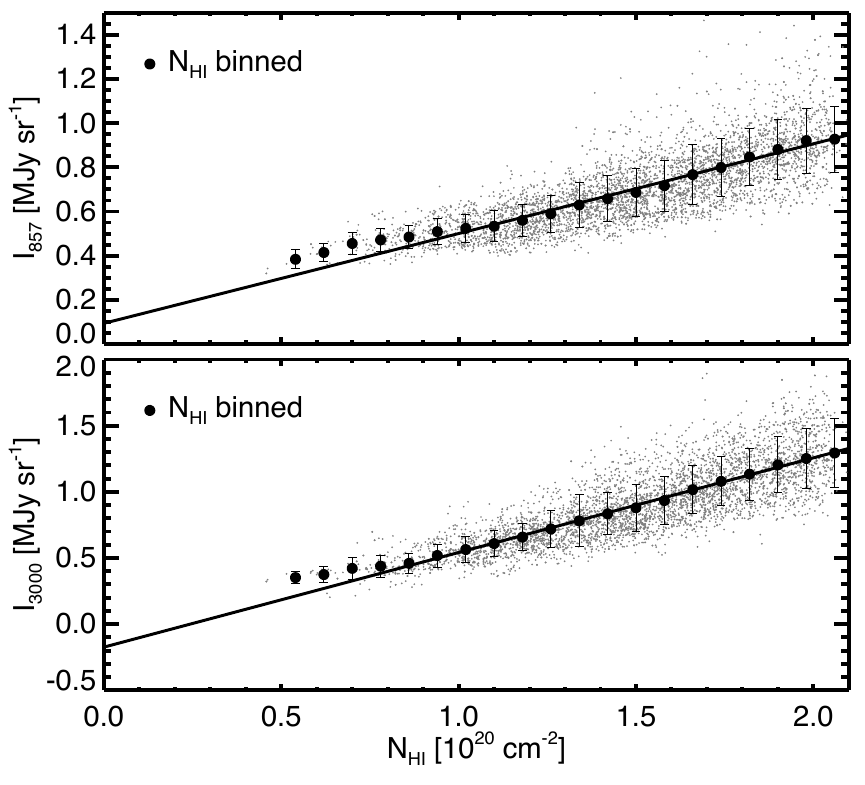}
\includegraphics[draft=false, angle=0]{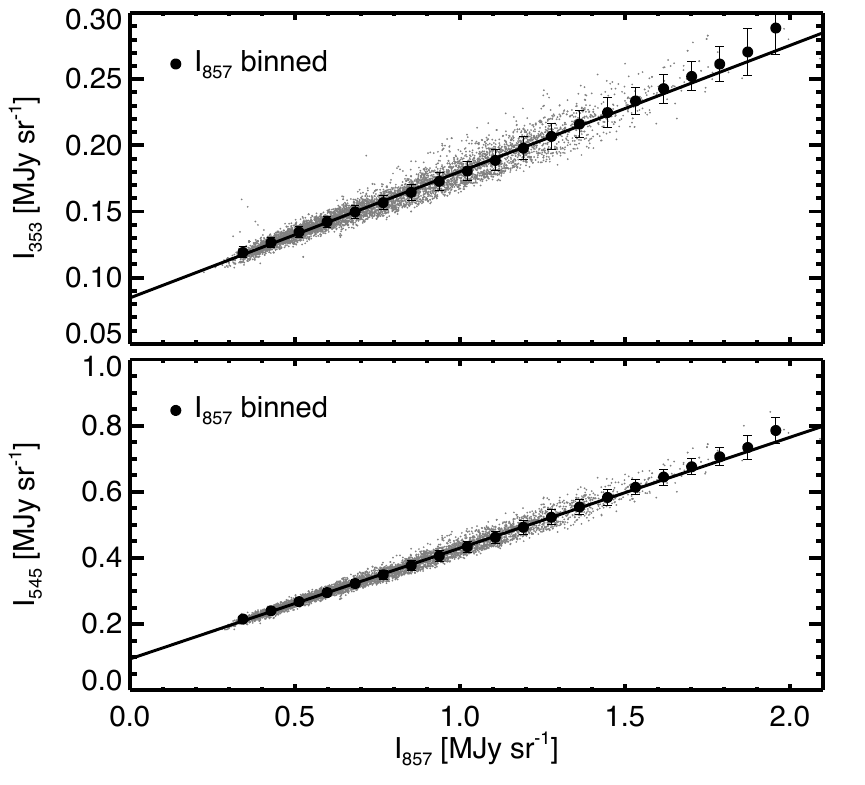}
\caption{\label{fig:offset} Correlation plots used to estimate the
Galactic zero levels of the \IRAS\ and \Planck\ maps (ZE subtracted).
\emph{Left:} Correlation of 857 and 3000\,GHz vs.\ \hi\ column density
obtained on the $N_\ion{H}{i}<2\times10^{20}$\,cm$^{-2}$ mask
(Fig.~\ref{fig:hi_mask} bottom left). \emph{Right:} Correlation of 353
and 545\,GHz vs.\ 857\,GHz obtained on the $N_\ion{H}{i}
<3\times10^{20}$\,cm$^{-2}$ mask (Fig.~\ref{fig:hi_mask} bottom
right). All maps in the analysis were smoothed to a common resolution
of 1\deg. The black circles and the associated bars are the average
and standard deviation of $I_\nu$ in bins of $N_\ion{H}{i}$ (\textit{left}) and
$I_{857}$ (\textit{right}).}
\end{figure*}

\subsection{Zero level}

The fit of the dust emission requires that the specific intensity at
each frequency and at each sky position is free of any other emission.
In particular the zero level of each map should be set in such a way
that it contains only Galactic dust emission. In order to set the zero
level of the maps to a meaningful Galactic reference we applied a
method based on a correlation with \hi, as described in
\citet{planck2013-p03f}.  

Some precautions need to be taken here because the ratio of dust to
\hi\ emission might vary locally due to variations of the radiation
field or of the dust optical properties. Locally the 21\,cm emission
might not be a perfect tracer of the column density due to \hi\
self-absorption effects or to the presence of ionized or molecular
gas. Nevertheless, the correlation between dust and \hi\ emission is
known to be tight in the diffuse ISM where most of the gas is
atomic. This correlation has been used several times to establish the
dust SED \citep{boulanger1996,planck2011-7.12}, to isolate
the cosmic infrared background 
\citep[CIB;][]{puget1996,planck2011-6.6,penin2012}, and to
establish a Galactic reference for dust maps
\citep{burstein1978,schlegel1998}.

The excess of dust emission with respect to the \hi\ correlation has
been used to reveal gas in molecular form, even in regions where CO
emission was not detected
\citep{desert1988,blitz1990,reach1998a,planck2011-7.0}.
Such an excess can be observed at column densities as low as $N_\ion{H}{i}
  = 2\times10^{20}$\,cm$^{-2}$.  Using this as an upper limit for
our correlation studies also ensures that self-absorption in the 21\,cm
line emission is not important.  
Note that this is also below the threshold at which significant H$_2$
is seen in the diffuse ISM
\citep{gillmon2006,wakker2006,rachford2002,rachford2009}.

To estimate the Galactic reference of the \IRAS\ and \Planck\
data, the maps were correlated against the 21\,cm LAB data
\citep{kalberla2005}, integrated in velocity.  The ranges are referred
to as LVC, low velocity gas with $|v_{\rm LSR}| < 35$\,km\,s$^{-1}$,
and IVC, intermediate velocity gas with $35< |v_{\rm LSR}| <
70$\,km\,s$^{-1}$ \citep{albert2004}.  HVC, high-velocity clouds with
$|v_{\rm LSR}| > 70$\,km\,s$^{-1}$ are excluded.
For LVC and IVC separately, column density maps $N_\ion{H}{i}$ assuming
optically thin emission are given in Fig.~\ref{fig:hi_maskhi} in the
all-sky Mollweide equal-area projection
and in Fig.~\ref{fig:fullres_polar_mapsm1} in a complementary
polar orthographic projection.\footnote{Each of these projections
  \citep{calabretta2002} covers the whole sky, but here we use
  the terminology ``all-sky'' and ``polar'' as shorthand for the two
  projections.  The polar view is most useful for the high-latitude
  sky, whereas the all-sky view is best for intermediate to low
  latitudes.}

For the correlation, all data sets were convolved to 1\deg\ resolution
and projected on an $N_{\rm side}=128$ grid.  The correlation was
performed using only pixels with the LVC $N_\ion{H}{i} <
2\times10^{20}$\,cm$^{-2}$, discarding pixels with detected IVC above
$0.1\times10^{20}$\,cm$^{-2}$.  
The resulting area covers 11.5\,\% of the sky, corresponding to more
than 4700\,deg$^2$.  This mask is presented in both the
all-sky view in Fig.~\ref{fig:hi_mask} (left panel) and the
polar view in Fig.~\ref{fig:fullres_polar_maps0}.  We refer to
this throughout as the ``low $N_\ion{H}{i}$ mask.''

The correlations are shown in Fig.~\ref{fig:offset}, left.
The values of the offset that were removed from the 3000 and 857\,GHz
maps are given in Table~\ref{tab:data}. We have checked that the
offsets estimated in that way are not sensitive to the resolution of the
\hi\ data or to the area of the sky selected. 
We have also checked that the assumption that the  21\,cm emission is optically thin does 
not introduce a significant bias in the analysis. For example, on assuming $T_{\rm spin}=80$\,K 
in converting 21\,cm emission to $N_\ion{H}{i}$ \citep{lockman2005}, 
the changes in the offsets at 857 and 3000\,GHz are only about 0.02\,MJy\,sr$^{-1}$. 
This is as expected because in the LAB data for the diffuse areas of the sky considered here 
values of the  21\,cm line brightness temperature higher than $10$\,K
are exceptional.
Compatible offsets, within the quoted uncertainties, were found using 16\arcm\ Galactic
All Sky Survey 21\,cm data \citep{mcclure-griffiths2009} of the area
around the Galactic south pole \citep{planck2013-XVII} and
9\arcm\ data obtained on smaller regions in the northern sky at the
Green Bank Telescope \citep{planck2011-7.12}. 
Finally, in Fig.~\ref{fig:offset} we note a systematic excess of the
dust emission at 857 and 3000\,GHz with respect to the correlation at
the lowest $N_\ion{H}{i}$.  This is also seen at 545 and 353\,GHz and it
is discussed further in Sect.~\ref{sec:WIM}.

The correlation of dust emission between \Planck\ frequencies is
observed to be tight (the correlation of \Planck\ is less tight with
the \IRAS\ 3000\,GHz map). We took advantage of this to estimate more
precisely the Galactic zero level of the 353 and 545\,GHz channels.
They were obtained by correlation with the 857\,GHz map on a larger mask
with total LVC plus IVC $N_\ion{H}{i} < 3\times10^{20}$\,cm$^{-2}$
(Fig.~\ref{fig:hi_mask}, right panel).
These correlations are shown in Fig.~\ref{fig:offset} (right).
The offset values obtained in this way (see Table~\ref{tab:data}) are
compatible within $1\sigma$ with the offset values deduced from the
\hi\ correlation in the smaller mask (0.104 and 0.088\,MJy\,sr$^{-1}$
respectively at 545 and 353\,GHz). At these frequencies we favour the
offset obtained with the correlation with 857\,GHz as it minimizes the
aforementioned effects in the correlation between dust and gas
emission. 

In the process, faint dipole residuals were identified in the 353 and
545\,GHz maps. The orientation of these residual dipoles coincides
with the solar dipole and their amplitudes (see Table~\ref{tab:data})
corresponds to $+7.6$\,\% and $-0.9$\,\% of the solar dipole amplitude
at 545 and 353\,GHz respectively, which is within the calibration
uncertainties at these frequencies. They were removed prior to the
dust SED fit.


\section{Model of the dust emission}
\label{sec:mbbfit}

\subsection{Dust emission observed by \Planck}

\label{sec:mbb}

The emission from interstellar dust in the far-infrared (FIR) to millimetre
range is dominated by the emission from the biggest grains that are in
thermal equilibrium with the local radiation field. Many studies and
reviews have been dedicated to this subject
\citep[e.g.,][]{draine2003,draine2007a,compiegne2011}.

In the optically thin limit, the SED of emission from a uniform
population of grains is well described, empirically, by a modified
blackbody (MBB):
\begin{equation}
I_\nu = \tau_\nu \, B_\nu(T)\,, 
\end{equation}
where $I_\nu$ is the specific intensity, $B_\nu(T)$ is the Planck
function for dust at temperature $T$, and $\tau_\nu$ is the
frequency-dependent dust optical depth modifying the blackbody shape
of the SED. 
The optical depth is the product of the dust opacity, $\sigma_{\mathrm{e}\,\nu}$,
and the gas column density, $N_{\rm H}$:
\begin{equation}
\label{eq:opacity}
\tau_\nu = \sigma_{\mathrm{e}\,\nu} \, N_{\rm H}\,.
\end{equation}
Alternatively, the optical depth is the product of the dust emissivity
cross section per unit mass, $\kappa_\nu$ (in cm$^2$\,g$^{-1}$), and
the dust mass column density, $M_{\rm dust}$:
\begin{equation}
\label{eq:tau_nu_def}
\tau_\nu = \kappa_\nu \, M_{\rm dust}\,,
\end{equation}
where $M_{\rm dust} = r \, \mu \, m_{\rm H} \, N_{\rm H}$, with $r$
being the dust-to-gas mass ratio, $\mu$ the mean molecular weight, and
$m_{\rm H}$ the mass of a hydrogen atom.
Note that $\kappa_\nu$ depends on the chemical composition and
structure of dust grains, 
but not the size for particles small compared to the wavelength, as
here. It is usually described as a power law $\kappa_\nu = \kappa_0
(\nu/\nu_0)^\beta$ \citep{hildebrand1983,compiegne2011}, where $\kappa_0$ is the
emission cross-section at a reference frequency $\nu_0$. Put together,
the emission of dust of a given composition and structure and in
thermal equilibrium is:
\begin{equation}
I_\nu = \kappa_0 \, (\nu/\nu_0)^\beta \, r \, \mu \, m_{\rm H} \, N_{\rm H} \, B_\nu(T)\,.
\end{equation}

In practice the shape of the observed SED depends on three main
parameters. First, the equilibrium temperature is set by the radiation
field strength, parametrized by the scaling factor, $U$, of the mean
interstellar radiation field (ISRF) in the solar neighbourhood from
\citet{mathis1983}; note that in dense regions, $U$ is decreased
because of attenuation. Second, the grain size distribution
\citep{mathis1977,weingartner2001} is important; exposed to the same
ISRF, bigger grains have a lower equilibrium temperature than smaller
ones.
Third, the dust structure and composition determine not only the
optical and UV absorption cross section, but also the emission
cross-section, the frequency-dependent efficiency to emit radiation,
usually modelled as above as a power law ($\kappa_0\,\nu^\beta$) but
possibly more complex depending on dust properties ($\beta$ could vary
with frequency and/or grain size and/or grain temperature).
In a given volume element along the line of sight, the distribution of
dust grain sizes will naturally create a distribution of equilibrium
temperatures. In addition, dust properties might vary along the line
of sight. Furthermore, $U$ might also change along some lines of
sight. Therefore, the observed dust SED is a mixture of emission
modified by these effects,
the sum of several different MBBs. 
Nevertheless, the simplification of fitting a single MBB is often adopted and indeed here, 
with only four photometric bands available, is unavoidable.  
The parametrization of the MBB for the empirical fit is:
\begin{equation}
\label{eq:mbb}
I_\nu = \tau_{\nu_0} \, B_\nu(T_{\rm obs}) \, \left( \frac{\nu}{\nu_0} \right)^{\beta_{\rm obs}}\,,
\end{equation}
where $\nu_0$ is a reference frequency at which the optical depth
$\tau_{\nu_0}$ is estimated ($\nu_0=353$\,GHz in our SED
applications in this paper). 

The main challenge is then to relate 
the parameters of the fit to physical quantities.
It has been shown by many authors
\citep{blain2003,schnee2007,shetty2009,kelly2012,juvela2012,juvela2012a,ysard2012}
that, in general, the values of $T_{\rm obs}$ and $\beta_{\rm obs}$
recovered from an MBB fit cannot be related simply to the mass-weighted
average along the line of sight of the dust temperature and spectral
index. Even for dust with a spectral index constant in frequency
(i.e., $\beta$ does not depend on $\nu$), the distribution of grain
sizes and the variations of $U$ along the line of sight could introduce
a broadening of the SED relative to the case of a single dust size and
single $U$. In addition, the dust luminosity is proportional to
$T^{4+\beta}$ and so dust that is hotter for any reason, including
efficiency of absorption, will contribute more to the emission at
all frequencies than colder dust. Therefore, the observed SED is not a
quantity weighted by mass alone.
The dust SED is wider than a single MBB due to the distribution of
$T$, and so the fit is bound to find a solution where $\beta_{\rm
  obs}<\beta$ and, in consequence, where $T_{\rm obs}$ is biased
toward higher values. This results in dust optical depth that is
generally underestimated: $\tau_{\rm obs} < \tau$. This effect is
somewhat mitigated when lower frequency data are included. In the
Rayleigh-Jeans limit the effect of temperature is low and the shape of
the spectrum is dominated by the true $\beta$. For $T=$15--25\,K dust,
this range is at frequencies lower than $\nu = k\, T / h
=$310--520\,GHz.

Models like the ones of \citet{draine2007a} and \citet{compiegne2011}
go beyond the simple MBB parametrization by
incorporating the variation of the equilibrium temperature of grains
due to the size distribution. The model of \citet{draine2007a} also
includes a prescription for the variation of $U$ along the line of
sight, but assumes fixed dust properties. 
Nevertheless, there are still many uncertainties in the properties 
of dust (the exact size distribution of big grains, the optical properties, and 
the structure of grains), in the evolution of these properties from diffuse to denser clouds,
and in the variation of the radiation field strength along the line of sight. 

Therefore, for our early exploration of the dust SED over the whole sky,
at 5\arcm\ resolution and down to 353\,GHz (850\,$\mu$m), we believe that it is useful to fit the
dust SED using the empirical MBB approach,
before attempting to use more physical models that rest on
specific hypotheses.
The three parameters $\tau_{\nu_0}$, $T_{\rm obs}$ and $\beta_{\rm obs}$ 
obtained from the MBB fit should be regarded as a way to fit the data empirically; 
the complex relationship between these recovered parameters and physical quantities needs to be investigated in 
detail with dedicated simulations \citep[e.g.,][]{ysard2012}, but is beyond the scope of this paper.

\subsection{Implementation of the SED fit}

\label{sec:implementation}

The fit of the dust SED with a MBB model has been carried out traditionally
using a $\chi^2$ minimization approach. 
Recently, alternative methods for fitting observational data with limited spectral 
coverage have been proposed, based on Bayesian or hierarchical models 
\citep{veneziani2010,kelly2012,juvela2013,veneziani2013}. These new
methods were developed specifically to limit the impact of instrumental 
noise on the estimated parameters. 
Even though these methods
offer interesting avenues, we developed our own strategy to fit the dust SED over the whole sky because of another challenge to be mitigated, 
arising from the cosmic infrared background anisotropies (the CIBA).  Although this
has been overlooked, it can be 
dominant in the faint diffuse areas of the sky, as we demonstrate.
We proceeded with a
method based on the standard $\chi^2$ minimization (see Appendix~\ref{sec:mpfit}) 
but implemented a two-step approach that limits the fluctuations of the estimated parameters 
at small angular scales induced by noise and the CIBA.
In developing the methodology 
we have explored using data degraded to lower
resolution and smaller $N_{\rm side}$.

\subsubsection{Frequency coverage}
 
One possible source of bias in the fit is the number of bands and 
their central frequency. 
The combination of \Planck\ 353 to 857\,GHz and \IRIS\ 3000\,GHz data
allows us to sample the low and high frequency sides of the dust SED.
For a typical temperature of 20\,K, the peak of the emission is at a
frequency of 2070\,GHz. This falls in a gap in the frequency coverage,
between 857\,GHz and 3000\,GHz. It is thus a concern that a fit of the
\Planck\ and \IRIS\ data might bias the recovered parameters $T_{\rm
  obs}$ and $\beta_{\rm obs}$. To explore this we combined the
\Planck\ data with the \DIRBE\ data at 1250, 2143, and 3000\,GHz (100,
140, and 240\,$\mu$m), all smoothed to 60\arcm, providing a better
sample of the dust SED near its peak. We found that the recovered dust
parameters are stable whether \DIRBE\ data are used or not; no bias is
observed in $T_{\rm obs}$, $\beta_{\rm obs}$, and $\tau_{353}$
compared with results obtained using just \Planck\ and \IRIS\ data.

We also evaluated the potential advantage of fitting the SED with only 
the \Planck\ 353 to 857\,GHz data, a more coherent dataset not relying 
on the \IRIS\ data.  However, because the Wien part of 
the SED is not sampled the results showed a clear bias 
of $T_{\rm obs}$, toward lower values;  consequently, when extrapolated 
to 100\,$\mu$m, the fits greatly 
underestimate the emission detected in the \IRIS\ data.
Therefore, in the following the $\chi^2$ minimization fit was carried out on the data described
in Sect.~\ref{sec:data}: the 857, 545, and 353\,GHz \Planck\ maps,
corrected for zodiacal emission, and the new 100\,$\mu$m map obtained by
combining the \IRIS\ and SFD maps.

\subsubsection{Noise and cosmic infrared background anisotropies}

\label{sec:noise_and_CIBA}

\begin{figure}
\centering
\includegraphics[draft=false, angle=0]{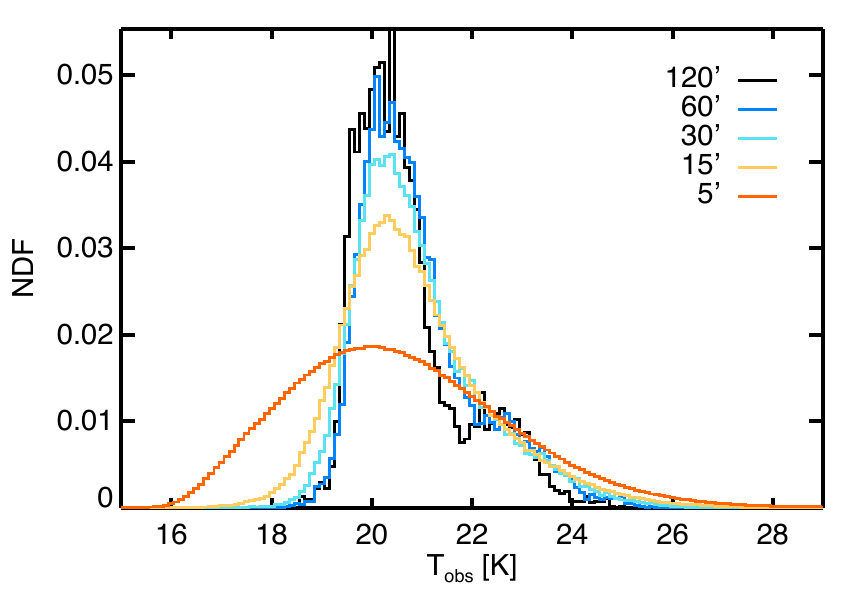}
\includegraphics[draft=false, angle=0]{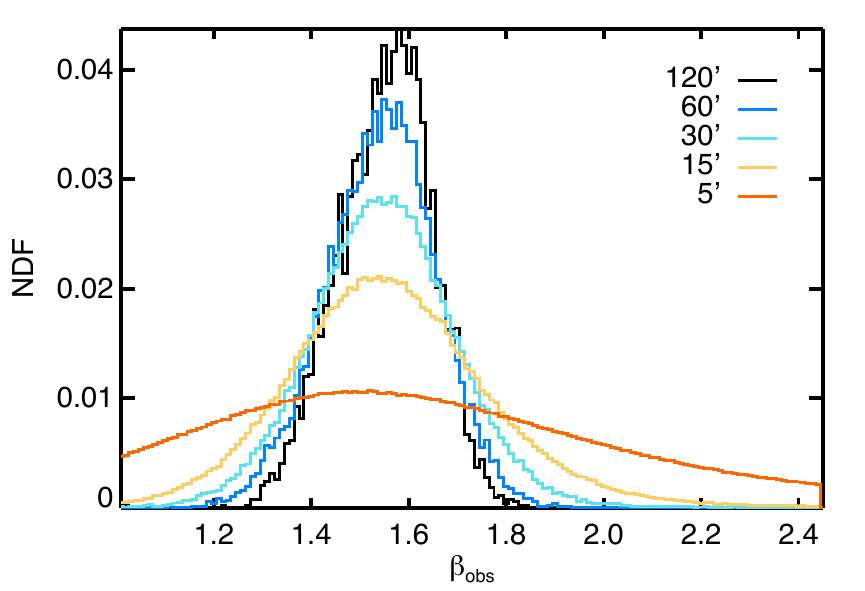}
\includegraphics[draft=false, angle=0]{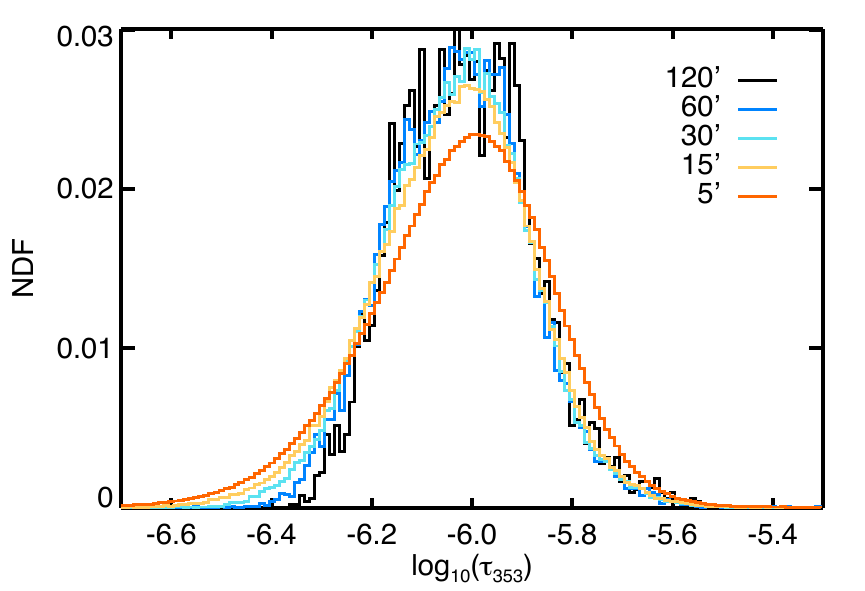}
\caption{\label{fig:t_beta_degeneracy} Normalized distribution
function (NDF) of $T_{\rm obs}$ (\textit{upper}), $\beta_{\rm obs}$ (\textit{middle}), and
$\tau_{353}$ (\textit{lower}) for data smoothed to different resolutions, from
5\arcm\ to 2\deg. The NDFs are shown only for the pixels corresponding
to the low $N_\ion{H}{i}$ mask of Fig.~\ref{fig:hi_mask}, left,
i.e., to the faintest 10\,\% of the sky pixels.}
\end{figure}

Degeneracy (anticorrelation) of the estimated $T_{\rm obs}$ and
$\beta_{\rm obs}$, inherent to the MBB fit of dust emission in the
presence of noise, has had dedicated specific study
\citep{shetty2009a,juvela2012}. As mentioned above, the CIBA is also 
a contaminating source in the estimate of the MBB parameters.

The CIBA is the result of the combined emission of distant unresolved
galaxies.  Its structure on the sky is diffuse.  The angular power
spectrum, with $C_\ell \propto \ell^{-1}$ for $100<\ell<2000$
according to \citet{planck2011-6.6} and
\citet{planck2013-pip56}, reveals the large-scale structure
of the Universe at high redshift. The zero levels of the maps were set
through correlation with \hi\ and so the data used in our study are
insensitive to the monopole of the CIBA.  However, the anisotropies, or
fluctuations, are present in the maps. Because the CIBA power spectrum
is flatter than that of interstellar dust emission 
\citep[$C_\ell \propto \ell^{-2.9}$, ][]{miville-deschenes2007a}, in relative terms
the CIBA is more visible at small scales.  Another feature of the CIBA
is that its structure on the sky is correlated in frequency, though
only partially because galaxies at different redshifts contribute to
the emission at different frequencies.  Because of this partial
correlation in frequency, the CIBA cannot be treated in the same way
as instrumental noise in the fit. But it cannot be included as another
component in the fitting function either. 
Nevertheless, the CIBA 
has an impact on the parameters of the fit; like the instrumental noise, 
the CIBA introduces an anticorrelation between $T_{\rm obs}$ and $\beta_{\rm obs}$.

One option to limit the effect of noise and the CIBA is to reduce the number
of free parameters in the fit. In that context we have examined the possibility 
of fitting the dust SED over the whole sky, at 5\arcm\ resolution,
using a fixed $\beta_{\rm obs}$ with 
values between 1.5 and 1.8. A value of $\beta_{\rm obs}=1.65$ provides the best 
fit with a reduced $\chi^2$ lower than unity everywhere on the sky but this is
mostly due to the fact that we took into account calibration
uncertainties in the fit (Appendix~\ref{sec:mpfit}). 
What is statistically significant is the
fact that on about 25\,\% of the sky the reduced $\chi^2$ is improved
by letting $\beta_{\rm obs}$ be a free parameter. 
This happens mostly in bright regions of the sky where the noise
is not an issue.
In molecular clouds and in the Galactic plane, there are variations 
in the shape of the SED that cannot be fit with only two parameters.

While fixing a parameter of the fit over the whole sky might be too strict, 
it might not be necessary to have all three parameters
at full resolution to describe the data. We have thus evaluated the possibility of
estimating one of the parameters at a lower resolution than the others. 
In the following we explore the impact of noise and the CIBA on the parameters of 
the fit as a function of angular resolution.

Both the noise and the CIBA have flatter power spectra than dust emission,
and so we expect the intrinsic dust parameters $T_{\rm obs}$,
$\beta_{\rm obs}$, and $\tau_{353}$ to have a smoother structure on
the sky than noise and the CIBA, 
except perhaps in bright photon-dominated regions 
where the shape of the dust SED might vary rapidly at small scales 
due to radiative transfer effects and potentially fast dust evolution.
Smoothing the maps by different amounts
before fitting on each pixel therefore offers the advantage of
revealing both this spatially smoother solution and the important
impact of noise and the CIBA on the result of the fit.  This is
illustrated in Fig.~\ref{fig:t_beta_degeneracy} where we present
normalized distribution functions (NDFs) of $T_{\rm obs}$, $\beta_{\rm
  obs},$ and $\tau_{353}$ obtained with data smoothed to 5\arcm,
15\arcm, 30\arcm, 60\arcm, and 120\arcm, selecting only pixels
corresponding to the low $N_\ion{H}{i}$ mask (Fig.~\ref{fig:hi_mask}) to
highlight a regime of relatively low signal-to-noise ratio. Smoothing
the data has no real impact on the average value of the parameters,
but the standard deviations of $T_{\rm obs}$ and $\beta_{\rm obs}$ go
down rapidly with smoothing; at 5\arcm\ resolution the standard
deviation of $T_{\rm obs}$ and $\beta_{\rm obs}$ is about twice as
large as with smoothed data. On the other hand, the dispersion of
$\tau_{353}$ is less affected by smoothing;
it is dominated instead by cosmic variance, the considerable range of
column densities even within this low $N_\ion{H}{i}$ mask.\footnote{The
  dispersions of quantities normalized by the column density, $\sigma_{\mathrm{e}\,353} =
  \tau_{353}/N_{\rm H}$ and the dust specific luminosity $L_{\rm H}$,
  are available only for lower resolutions; at 30\arcm\ resolution for
  this mask (see Table~\ref{tab:summary2} in
  Sect.~\ref{sec:discussion} below), they are considerably lower in
  fractional terms
  than the dispersion of $\tau_{353}$ in
  Fig.~\ref{fig:t_beta_degeneracy}.}
Of course the dust parameters might also vary at small scales and so a
trade-off needs to be found. 

To explore and quantify the impact of noise and the CIBA on the fit at
different angular resolutions, for later comparison with the actual
dispersion, we used Monte Carlo simulations of the SED, including dust
emission, noise, and the CIBA. The details of the Monte Carlo simulations,
including the information on the inter-frequency coherence, are
described in Appendix~\ref{sec:MC_CIBA}. We considered five different
angular resolutions of the data: 5\arcm, 15\arcm, 30\arcm, 60\arcm, and
120\arcm. The noise and CIBA levels used for each resolution are given
in Table~\ref{tab:CIB_levels}.
Here we present results for an SED appropriate to
Fig.~\ref{fig:t_beta_degeneracy} by adopting the median dust
parameters found in the low $N_\ion{H}{i}$ mask 
that corresponds to the faintest 10\,\% of
the sky.
We simulated $10^5$ realizations of this SED to which noise and the CIBA
were added. For each realization the three parameters $\tau_{353}$,
$T_{\rm obs}$, and $\beta_{\rm obs}$ were estimated as in
Appendix~\ref{sec:mpfit}.  The $1\sigma$ dispersions of $T_{\rm obs}$,
and $\beta_{\rm obs}$ obtained at each resolution are given in
Table~\ref{table:MC_results}, for noise and the CIBA separately.  The
simulated effect of smoothing on the $\beta_{\rm obs}$ -- $T_{\rm obs}$
anticorrelation is shown in Fig.~\ref{fig:T_beta_noise_and_CIBA}.

At full resolution the noise is the dominant source of error on the
retrieved parameters. For the specific faint dust spectrum considered
here, the noise produces an uncertainty of 2.1\,K while the
uncertainty due to the CIBA is only 0.39\,K. The same is true for
$\beta_{\rm obs}$: the uncertainties are 0.49 and 0.11 for the noise
and the CIBA, respectively. However, even with moderate smoothing of
the data, the impact of noise on the fit reduces sharply, whereas the
reduction of the impact of the CIBA is less dramatic.  This arises
because the CIBA has a power spectrum that is steeper than that of
typical (white) noise.  In addition, unlike the CIBA, noise has power
up to the pixel scale (i.e., it is not attenuated by the beam).  For
example, as seen in Table~\ref{table:MC_results}, for data smoothed to
30\arcm, the noise levels of the \Planck\ and \IRAS\ data go down by a
factor $18.5$ while the CIBA standard deviation decreases only by a
factor $2.5$.  As a result, our simulations show that for data
smoothed to resolution larger than 15\arcm\ the CIBA becomes the main
source of error.

Similar relative effects as a function of resolution are seen in the
$\beta_{\rm obs}$ -- $T_{\rm obs}$ anticorrelation in
Fig.~\ref{fig:T_beta_noise_and_CIBA}.  At all resolutions the
estimates of $T_{\rm obs}$ and $\beta_{\rm obs}$ lie within an
ellipse in $\beta_{\rm obs}$ -- $T_{\rm obs}$ space.  The orientation and
extent of the ellipse depends on the amplitudes of the noise and of the
CIBA, which are both different and in a different ratio at each
resolution. In all cases the ellipse is centred on the input values,
demonstrating that noise and the CIBA do not bias the estimate of $T_{\rm
  obs}$ and $\beta_{\rm obs}$. This is the case even though the CIBA
has a flatter (broader) SED than interstellar dust.  Because the CIB
monopole was removed in the data and therefore not included in the
simulation, the CIBA produces as many negative as positive CIB
fluctuations on the sky at each frequency. Because they are
(partially) correlated in frequency, positive CIB fluctuations bias
the SED and descriptive dust parameters toward a flatter SED while
negative CIB fluctuations have the opposite effect, toward a steeper
SED.

\subsubsection{The two-step approach}

\label{sec:two-step}

\begin{table}
\caption{\label{table:MC_results} Results of Monte Carlo simulations
  of three-parameter fits:
$1\sigma$ uncertainties of $T_{\rm obs}$ and $\beta_{\rm obs}$ due to
noise and the CIBA, for data at different resolutions. }
\begin{center}
\begin{tabular}{c D{.}{.}{1.3} c D{.}{.}{1.4} D{.}{.}{1.3}} \specialrule{\lightrulewidth}{0pt}{0pt} \specialrule{\lightrulewidth}{1.5pt}{\belowrulesep}
$\theta$ & 
\multicolumn{1}{c}{$\delta_{\rm noise}(T_{\rm obs})$} & 
$\delta_{\rm CIBA}(T_{\rm obs})$ & 
\multicolumn{1}{c}{$\delta_{\rm noise}(\beta_{\rm obs})$} & 
\multicolumn{1}{c}{$\delta_{\rm CIBA}(\beta_{\rm obs})$} \\ 
\,[arcmin] & \multicolumn{1}{c}{[K]} & [K] \\\midrule
$\:\:\:\,\,5$ & 2.1 & 0.39 & 0.49 & 0.11\\
$\:\:15$ & 0.32 & 0.32 & 0.054 & 0.070\\
$\:\:30$ & 0.15 & 0.23 & 0.026 & 0.049\\
$\:\:60$ & 0.075 & 0.16 & 0.013 & 0.035\\
120 & 0.037 & 0.11 & 0.0064 & 0.024\\
\bottomrule[\lightrulewidth]
\end{tabular}
\end{center}
{\bf Note:} The simulation was done for a single dust SED typical of
the 10\,\% faintest area of the sky, whose parameters are the median
values found in the low $N_\ion{H}{i}$ mask: $T_{\rm obs}=20.8$\,K,
$\beta_{\rm obs}=1.55$ and $\tau_{353}=9.6\times 10^{-7}$
(Table~\ref{tab:summary}).  The CIBA was modelled assuming partial
correlation in frequency (see Appendix~\ref{sec:MC_CIBA} for details).
The noise and CIBA levels used for each resolution are given in
Table~\ref{tab:CIB_levels}.  The values given here are the standard
deviations of the parameters $T_{\rm obs}$ and $\beta_{\rm obs}$
obtained from three-parameter SED fits of $10^5$ realizations;
$\delta_{\rm noise}$ and $\delta_{\rm CIBA}$ represent the separate
contributions of noise and the CIBA to the total standard deviation.
\end{table}

\begin{figure}
\centering
\includegraphics[draft=false, angle=0]{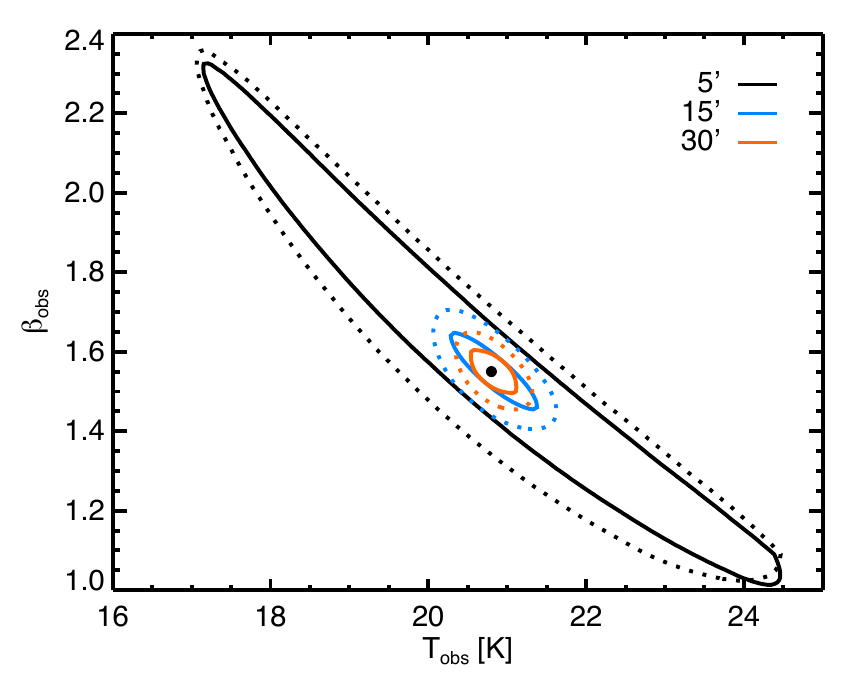}
\caption{\label{fig:T_beta_noise_and_CIBA} $\beta_{\rm obs}$ -- $T_{\rm
    obs}$ diagram showing the $1\sigma$ contour of the fit of a single
  SED (sum of dust emission, noise, and the CIBA) for simulated data with
  $\tau_{353}=9.6\times 10^{-7}$ and noise and the CIBA levels for each
  resolution from Table~\ref{tab:CIB_levels}: 5\arcm\ (black),
  15\arcm\ (blue) and 30\arcm\ (red). Contours are shown for noise only (solid) 
  and for noise and the CIBA (dotted). The other two parameters of
  the dust emission used in this simulation, marked by a black dot,
  are, like $\tau_{353}$, the median values found in the low
  $N_\ion{H}{i}$ mask: $T_{\rm obs}=20.8$\,K and $\beta_{\rm obs}=1.55$. }
\end{figure}

Given the impact of noise and the CIBA on
the recovered parameters, described in the previous section, we have
chosen to fit the data in two steps. 
First, we fit the data smoothed to 30\arcm\ (but on the $N_{\rm
  side}=2048$ grid). As shown in Figs.~\ref{fig:t_beta_degeneracy} 
and \ref{fig:T_beta_noise_and_CIBA} this
greatly reduces the effect of noise on the estimate of $\beta_{\rm
  obs}$ and $T_{\rm obs}$. 
Second, we fit the data at 5\arcm\ resolution with a fixed
$\beta_{\rm obs}$ taken from the map of $\beta_{\rm obs}$ obtained with data at 30\arcm\
resolution. That way two degrees of freedom ($\tau_{353}$ and
$T_{\rm obs}$ -- see Eq.~\ref{eq:mbb}) are still available to capture
the variations of the dust SED at full resolution while limiting the
effect of the $\beta - T$ degeneracy due to noise.

This two-step approach is in the same spirit as the one implemented in 
the {\tt Commander-Ruler} algorithm \citep{planck2013-p06}. 
The advantage of such methods arises by favouring a spatially smoother solution
for parameters that are not expected to vary strongly at small scale.
In the second fit we chose to fix $\beta_{\rm obs}$ rather than $T_{\rm obs}$.
It is not yet clear how the actual spectral index of the grain opacity,
$\beta$, might vary on small scales 
\citep[some models even assume that it is constant:][]{draine2007a,compiegne2011}.
However, the dust temperature is expected to vary on small scales,
especially in dense regions of the ISM due to the attenuation of the
radiation field. %

\begin{figure*}
\centering
\includegraphics[draft=false, angle=0]{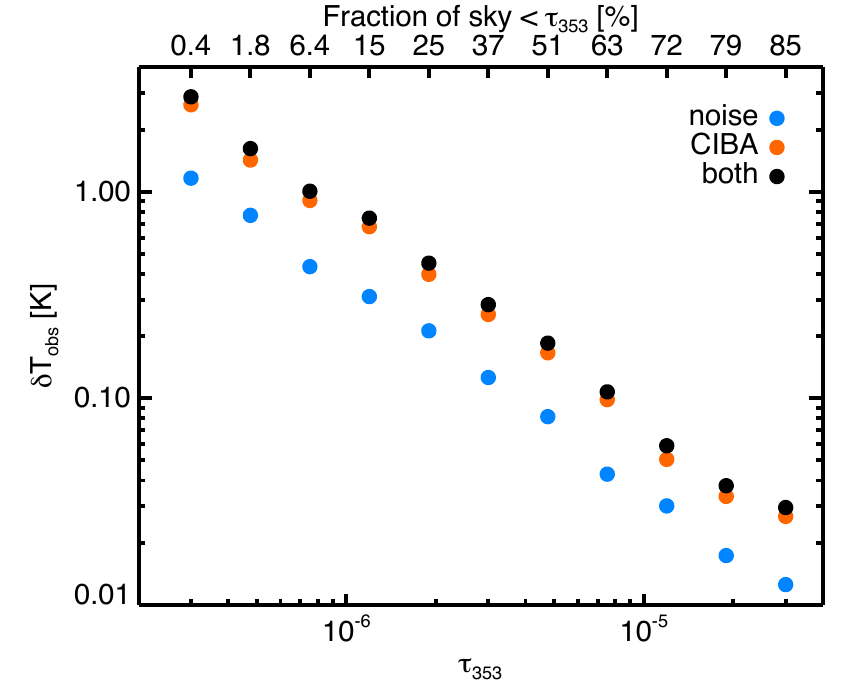}
\includegraphics[draft=false, angle=0]{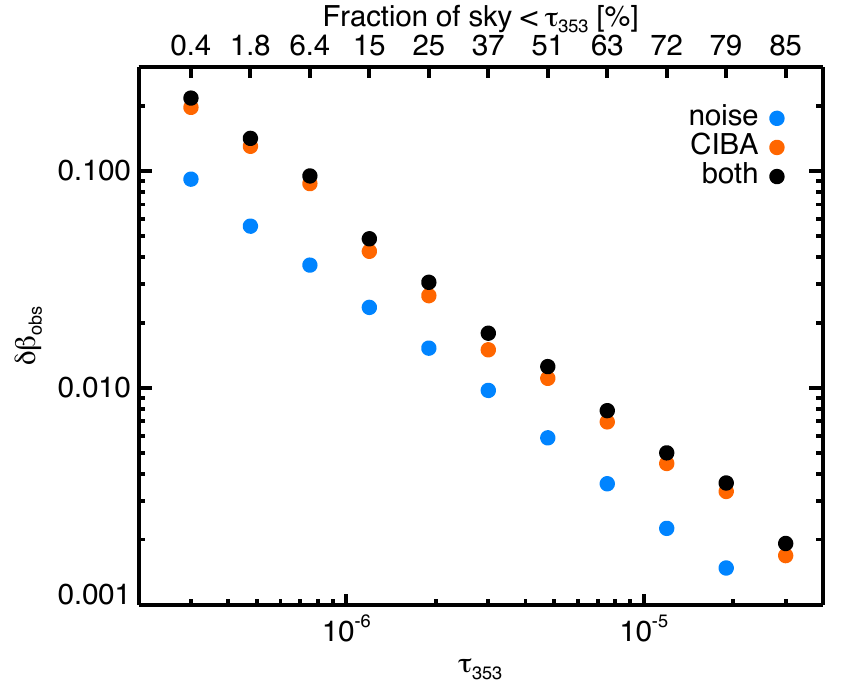}
\caption{\label{fig:MC_vs_tau} Estimate of the uncertainties of
$T_{\rm obs}$ (\textit{left}) and $\beta_{\rm obs}$ (\textit{right})  due to noise and
the CIBA for a two-step fit to single dust SEDs with $T_{\rm obs}=20.8$\,K
and $\beta_{\rm obs}=1.55$ but increasing $\tau_{353}$. These results
were obtained using Monte Carlo simulations (see Appendix~\ref{sec:MC_CIBA}).
The fits for $\beta_{\rm obs}$ were carried out assuming noise and CIBA
levels at 30\arcm\ and a free $T_{\rm obs}$.
The fits for $T_{\rm obs}$ were carried out assuming noise and CIBA levels at
5\arcm\ and a fixed $\beta_{\rm obs}=1.55$.
The dots give the contribution of the noise (blue) and the CIBA (orange)
to the uncertainties, and the quadratic sum of the two (black).
The top axes of both plots indicate the fraction of sky
that has $\tau_{353}$ lower than the value on the lower axis.}
\end{figure*}

We performed Monte Carlo simulations to evaluate the contributions of noise
and the CIBA (partly correlated in frequency -- see
Appendix~\ref{sec:MC_CIBA}) to variations, whence uncertainties, of the 
recovered $T_{\rm obs}$ and $\beta_{\rm obs}$ 
for the specific case of the adopted two-step fit (30\arcm\ and
5\arcm). Fig.~\ref{fig:MC_vs_tau} illustrates the uncertainties of
$T_{\rm obs}$ and $\beta_{\rm obs}$ arising from noise and the CIBA for dust
SEDs on lines of sight with increasing $\tau_{353}$.

For the typical SED  ($\tau_{353} = 9.6 \times 10^{-7}$, 
$T_{\rm obs}=20.8$\,K, and $\beta_{\rm obs}=1.55$) corresponding to the
10\,\% faintest area of the sky, we made a comparison of the Monte-Carlo 
results for the two-step fit and the direct three-parameter fit (Sect.~\ref{sec:noise_and_CIBA}).
The uncertainties of the direct fit at 5\arcm\ are
$\delta(T_{\rm obs})=2.1$\,K and $\delta(\beta_{\rm obs})=0.50$,
adding the contributions of noise and CIBA in quadrature
(see Table~\ref{table:MC_results}).
For the same SED parameters, the uncertainties of the two-step fit 
are $\delta(T_{\rm obs})=0.8$\,K and $\delta(\beta_{\rm obs})=0.06$ 
(see Fig.~\ref{fig:MC_vs_tau}).
In addition, 
the results of the two-step fit simulations indicate that for
both $T_{\rm obs}$ at 5\arcm\ and $\beta_{\rm obs}$ at 30\arcm\ the
CIBA has a greater contribution than the noise, contrary to the
situation for the direct three-parameter fit. 

For the faintest 0.4\,\% of the sky (top axis in Fig.~\ref{fig:MC_vs_tau}), 
the results of the simulations indicate that the combined effects of 
noise and the CIBA produce variations $\delta T_{\rm obs}=3.0$\,K and
$\delta\beta_{\rm obs}=0.2$.
On the other hand, for about 93\,\% of the sky the variations are much smaller, 
$\delta T_{\rm obs}<1.0$\,K and $\delta\beta_{\rm obs}<0.1$, 
i.e., $<5$\,\% and $<6$\,\% fractional error, respectively. 
This is in accordance with the fact that at 353\,GHz, where the CIBA is the
strongest contaminant, about 93\,\% of the sky has $I_{353} >
3\sigma_{\rm CIBA}(353)$.

\subsection{Parameters and uncertainties}
\label{sec:fullskymaps}

\begin{figure*}
\centering
\includegraphics[draft=false, angle=0]{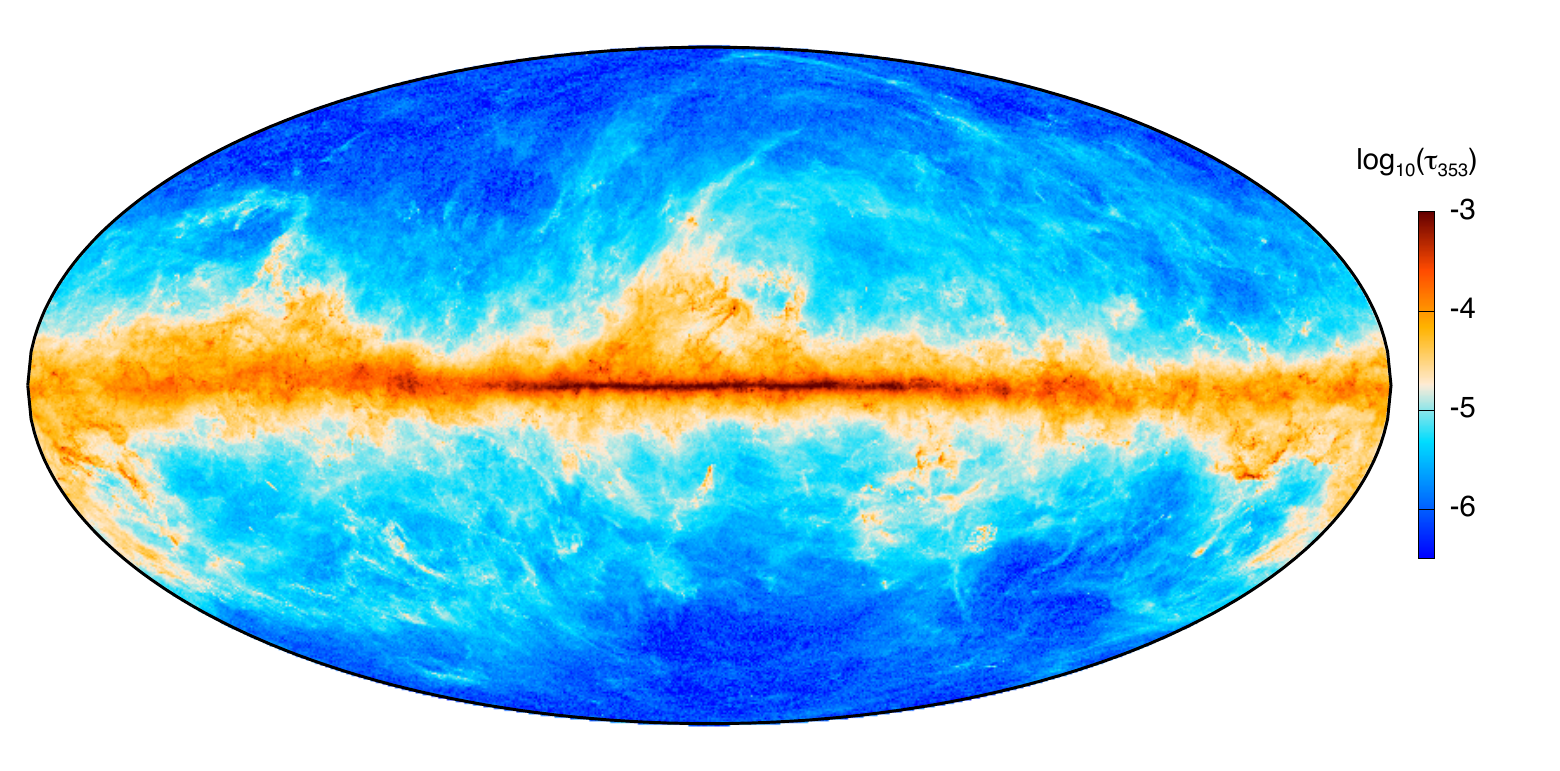}
\includegraphics[draft=false, angle=0]{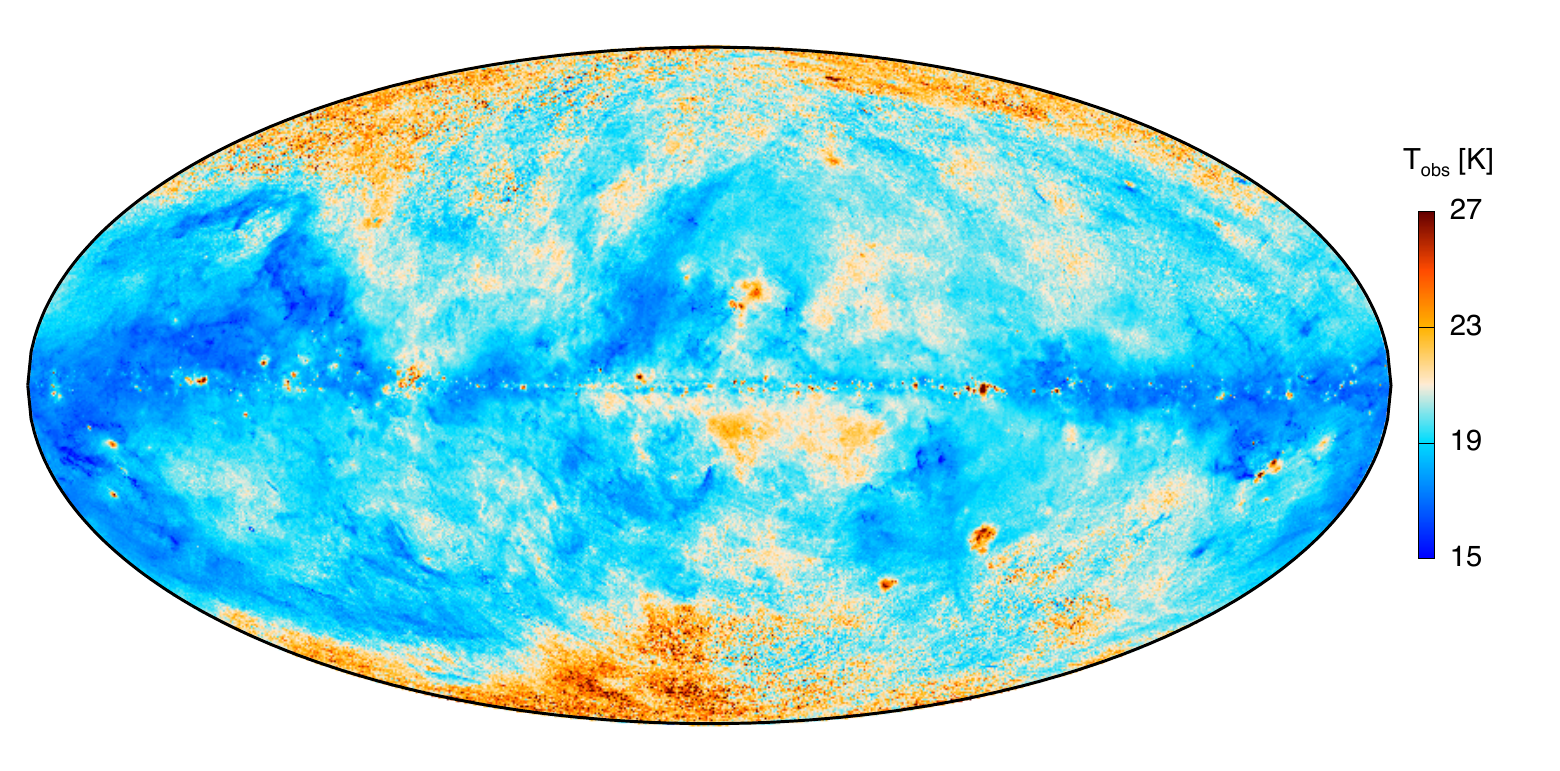}
\includegraphics[draft=false, angle=0]{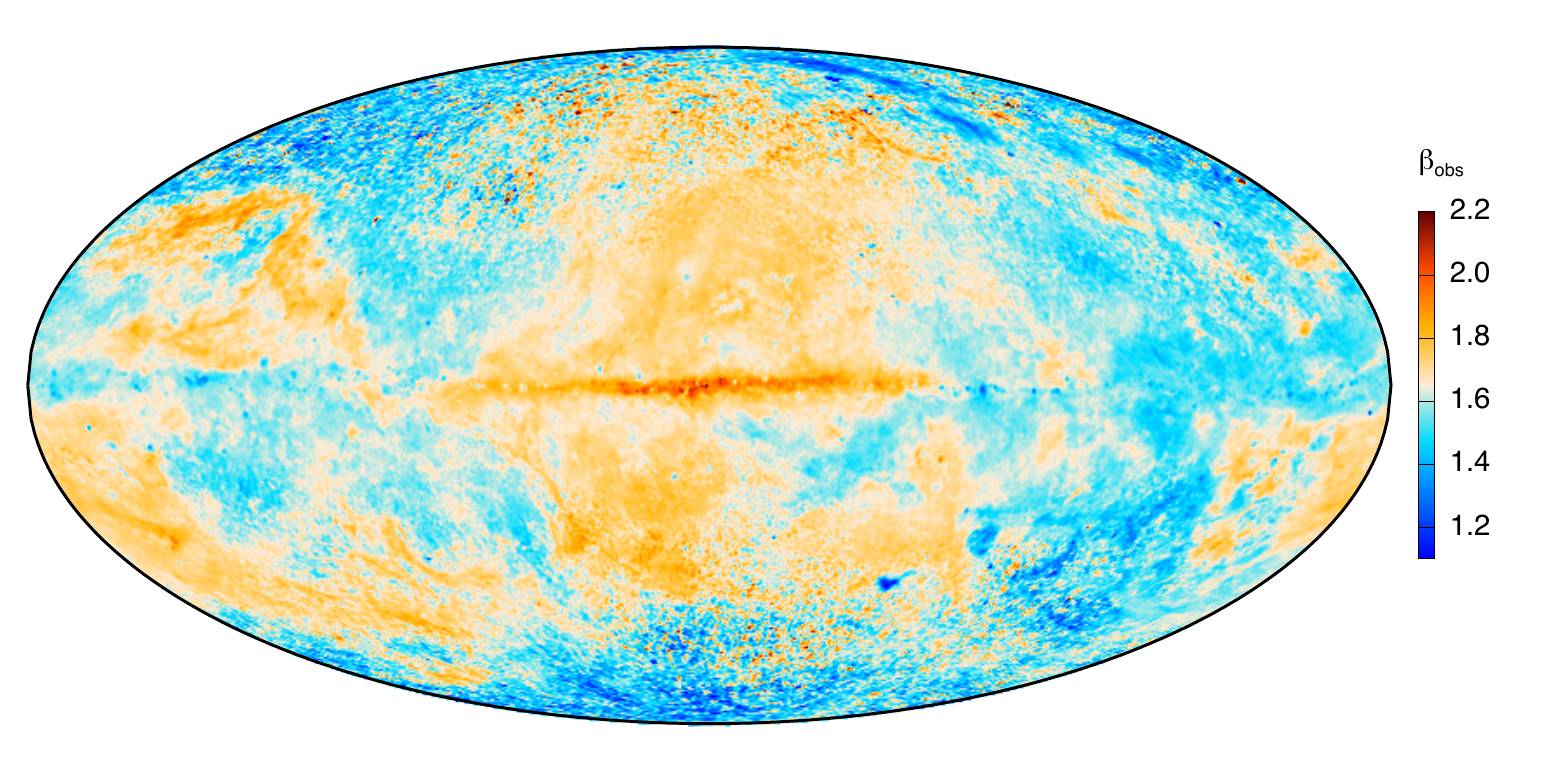}
\caption{\label{fig:fullres_maps} All-sky maps of the parameters of
the MBB fit of \Planck\ 353, 545, and 857\,GHz and \IRAS\ 100\,$\mu$m data.
\textit{Upper:} optical depth at 353\,GHz, $\tau_{353}$, at 5\arcm\
resolution, displayed logarithmically (the range shown corresponds to $-6.5 <
\log_{10}(\tau_{353}) < -3 $).  \textit{Middle:} observed dust
temperature, $T_{\rm obs}$, at 5\arcm\ resolution, in kelvin.  
\textit{Lower:} observed dust spectral index, $\beta_{\rm obs}$, at 30\arcm\
resolution.}
\end{figure*}

\begin{figure*}
\centering
\includegraphics[draft=false, angle=0]{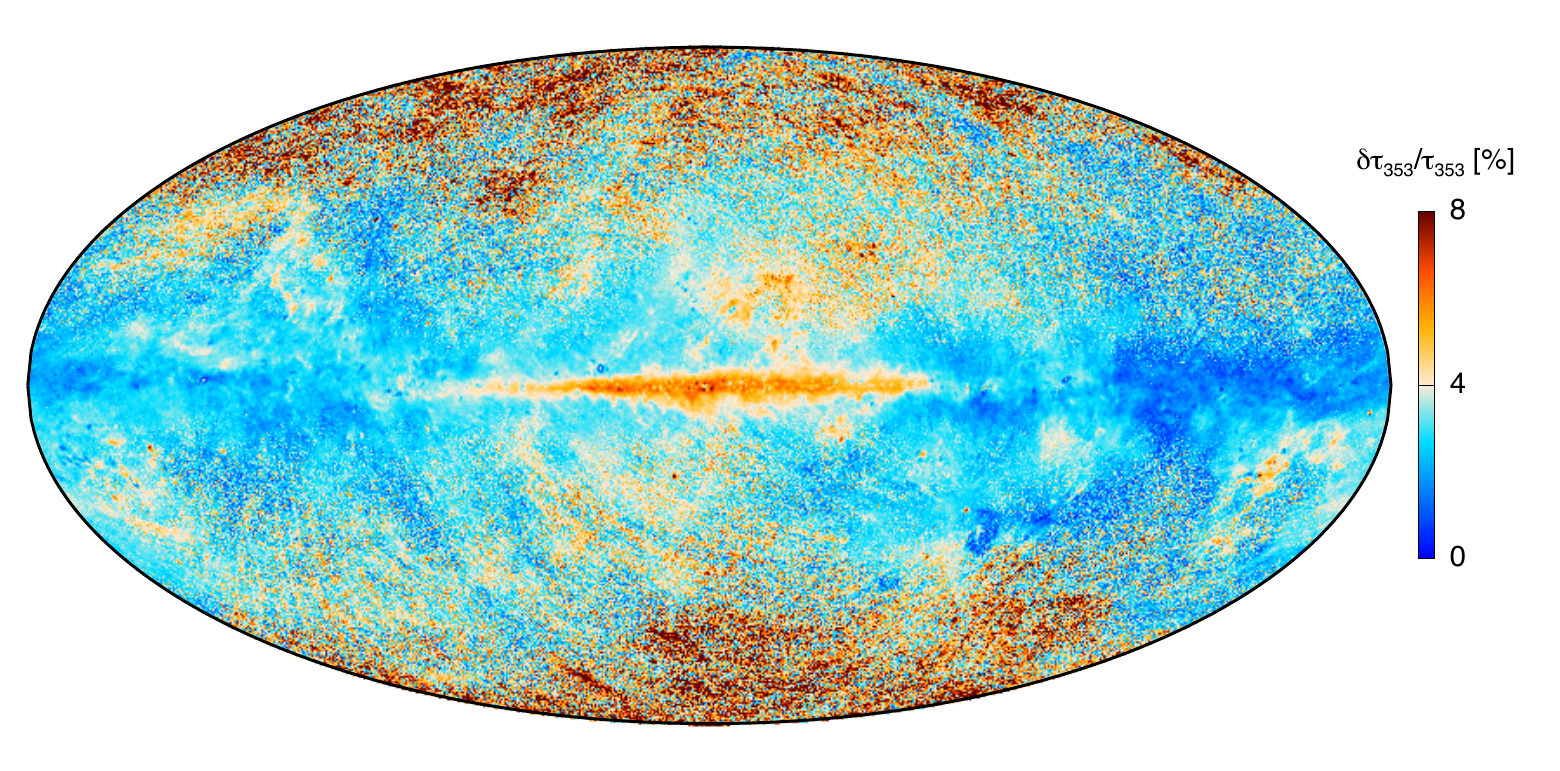}
\includegraphics[draft=false, angle=0]{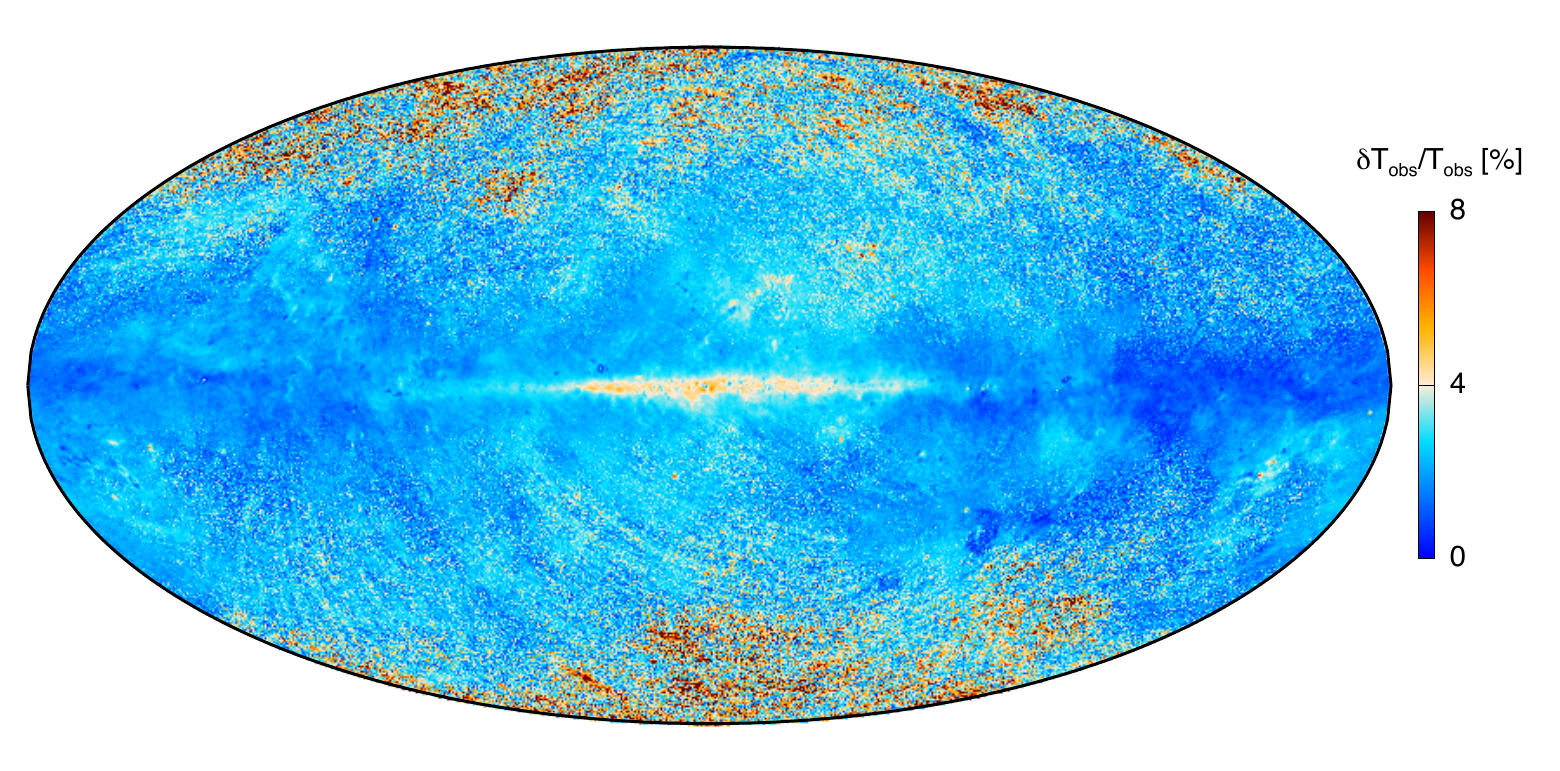}
\includegraphics[draft=false, angle=0]{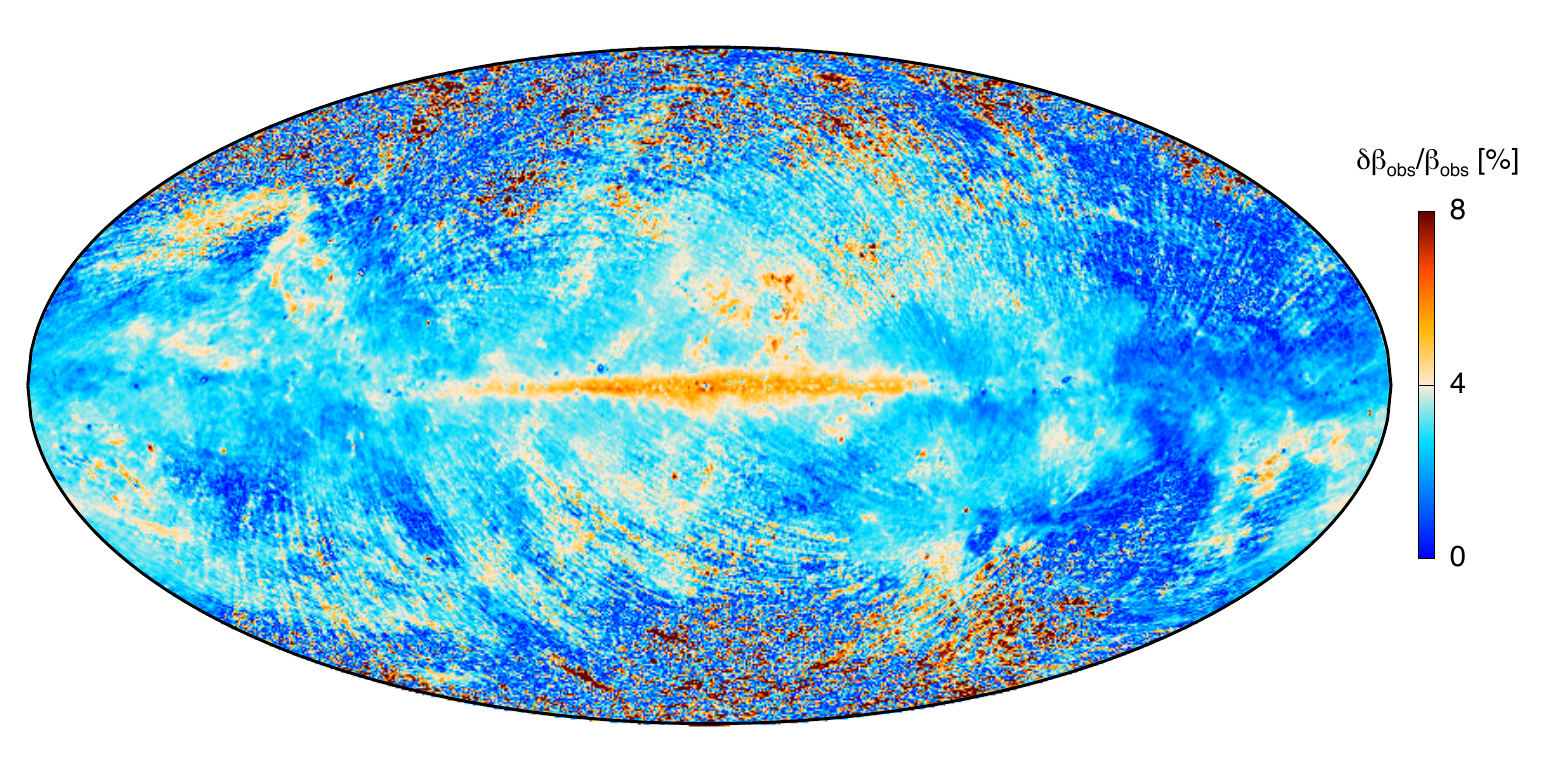}
\caption{\label{fig:fullres_error_maps} All-sky maps of the fractional
uncertainty (in percent) of the parameters of the MBB fit of \Planck\ 353,
545, and 857\,GHz and \IRAS\ 100\,$\mu$m data.  \textit{Upper:} optical
depth at 353\,GHz, $\tau_{353}$, at 5\arcm\ resolution.  \textit{Middle:}
observed dust temperature, $T_{\rm obs}$, at 5\arcm\ resolution.  
\textit{Lower:} observed dust spectral index, $\beta_{\rm obs}$, at 30\arcm\
resolution. }
\end{figure*}

The all-sky maps of the dust parameters, $T_{\rm obs}$, $\beta_{\rm
  obs}$, and $\tau_{353}$, and of their fractional uncertainties are
presented in Figs.~\ref{fig:fullres_maps} and
\ref{fig:fullres_error_maps}, respectively. The precision of the three
parameters is of the order of a few percent on most of the sky. The
uncertainties shown here are based on the statistical ones returned by
the $\chi^2$ minimization fit assuming that the model is a good
representation of the data. For $\beta_{\rm obs}$ the uncertainty is
from the 30\arcm\ fit. For $T_{\rm obs}$ we added quadratically the
fractional uncertainties from the 30\arcm\ and 5\arcm\ fits to include
the covariance between $T_{\rm obs}$ and $\beta_{\rm obs}$, whence
\begin{equation}
\delta T_{\rm obs} = T_{\rm obs} \, \sqrt{ \left( \frac{ \delta T_{\rm obs,30} }{ T_{\rm obs,30}} \right)^2 + 
\left( \frac{ \delta T_{\rm obs,5} }{ T_{\rm obs,5}} \right)^2 }\,,
\end{equation}
where the subscripts 5 and 30 refer to the parameter or uncertainty
maps obtained at 5\arcm\ and 30\arcm, respectively. Similarly, the
uncertainty of $\tau_{353}$ is from the quadratic sum of the
fractional uncertainties of $I_{353}^{\rm m}$ and $B_{353}(T_{\rm obs})$
where $I_{353}^{\rm m}$ is the reconstructed model of the emission at
353\,GHz. To estimate the uncertainty of $B_{353}(T_{\rm obs})$, we
simply computed $\max|B_{353}(T_{\rm obs}\pm\delta T_{\rm
  obs})-B_{353}(T_{\rm obs})|$.

The three uncertainty maps have a similar spatial structure. In
general the fractional uncertainties are higher in the most diffuse
areas of the sky (where the noise and the CIBA have a more important
contribution) and in the inner Galaxy region. Striping patterns are
visible, especially in the $T_{\rm obs}$ uncertainty map; these are
likely to be coming from the \IRAS\ data. The uncertainty of $T_{\rm
  obs}$ is of the order of 1--3\,\% in bright areas, with a noticeable
increase in the inner Galaxy and rising to 5--8\,\% in the most
diffuse areas of the sky. The same general trend is seen for
$\tau_{353}$ but with higher values: 2--5\,\% in bright areas and up
to 10\,\% in diffuse areas. The uncertainty of $\beta_{\rm obs}$,
based on analysis at 30\arcm\ resolution, has a slightly different
spatial structure.  It is typically of 3--4\,\% with a smaller
decrease in bright areas and a noticeable increase in the inner Galaxy
to 6--8\,\%.

\begin{figure}
\centering
\includegraphics[draft=false, angle=0]{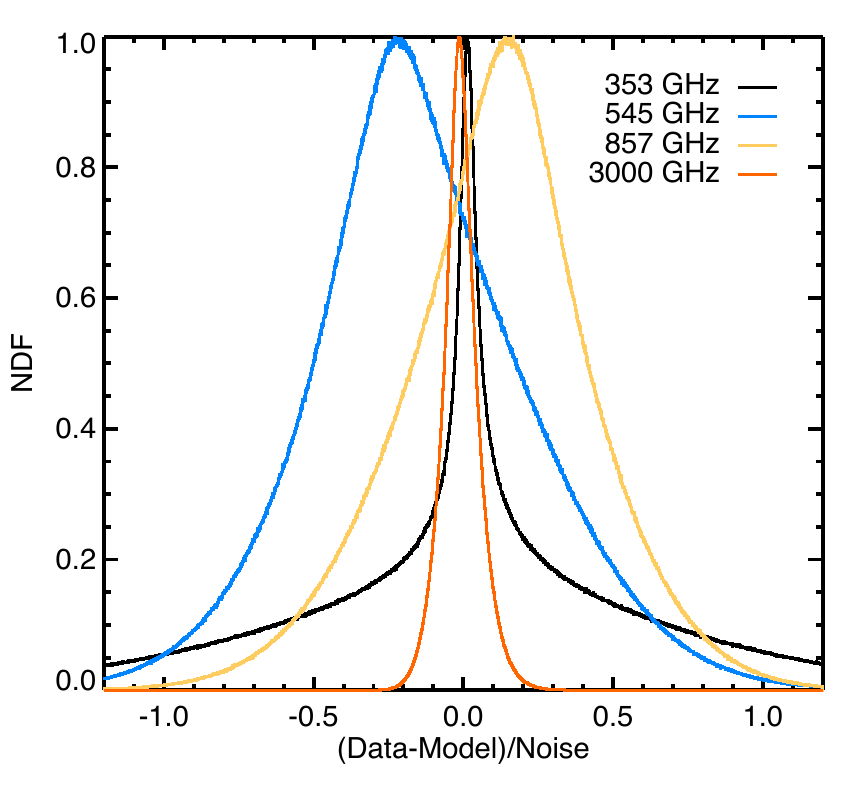}
\caption{\label{fig:data_model_noise} Normalized distribution function
of $(\textrm{Data}-\textrm{Model})/\textrm{Noise}$ for the four frequencies used in the fit.}
\end{figure}

The reduced $\chi^2$ of the fit is much smaller than unity over most
of the sky, due to the fact that calibration uncertainties are taken
into account in the fit to give less weight to data points with less
precise calibration (Appendix~\ref{sec:mpfit}). To illustrate this,
Fig.~\ref{fig:data_model_noise} shows the distribution function of
$(\textrm{Data}-\textrm{Model})/\textrm{Noise}$ for each frequency 
used in the fit. The noise used
here follows the definition of Eq.~\ref{eq:noise}; it takes into
account instrumental noise and the uncertainties of the calibration,
the zero level, and the CMB subtraction. The range adopted in
Fig.~\ref{fig:data_model_noise} corresponds to only $\pm 1 \sigma$. 
At 353 and 3000\,GHz, for most of the sky pixels the data are fitted
more tightly (to better than $0.1\sigma$) than at 545 and
857\,GHz. This implies that 353 and 3000\,GHz have a lot of weight in
the estimation of the parameters.  The 3000\,GHz band provides the
only data point on the Wien part of the MBB and therefore strongly
influences the determination of $T_{\rm obs}$.  On the other hand, the
353\,GHz band strongly influences the determination of $\tau_{353}$
and $\beta_{\rm obs}$ because it is the closest to the Rayleigh-Jeans
part of the spectrum.
The compensating small offsets of the distributions at the other two
frequencies might suggest that the adopted model does not adequately describe
the data. However, these offsets are well within the 
calibration uncertainties of the data; the overall reduced $\chi^2$ is 
lower than unity and these offsets might be removed by a small systematic change 
in the relative calibration of the data. Given the actual precision of
the calibration, it would be premature to conclude that a more complex model 
is required to fit the data.

We compared the reduced $\chi^2$ with that from a fit of the data
at 5\arcm\ with $\tau_{353}$, $T_{\rm obs}$, and $\beta_{\rm obs}$ as
free parameters. We were looking for pixels on the sky for which the
two-step fitting procedure provides a reduced $\chi^2$ greater than
unity (i.e., a relatively bad fit) while fitting the three parameters
simultaneously at full resolution would provide a better solution with
a lower reduced $\chi^2$.  This occurred for only 0.3\,\% of the
pixels.  These pixels, possibly dominated by galaxies, are grouped in
small-scale structures located at high Galactic latitude and away
from bright interstellar areas.

\subsection{Dust radiance}
\label{sec:luminosity}

\begin{figure*}
\centering
\includegraphics[draft=false, angle=0]{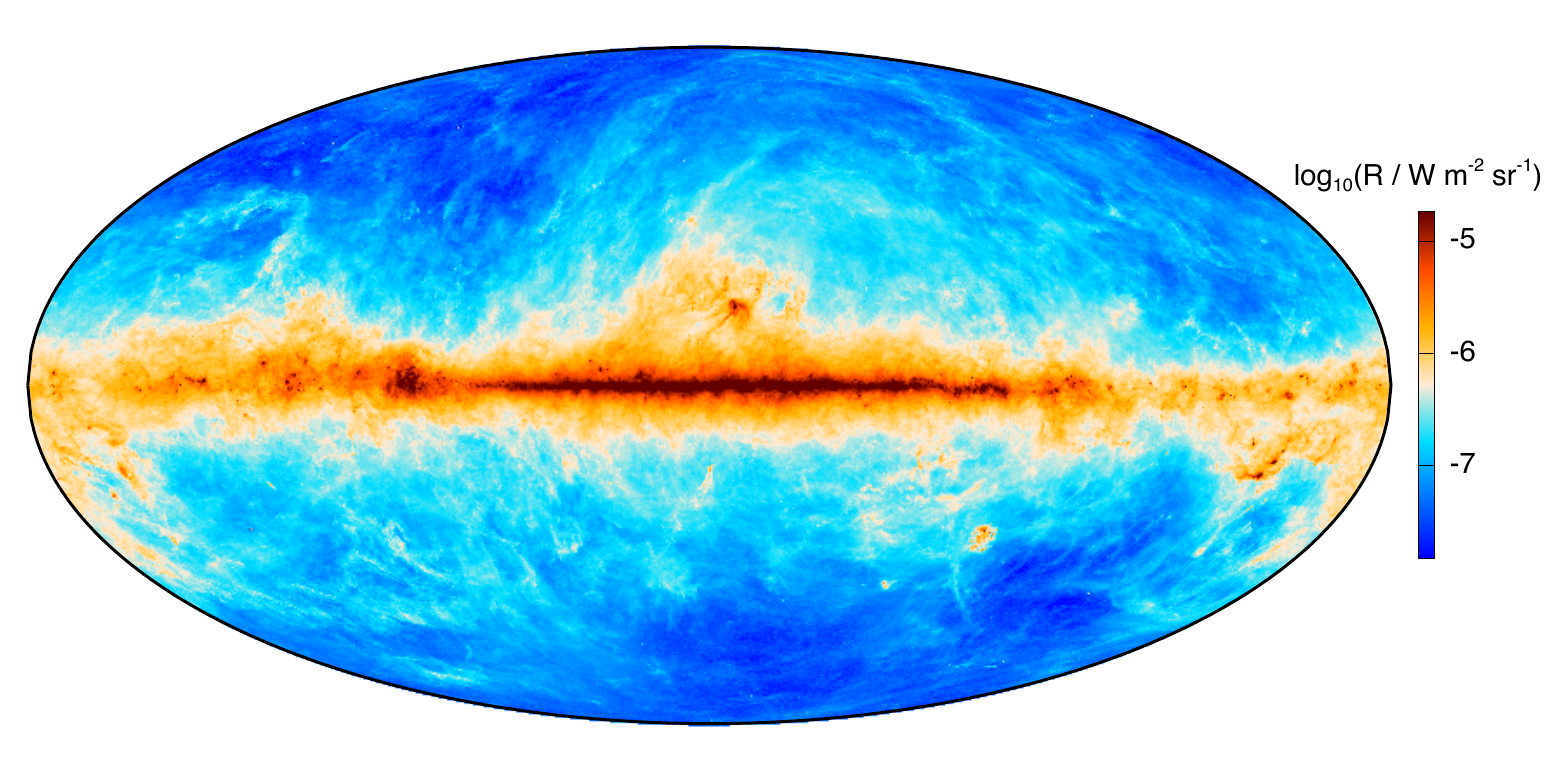}
\caption{\label{fig:L_allsky} All-sky map of dust radiance at 5\arcm\
resolution. The range shown corresponds to $-7.8 <
\log_{10}({\cal{R}}) < -4.7$.
}
\end{figure*}

In the previous sections we have described the properties of the
parameters that define the shape of the dust SED.  Now we examine the
dust radiance or dust integrated intensity defined as
\begin{equation}
\label{eq:radiance}
{\cal{R}} = \int_\nu I_\nu \, d\nu.
\end{equation}
Because the grains are in thermal equilibrium, this also represents
the energy absorbed.
Here we estimate \radiance\ at each sky position by integrating the MBB fit:
\begin{equation}
\label{eq:mbbluminosity}
{\cal{R}} = \int_\nu \tau_{353} \, B_\nu(T_{\rm obs}) \, \left( \frac{\nu}{353} \right)^{\beta_{\rm obs}} d\nu.
\end{equation}
This can be done analytically in terms of the Gamma ($\Gamma$) and
Riemann zeta functions ($\zeta$):
\begin{equation}
\label{eq:L_analytical}
{\cal{R}} = \tau_{353} \, \frac{\sigma_S}{\pi}\,  T_{\rm obs}^4\, \left(\frac{kT_{\rm obs}}{h \nu_0}\right)^{\beta_{\rm obs}} 
\frac{\Gamma(4+\beta_{\rm obs})\, \zeta(4+\beta_{\rm obs})}{\Gamma(4)\, \zeta(4)}\, ,
\end{equation}
where $\sigma_S$ is the Stefan-Boltzmann constant, $k$ is the
Boltzmann constant, $h$ is the Planck constant, and $\nu_0 = 3.53
\times 10^{11}$\,Hz.
Using the fit parameters described above we produced the all-sky map of
\radiance\ shown in Fig.~\ref{fig:L_allsky}, expressed in units of
W\,m$^{-2}$\,sr$^{-1}$.

Note that even though the calculation of \radiance\ uses the dust parameters ($T_{\rm obs}$, 
$\beta_{\rm obs}$, $\tau_{353}$), \radiance\ does not suffer from any degeneracy in the fit parameters.
In this context the MBB should be seen as an interpolating function; \radiance\ is not very sensitive to the assumptions 
made in fitting the SED as long as the fit accounts for the data, including the high-frequency turnover.
The uncertainty of \radiance\ arises mostly from the calibration uncertainty of the data and, 
to a lesser extent, from the limited number of bands used in the 
fit.\footnote{For example, a larger number of bands could reveal that a 
single-temperature MBB is not an adequate fitting function, a conclusion 
that cannot be reached with the four bands used here.}

In thermal equilibrium, \radiance\ is equal to the amount of light
absorbed by dust. Assuming constant 
properties along the line of sight, including the dust-to-gas ratio,
\begin{equation}
\label{eq:absorb}
{\cal{R}} \propto  U\, \overline{\sigma_a}\, N_{\rm H},
\end{equation}
where $\overline{\sigma_a}$ is the absorption opacity defined similarly to
the emission opacity in Eq.~\ref{eq:opacity}, averaged over the size
distribution and also, in this case, over the spectrum of the
ISRF.

This is complementary to $\tau_{353}$, which is also used as a proxy for $N_{\rm H}$:
\begin{equation}
\label{eq:divide}
\tau_{353} = \frac{I_{353}}{B_{353}(T_{\rm obs})} = \sigma_{\mathrm{e}\,353} \, N_{\rm H}.
\end{equation}
Division by the Planck function factors out any effects due to spatial 
variations of the dust temperature (potentially linked to spatial variations of 
$U$), but $\tau_{353}$ is only proportional to $N_{\rm H}$ if the dust opacity 
\opacity\ is constant. This limitation does not apply to \radiance, which is 
independent of \opacity\ because of thermal equilibrium; \radiance\ is simply 
the energy emitted by dust (Eq.~\ref{eq:radiance}), whatever the shape of 
the SED and regardless of how efficient the grain cooling is. Thus \radiance\ is 
closer to a measured quantity, while $\tau_{353}$ is a parameter deduced from 
a model. 

\begin{figure*}
\centering
\includegraphics[draft=false, angle=0]{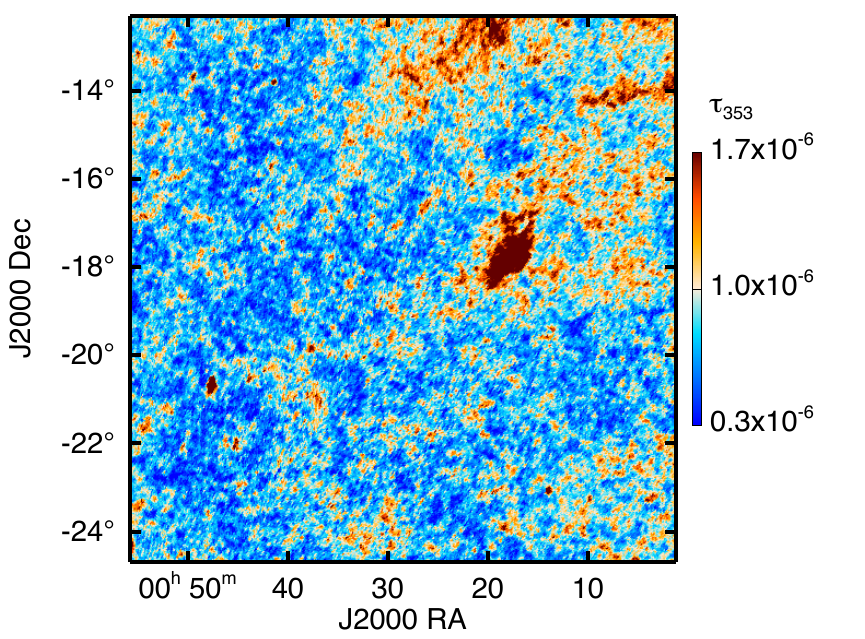}
\includegraphics[draft=false, angle=0]{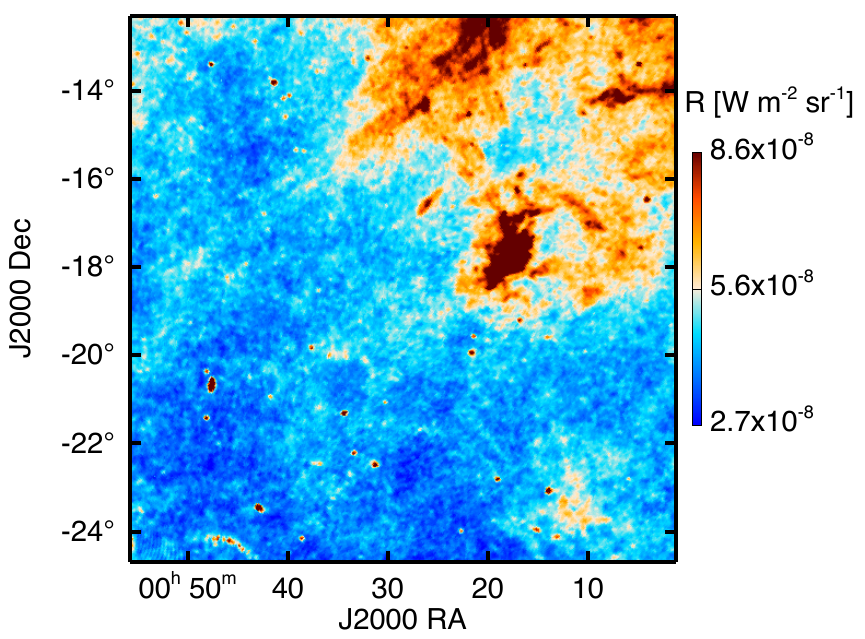}
\caption{\label{fig:maps_tau_vs_L} Maps of $\tau_{353}$ (\textit{left}) and
\radiance\ (\textit{right}) of a diffuse area of the sky, centred on
$l=90$\deg, $b=-80$\deg.}
\end{figure*}

At high Galactic latitudes, where the spatial variations of $U$ and 
\opacity\ are expected to be minimal so that both $\tau_{353}$ and \radiance\  should be 
proportional to dust column density, comparison of maps of $\tau_{353}$ and \radiance\ 
reveals another fundamental difference, as illustrated in Fig.~\ref{fig:maps_tau_vs_L} for 
one of the faintest areas in the sky: the map of $\tau_{353}$ shows 
surprisingly strong small-scale fluctuations that are absent in the map of \radiance. 

This significant difference is due to the impact of the CIBA, especially its decorrelation in frequency. 
On the one hand, $\tau_{353}$ is the division of $I_{353}$ by $B_{353}(T_{\rm obs})$ (Eq.~\ref{eq:divide}) and so
is contaminated by the CIBA at not only 353\,GHz\ but also  3000\,GHz; i.e., 
because the 3000\,GHz band is the only one in the Wien range, it has a 
strong weight in the determination of $T_{\rm obs}$. 
Furthermore, the CIBA at 3000\,GHz and the CIBA in the \Planck\ bands are weakly correlated, 
so that $T_{\rm obs}$ 
contains most of the information on the CIBA at 3000\,GHz. Therefore, 
through $I_{353}$ and $B_{353}(T_{\rm obs})$, the map of $\tau_{\rm 353}$ 
is affected by the CIBA on both the Rayleigh-Jeans and Wien sides, respectively, resulting 
in strong small scale fluctuations. On the other hand, because \radiance\ is obtained 
by integrating $I_\nu$ over frequency, it benefits from the fact that 
the CIBA decorrelates in frequency; i.e., the integral over frequency
of the CIBA is close to zero.

In order to relate \Planck\ dust emission to Galactic reddening
(Sect.~\ref{sec:extinction}), we also made a fit of the dust model to a
version of the \Planck\ and \IRAS\ data from which point sources had
been removed (Appendix~\ref{sec:pointsources}).  From this fit we
have also made maps of $\tau_{353}$ and \radiance.

\section{The Galactic dust emission observed by \Planck}
\label{sec:discussion}

\begin{table*}
\caption{\label{tab:summary} Summary of the mean and standard
deviation of the dust parameters for specific masks. }
\begin{center}
\begin{tabular}{lccccccccc}\specialrule{\lightrulewidth}{0pt}{0pt} \specialrule{\lightrulewidth}{1.5pt}{\belowrulesep}
Mask & Coverage &
$\langle T_{\rm obs}\rangle$ & 
$\sigma(T_{\rm obs})$ & 
$\langle \beta_{\rm obs}\rangle$ &
$\sigma(\beta_{\rm obs})$ &
$\langle \tau_{353}\rangle$ &
$\sigma(\tau_{353})$ &
$\langle \cal{R} \rangle$ &
$\sigma(\cal{R})$ \\
& [\%] & [K] & [K] & & & & & [W\,m$^{-2}$\,sr$^{-1}$] & [W\,m$^{-2}$\,sr$^{-1}$] \\ \midrule
Whole sky & $100$ & 19.7 & 1.4 & 1.62 & 0.10 & $45.0\times 10^{-7}$ & $53.3\times 10^{-7}$ & $15.2\times 10^{-8}$ & $16.9\times 10^{-8}$  \\ 
G56 & $\:\:57$ & 20.2 & 1.2 & 1.60 & 0.12 & $21.7\times 10^{-7}$ & $16.9\times 10^{-7}$ & $\:\:8.0\times 10^{-8}$ & $\:\:5.3\times 10^{-8}$  \\ 
$|b| > 15$\deg & $\:\:50$ & 20.3 & 1.3 & 1.59 & 0.12 & $18.5\times 10^{-7}$ & $13.2\times 10^{-7}$ & $\:\:7.1\times 10^{-8}$ & $\:\:4.1\times 10^{-8}$  \\ 
G45 & $\:\:47$ & 20.3 & 1.3 & 1.59 & 0.12 & $17.5\times 10^{-7}$ & $12.7\times 10^{-7}$ & $\:\:6.9\times 10^{-8}$ & $\:\:4.0\times 10^{-8}$  \\ 
G35 & $\:\:37$ & 20.5 & 1.3 & 1.57 & 0.13 & $13.8\times 10^{-7}$ & $\:\:8.8\times 10^{-7}$ & $\:\:5.7\times 10^{-8}$ & $\:\:2.8\times 10^{-8}$  \\ 
South cap & $\:\:17$ & 20.5 & 1.4 & 1.59 & 0.13 & $14.5\times 10^{-7}$ & $10.7\times 10^{-7}$ & $\:\:6.2\times 10^{-8}$ & $\:\:3.6\times 10^{-8}$  \\ 
Low $N_\ion{H}{i}$ & $\:\:11$ & 20.8 & 1.4 & 1.55 & 0.15 & $\:\:9.6\times 10^{-7}$ & $\:\:4.1\times 10^{-7}$ & $\:\:4.1\times 10^{-8}$ & $\:\:1.2\times 10^{-8}$  \\ 
Lowest 1\,\% & $\:\:\:\:1$ & 20.9 & 1.7 & 1.51 & 0.18 & $\:\:6.4\times 10^{-7}$ & $\:\:2.9\times 10^{-7}$ & $\:\:2.5\times 10^{-8}$ & $\:\:0.5\times 10^{-8}$  \\ 
\bottomrule[\lightrulewidth]
\end{tabular}
\end{center}
{\bf Note: } The angular resolution of all quantities is 5\arcm\ except for $\beta_{\rm obs}$
which is at 30\arcm. The $|b| > 15$\deg\ mask also includes the
restriction $N_\ion{H}{i} < 5.5 \times 10^{20}\,$cm$^{-2}$. The {Low
$N_\ion{H}{i}$} mask is the one shown in Fig.~\ref{fig:hi_mask},
left. The {South cap} mask corresponds to that developed for the
analysis in \citet{planck2013-XVII}.
The {Lowest 1\,\%} mask corresponds to the lowest 1\,\% $N_\ion{H}{i}$ column
density estimated using the LAB data.  
The remaining masks (G35, G45, and G56) are among those used in the \Planck\ 
cosmology papers \citep[e.g., ][]{planck2013-p08}, based in part on 
thresholding the \Planck\ 353\,GHz temperature map.
\end{table*}

\subsection{Spatial variations of the dust parameters and \radiance}
\label{sec:paramsr}

\begin{figure*}
\centering
\includegraphics[draft=false, angle=0]{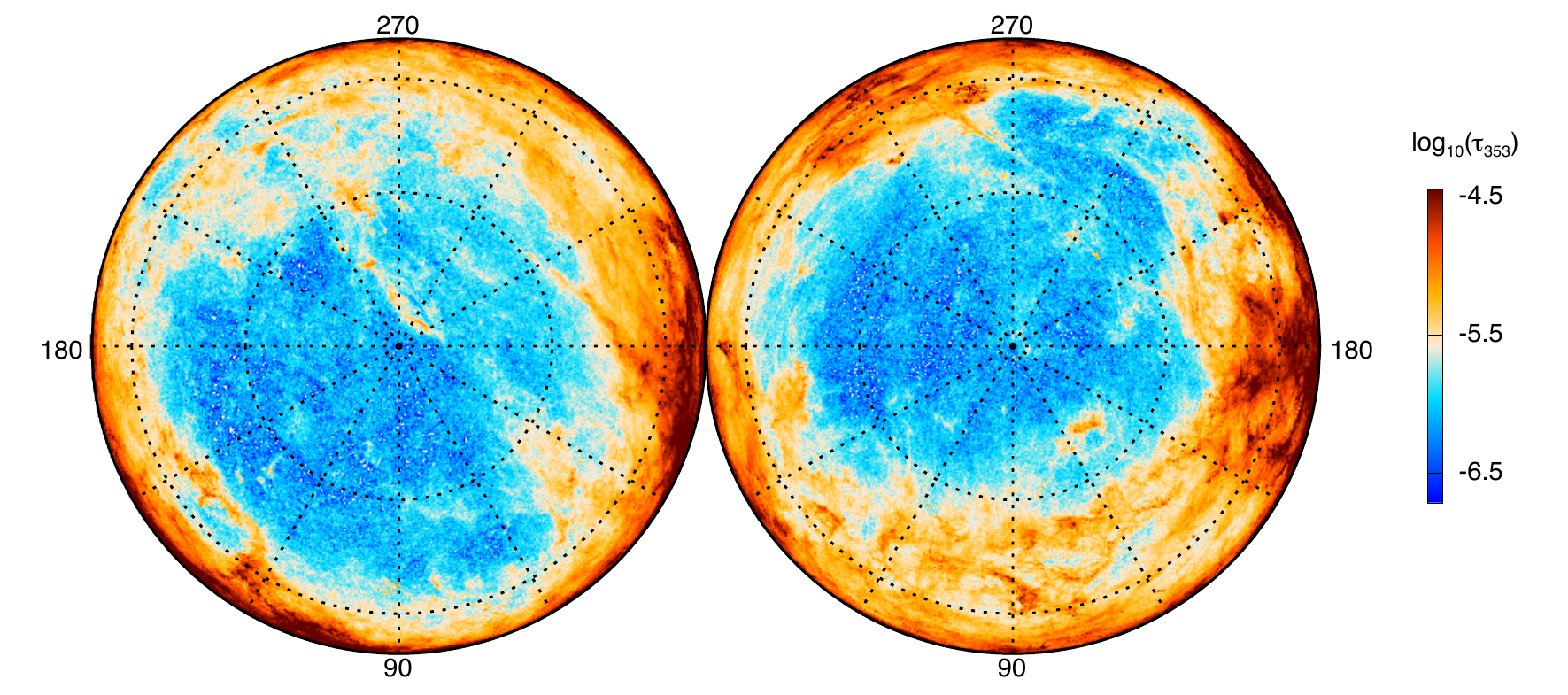}
\includegraphics[draft=false, angle=0]{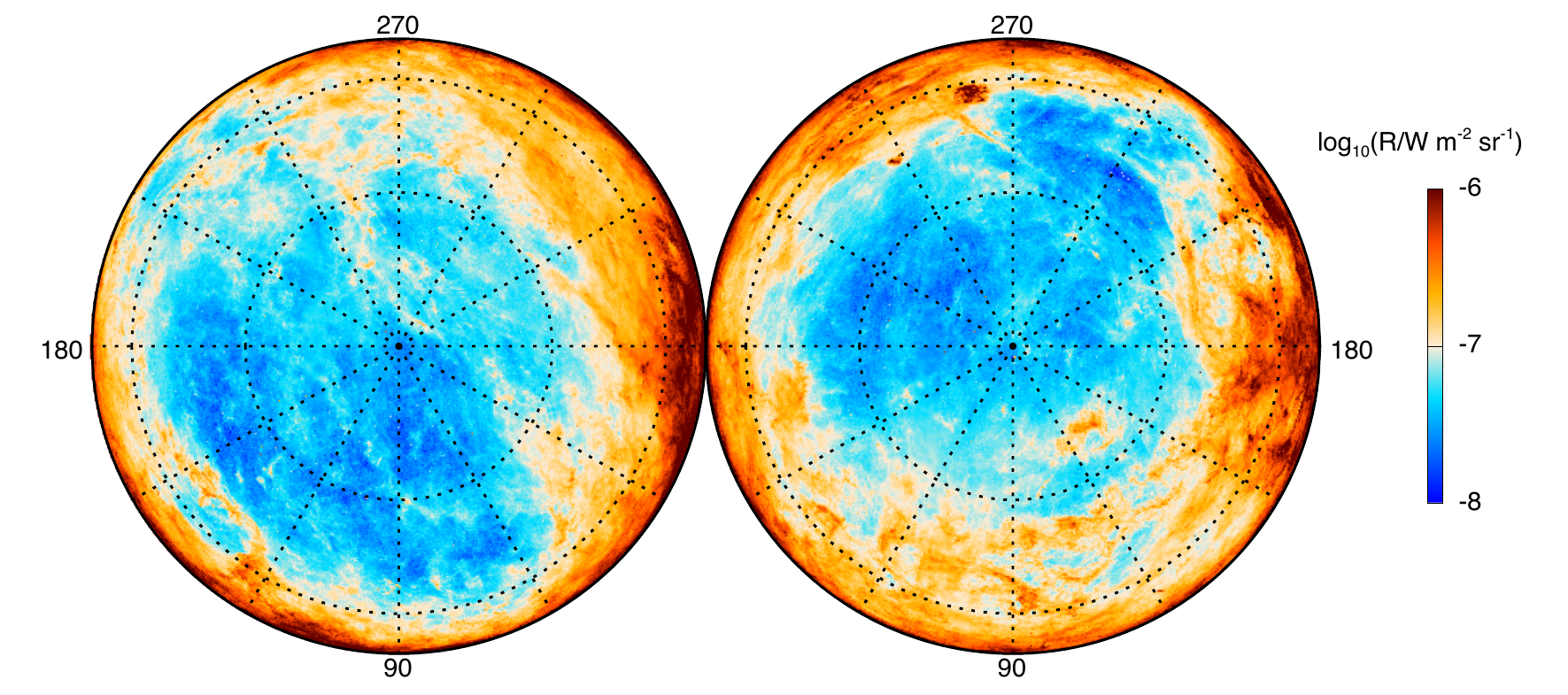}
\caption{\label{fig:fullres_polar_maps1} Polar views of $\log_{10}(\tau_{353})$
  (\textit{upper}) and $\log_{10}(\cal{R})$ (\textit{lower}).}
\end{figure*}

\begin{figure*}
\centering
\includegraphics[draft=false, angle=0]{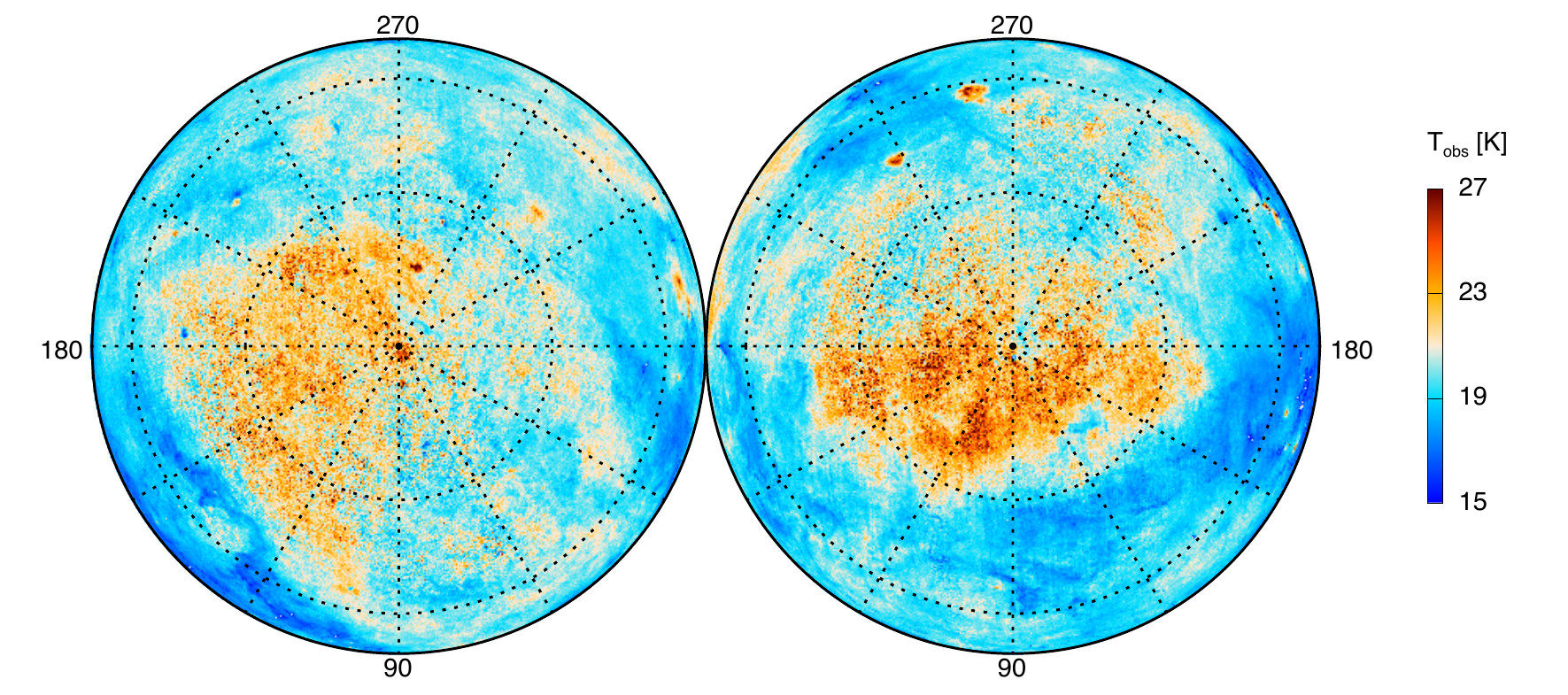}
\includegraphics[draft=false, angle=0]{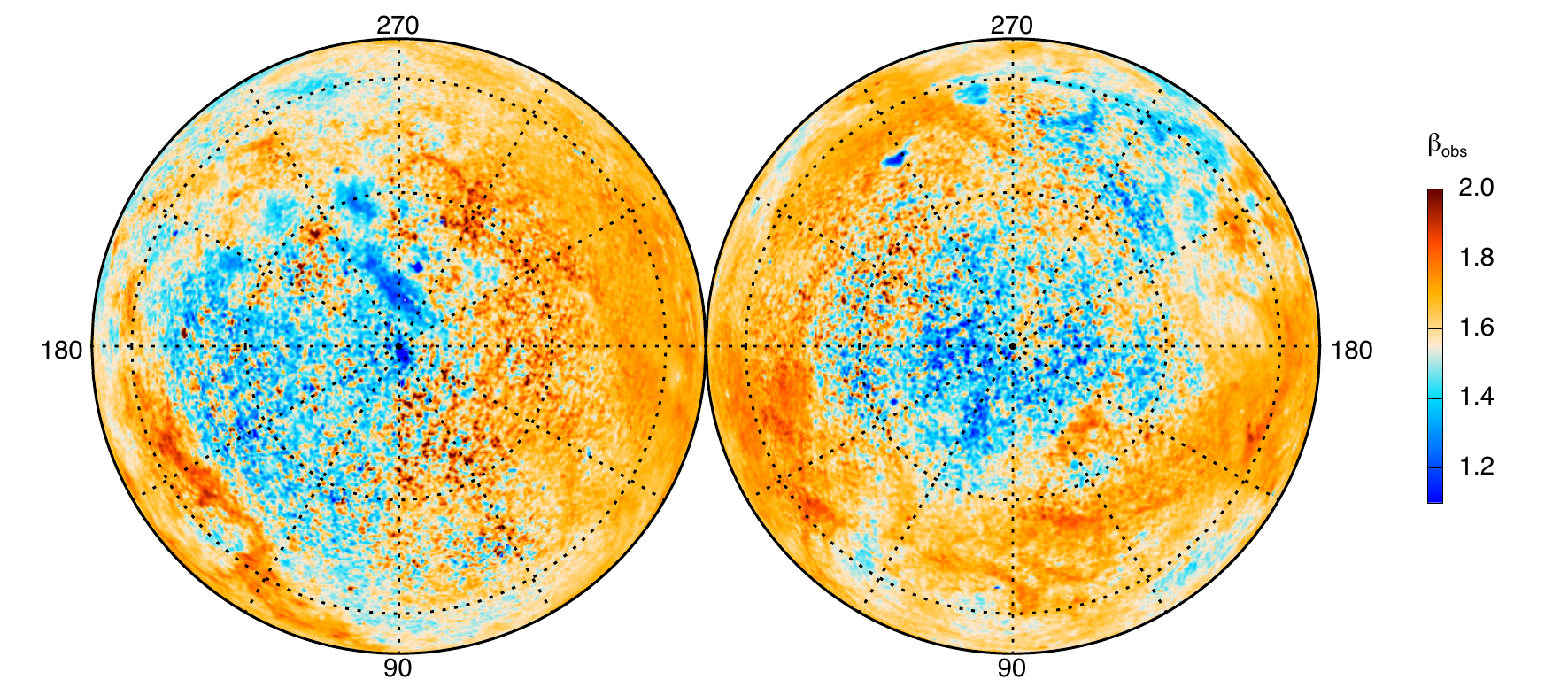}
\caption{\label{fig:fullres_polar_maps2} Polar views of $T_{\rm obs}$
  (\textit{upper}) and $\beta_{\rm obs}$ (\textit{lower}).}
\end{figure*}

The all-sky maps of $T_{\rm obs}$, $\beta_{\rm obs}$, and $\tau_{353}$
in Fig.~\ref{fig:fullres_maps} -- represented as well in a polar
orthographic projection  in
in  Figs.~\ref{fig:fullres_polar_maps1} and
\ref{fig:fullres_polar_maps2} to show details in the high-latitude sky -- 
represent the first attempt to fit these
three parameters at the same time over the whole sky. 
Together, these maps of the fit parameters provide information 
on the dust SED and, quite likely, on the dust properties and their 
variations with interstellar environment.
They are complemented by the map of \radiance\ in Figs.~\ref{fig:L_allsky} and \ref{fig:fullres_polar_maps1}.
Here we discuss only some broad features of these maps, 
leaving more detailed analysis to future work.

The mean and standard deviation of $T_{\rm obs}$, $\beta_{\rm obs}$,
$\tau_{353}$, and \radiance\ are given in Table~\ref{tab:summary} for
several different masks ranked in order of decreasing dust
contamination, using $\sigma(\tau_{353})$ as a proxy, and so (mostly)
of decreasing sky coverage.  These include some masks used in \Planck\
cosmology papers  \citep[e.g., ][]{planck2013-p08}. 
Note how the ranking is reflected in the means and
standard deviations listed.

Over the whole sky, the mean of $\beta_{\rm obs}$ is 1.62 and its
standard deviation is 0.10. The mean of $T_{\rm obs}$ is 19.7\,K and
its standard deviation is 1.4\,K. The distribution function of $T_{\rm
  obs}$ is slightly positively skewed with a high tail that extends up
to 60\,K. Only about 100 out of the more than $50$~million pixels of the
$N_{\rm side}=2048$ map have $T_{\rm obs}<13$\,K.\footnote{This is not in
contradiction with the cold clumps detected in the \Planck\ data 
\citep{planck2011-7.7b}, some with temperature as low as 7\,K.
These clumps were identified after removing a hotter
background/foreground emission. We do not obtain such low values of
$T_{\rm obs}$ because we model the observed specific intensity on each
line of sight.}

The maps of $T_{\rm obs}$ and $\tau_{353}$ presented here should be
compared with the ones published as \Planck\ early results by
\citet{planck2011-7.0}.  Apart from the
facts that we use a more recent release of \Planck\ data (with a
different calibration of the 545 and 857\,GHz and with ZE removed) and
a slightly different approach to  the offset
determination,\footnote{Both studies use the correlation with \hi\ to
  set the offsets but \citet{planck2011-7.0} used a higher
  threshold in column density ($N_\ion{H}{i} < 1.2 \times
  10^{21}$\,cm$^{-2}$) than adopted here ($N_\ion{H}{i} < 2 \times
  10^{20}$\,cm$^{-2}$).}
the main difference is that we fit for $\beta_{\rm obs}$ while
\citet{planck2011-7.0} used a fixed value, $\beta_{\rm
  obs}=1.8$, a convention shared among all the \Planck\ Early Papers
dedicated to dust emission
\citep{planck2011-7.13,planck2011-7.12,planck2011-7.3}.
Even with these differences in data and methodology, the maps of
$T_{\rm obs}$ are remarkably similar.  The map of $T_{\rm obs}$
presented here is higher by about 1\,K than that of
\citet{planck2011-7.0}, due principally to the modification
of the calibration of the 545 and 857\,GHz channels.

Like in \citet{planck2011-7.0}, the lowest $T_{\rm obs}$
values are found in the outer Galaxy and in molecular clouds. In
general the well-known molecular clouds have a lower $T_{\rm obs}$
(15--17\,K) and higher $\beta_{\rm obs}$ (around 1.8) than in the
diffuse ISM. This trend is compatible with the result of
\citet{planck2011-7.13} who reported a steepening of the SED
from diffuse to molecular areas in the Taurus molecular cloud.

Small-scale regions of higher $T_{\rm obs}$ are seen along the
Galactic plane and in many of the Gould Belt clouds, most probably
related to the local production of dust-heating photons in Galactic
star forming regions. The Magellanic Clouds are clearly visible in the
parameter maps with a higher $T_{\rm obs}$ and lower $\beta_{\rm obs}$
\citep{planck2011-6.4b}.

The main noticeable difference with respect to the early results of
\citet{planck2011-7.0} is the lower $T_{\rm obs}$ found
here in the inner Galactic plane. This is due to the fact that we fit
for $\beta_{\rm obs}$, which appears to have a systematically higher
value in the inner Galactic plane, in the range $1.8$--$2.0$. The
impact of noise and the CIBA is obviously negligible in this bright area
of the sky.  The higher $\beta_{\rm obs}$ found here clearly provides
a better representation of the SED, as shown also by \citet{planck2013-XIV}. 
This steepening of the dust SED in
the inner Galactic plane is also compatible with the analysis of
\Herschel\ observations of that region by \citet{paradis2012a}.

One striking feature of the $T_{\rm obs}$ (polar) map is the increase toward
both Galactic poles. 
Selecting the pixels corresponding to the lowest 1\,\% $N_\ion{H}{i}$,
the mean $T_{\rm obs}$ is 20.9\,K and the mean $\beta_{\rm obs}$ is
$1.51$. This systematic increase of $T_{\rm obs}$ was also visible in
the early all-sky map of \citet{planck2011-7.0} that used a
constant $\beta_{\rm obs}$, different offsets, a different 3000\,GHz
map and no ZE removal for the \Planck\ data.
In addition, the values we report for the south Galactic pole mask
(mean $T_{\rm obs} = 20.5$\,K, $\sigma(T_{\rm obs})= 1.4$\,K, mean
$\beta_{\rm obs}=1.59$, $\sigma(\beta_{\rm obs})=0.13$) are compatible
with the ones reported by \citet{planck2013-XVII} using a
correlation method that is insensitive to offsets and ZE removal.
The nature of this increase of $T_{\rm obs}$ over a large scale
in the most diffuse areas
at high Galactic latitudes, incidentally correlated with lower values
of $\beta_{\rm obs}$ (see Fig.~\ref{fig:fullres_polar_maps2}), is
still to be understood (see Sect.~\ref{sec:radiation_field}) but it is
unlikely to be caused by a bias 
by instrumental noise or the CIBA, which both create small-scale fluctuations.

\subsection{$\beta_{\rm obs}$ -- $T_{\rm obs}$ relation}
\label{sec:betaT}

\begin{figure}
\includegraphics[draft=false, angle=0]{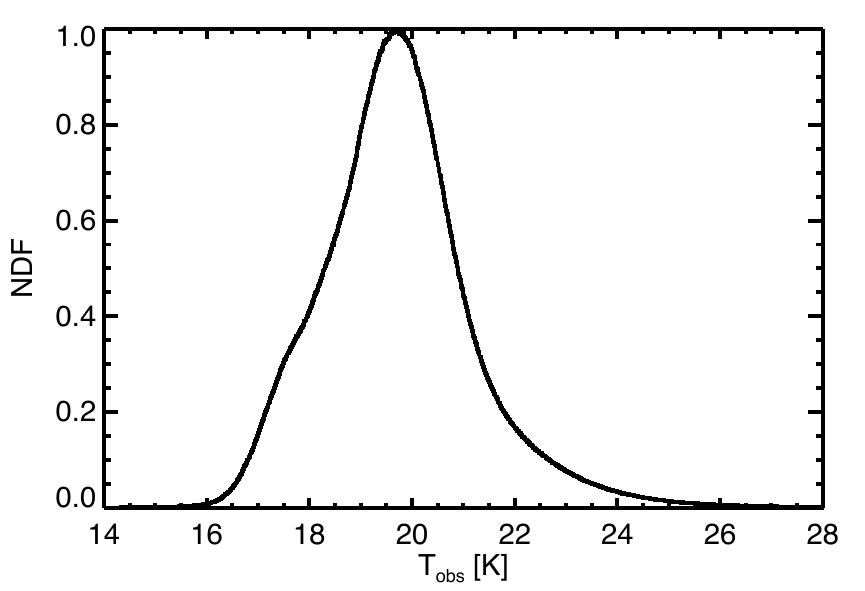}
\includegraphics[draft=false, angle=0]{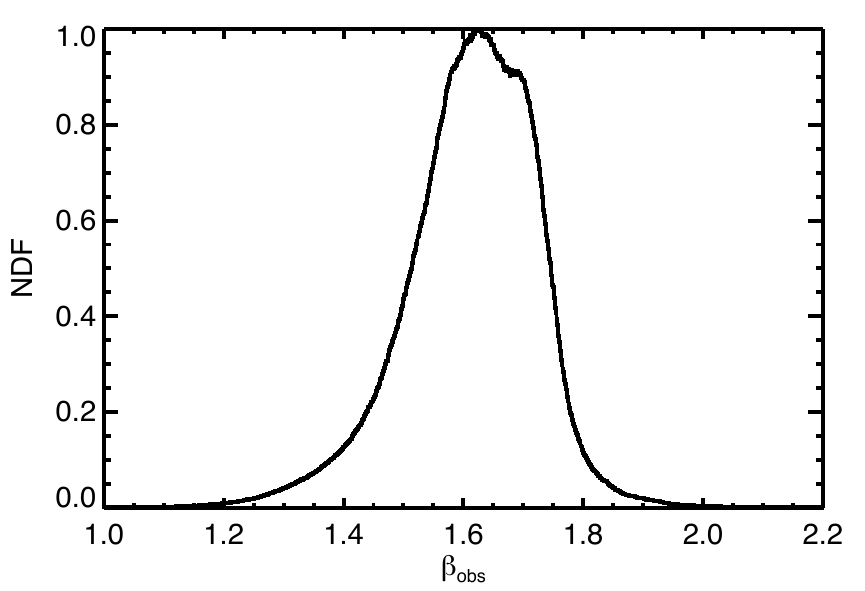}
\includegraphics[draft=false, angle=0]{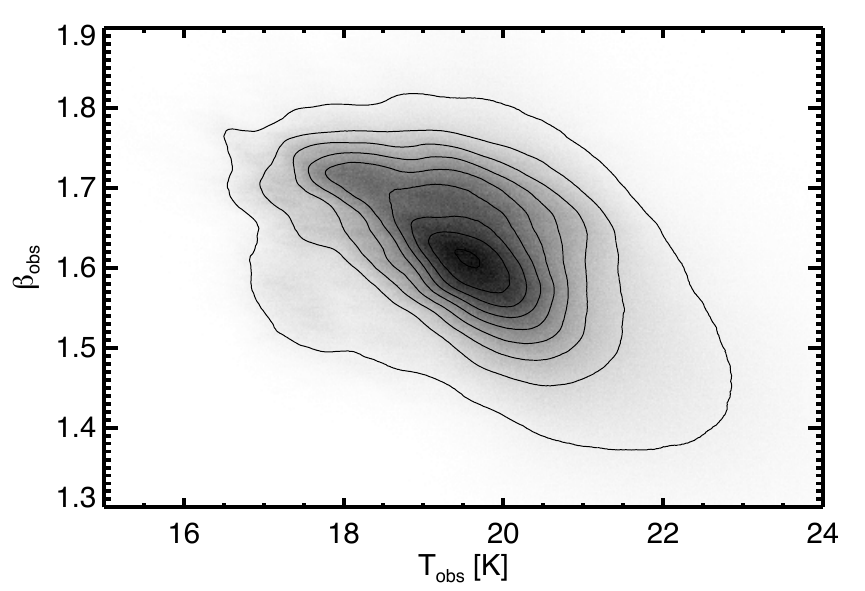}
\caption{\label{fig:T_Beta}
Normalized distribution function of 
$T_{\rm obs}$ (\textit{upper}) and $\beta_{\rm obs}$ (\textit{middle}) for the whole sky. 
The \textit{lower} panel shows the density plot of $\beta_{\rm obs}$ vs.\
$T_{\rm obs}$, revealing an overall anticorrelation. 
The grey scale and the black contours
show the density of points on a linear scale with contours equally
spaced. The dispersions are $\sigma(\beta_{\rm obs}) = 0.1$ and 
$\sigma(T_{\rm obs})= 1.4$\,K (Table~\ref{tab:summary}). 
}
\end{figure}

The all-sky maps of the fit parameters (Fig.~\ref{fig:fullres_maps})
reveal some spatial correlation between the parameters.  This is
especially clear between $\beta_{\rm obs}$ and $T_{\rm obs}$ as
illustrated in Fig.~\ref{fig:T_Beta}, lower, using results for all 
pixels on the sky. Because it includes so many different regions, this
two-dimensional histogram can reveal only global trends, here the
general anticorrelation.

This anticorrelation is visible in the faintest parts of the sky, at both small
and large scales. It is
also seen at the scale of clouds; the Gould Belt clouds have a low
$T_{\rm obs}$ (15--16\,K) and high $\beta_{\rm obs}$ ($\sim 1.8$). 
Several other studies have highlighted similar $\beta_{\rm obs} - T_{\rm obs}$
anticorrelations from observations of specific regions on the sky
\citep{dupac2003a,desert2008,paradis2010,planck2011-7.13}.
On the other hand, this behaviour does not extend to the
Galactic plane where the two parameters seem to be more correlated
than anticorrelated.  

As pointed out in Sect.~\ref{sec:implementation} \citep[see also][]{shetty2009}, 
instrumental noise is an obvious 
candidate that might create a $\beta - T$ anticorrelation.
However, the fractional variations of $\beta_{\rm obs}$ and $T_{\rm obs}$ observed 
here over most of the sky significantly exceed the statistical uncertainties 
of these parameters taking into account noise and calibration uncertainties
(see Fig.~\ref{fig:fullres_error_maps}).

On the other hand, as shown in Sect.~\ref{sec:noise_and_CIBA} and
Appendix~\ref{sec:MC_CIBA}, for faint dust emission the CIBA can
produce significant variations of $T_{\rm obs}$ and $\beta_{\rm obs}$
at small scales, and although this effect is in fact observed, it is
not accounted for in the error budget.  To be quantitative, in
the pixels corresponding to the lowest 1\,\% values of $N_\ion{H}{i}$ the
observed standard deviations of these parameters are the largest --
$\sigma(T_{\rm obs})=1.7$\,K and $\sigma(\beta_{\rm obs})=0.18$ --
while over the whole sky $\sigma(T_{\rm obs})=1.4$\,K and
$\sigma(\beta_{\rm obs})=0.10$ (see Table~\ref{tab:summary}).  
Based on the Monte-Carlo simulations presented in Sect.~\ref{sec:two-step},
for values of $\tau_{353}$ typical of the faintest
1\,\% pixels of the sky the noise and CIBA produce
fluctuations of $T_{\rm obs}$ and $\beta_{\rm obs}$ of the order of
$\delta T_{\rm obs}=1.7$\,K and $\delta\beta_{\rm obs}=0.15$, providing
a credible explanation for the magnitude of the small-scale variations
observed in that mask (Table~\ref{tab:summary}). 

Even though noise and the CIBA seem to be responsible for the $\beta - T$ anticorrelation
in the most diffuse areas of the sky, they can cause only small-scale fluctuations
because of their flat power spectra.  
Because the monopole of the CIB was removed from the map, the CIBA
does not bias $\beta_{\rm obs}$ and $T_{\rm obs}$ globally on the sky, and 
cannot produce large-scale variations like the increase of $T_{\rm obs}$ 
toward the Galactic poles. 
We have also checked that these results are largely unaffected by the ZE removal 
(Appendix~\ref{sec:PIP82}).

In brighter regions, our Monte-Carlo simulations (Sect.~\ref{sec:two-step}; )
indicate that noise and the CIBA introduce variations in 
$T_{\rm obs}$ and $\beta_{\rm obs}$ (Fig.~\ref{fig:MC_vs_tau})
that are below the observed dispersions (Fig.~\ref{fig:T_Beta}, lower).
This is true for more than 90\% of the sky.
One can appreciate these results by looking directly at the parameter maps
(Fig.~\ref{fig:fullres_maps}).  Away from the most diffuse areas of
the sky, where $T_{\rm obs}$ and $\beta_{\rm obs}$ vary at small scale
mostly due to the CIBA, the main clouds and interstellar structures
that are seen in $I_\nu$ and in $\tau_{353}$ can be recognized in the
maps of $T_{\rm obs}$ and $\beta_{\rm obs}$.  

The broad spectral coverage, the high signal-to-noise (and high
signal-to-CIBA) ratio of the data on more than 90\,\% of the sky, and
the methodology used to minimize the effects of noise and the CIBA on the
fit parameters, combine to produce values of $T_{\rm obs}$ and
$\beta_{\rm obs}$ with uncertainties of a few percent, much smaller
than their dispersions over the sky (Table~\ref{tab:summary}).
We conclude that on most of the sky, the relation between 
$\beta_{\rm obs}$ and $T_{\rm obs}$ is not an artifact of the data processing 
(zero levels, ZE correction) or due to noise or the CIBA. This conclusion also
holds for the large-scale variations of $\beta_{\rm obs}$ and $T_{\rm obs}$ at
high Galactic latitude.
On most of the sky, the systematic variations of $T_{\rm obs}$ and $\beta_{\rm obs}$ are related
to real changes in the shape of SED of the interstellar dust emission.  

Even with data-related effects mitigated, the interpretation of
the relationship between the MBB parameters is complex.  In particular at this point
it is difficult to be definitive about the origin of the relationship between
 $T_{\rm obs}$ and $\beta_{\rm obs}$.  
It depends on details of radiative transfer, of variations in $U$ along 
the line of sight, and of variations in grain structure and size distribution.  
To identify the relative roles of dust evolution and line-of-sight integration
effects in this observed phenomenon, detailed studies of specific spatially-coherent
objects in various interstellar environments and at all scales are
needed.

\subsection{Dust SED in the diffuse ISM}

As described in \citet{planck2013-p03f}, the calibration scheme for
the 545 and 857\,GHz data has changed since the \Planck\ Early
Results. These channels are no longer calibrated using the \FIRAS\ data,
but instead rely on observations of planets as for \IRAS, \DIRBE, and
\Herschel. Compared to the previous situation, the calibration
factor has been divided by 1.15 at 545\,GHz and 1.07 at 857\,GHz
\citep{planck2013-p03f}, so that the specific intensities are now
lower.

There are two main impacts on dust modelling. First, the shape of the
dust SED is modified, changing the average $T_{\rm obs}$ and
$\beta_{\rm obs}$. 
The \FIRAS\ average dust SED of the diffuse ISM mask ($|b|>15$\deg\ and
$N_\ion{H}{i} < 5.5 \times 10^{20}\,$cm$^{-2}$, following the definition
of \citet{compiegne2011}) was modelled with $T_{\rm obs}=17.9$\,K and
$\beta_{\rm obs}=1.84$ by \citet{planck2011-7.12},
compatible with the average SED that they found in selected high
Galactic latitude fields using \IRAS\ and the early \Planck\ data.\footnote{This
  was expected because the 857, 545, and 353\,GHz data used in that study
  were calibrated on \FIRAS.} 
With the new calibration, the mean
values found for the same mask are significantly different: $\langle
T_{\rm obs}\rangle=20.3$\,K and $\langle \beta_{\rm obs} \rangle=1.59$
(see Table~\ref{tab:summary}).
The dust parameters  found here are similar to those found in external galaxies
with \Herschel,\footnote{
The calibration of each of \Herschel\ and \Planck\ at 
545 and 857\,GHz is based on observations of planets and uses the same 
model of planetary emission \citep{planck2013-p03f}.
}
even though the 
\Herschel\ frequency coverage is not as extensive \citep[e.g., ][]{dale2012}. 

The second impact is on the value of the dust opacity
$\sigma_{\mathrm{e}\,\nu}=\tau_{\nu}/N_{\rm H}$. The increase in $T_{\rm obs}$ due to
the recalibration lowers $\tau_\nu$ and the opacity. At 250\,$\mu$m
(1\,200\,GHz), a reference wavelength often used, \citet{boulanger1996}
obtained $\sigma_{\mathrm{e}\,1200} = 1.0\times10^{-25}$\,cm$^2$ while here for
the $|b| > 15$\deg\ mask we obtain $\sigma_{\mathrm{e}\,1200} =
0.49\times10^{-25}$\,cm$^2$ (from $\sigma_{\mathrm{e}\,353}$ in
Table~\ref{tab:summary2} and $\beta_{\rm obs} =1.59$).

Changing $\beta_{\rm obs} $ directly affects the assessment of the
material needed to explain the observed thermal emission; in a MBB fit
to the SED, a lower $\beta_{\rm obs}$ leads to a higher $T_{\rm obs}$
and therefore to a lower optical depth, which in turn could be
interpreted as a lower column density (or mass), or a lower
opacity.
We also note that the mean value of $\beta_{\rm obs}$ found is lower
than used for some components in dust models, like graphite in
\citet{draine2007a} where $\beta = 2$; when fitting with such a model,
a higher radiation field strength $U$ would be needed.

\begin{figure*}
\centering
\includegraphics[draft=false, angle=0]{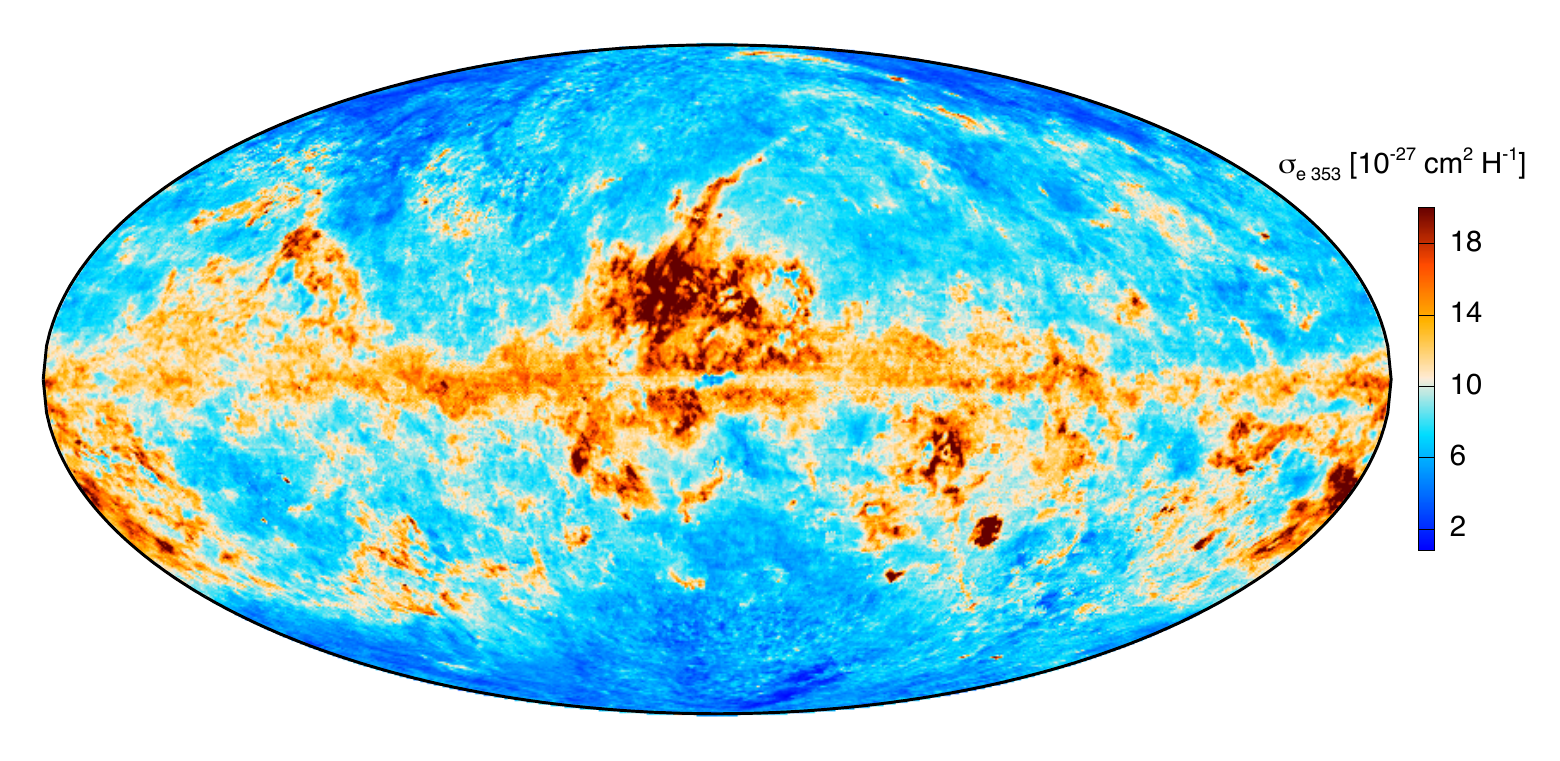}
\includegraphics[draft=false, angle=0]{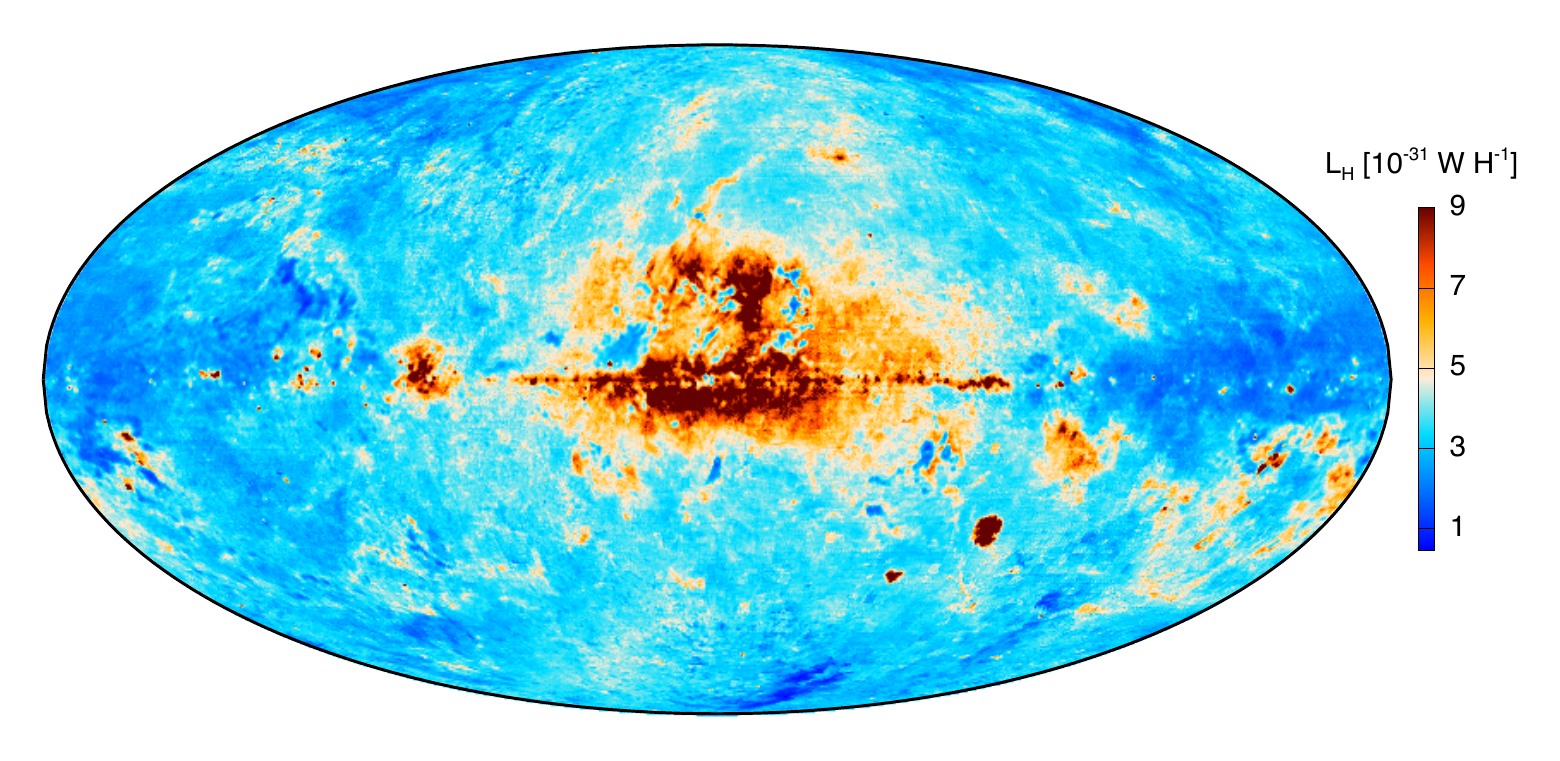}
\caption{\label{fig:allsky_tau_and_L_overNH} All-sky maps of \opacity\
(\textit{upper}) and \Ldust\ (\textit{lower}). The gas column density $N_{\rm H}$ is
$N_\ion{H}{i}+2\,X_{\rm CO}\,W_{\rm CO}$ where $N_\ion{H}{i}$ is from the
LAB data, $W_{\rm CO}$ is from the \Planck\ $^{12}$CO $J$=1$\rightarrow$0 map (type
3), and $X_{\rm CO}=2\times 10^{20}$\,\XCOunits.}
\end{figure*}

\begin{figure*}
\centering
\includegraphics[draft=false, angle=0]{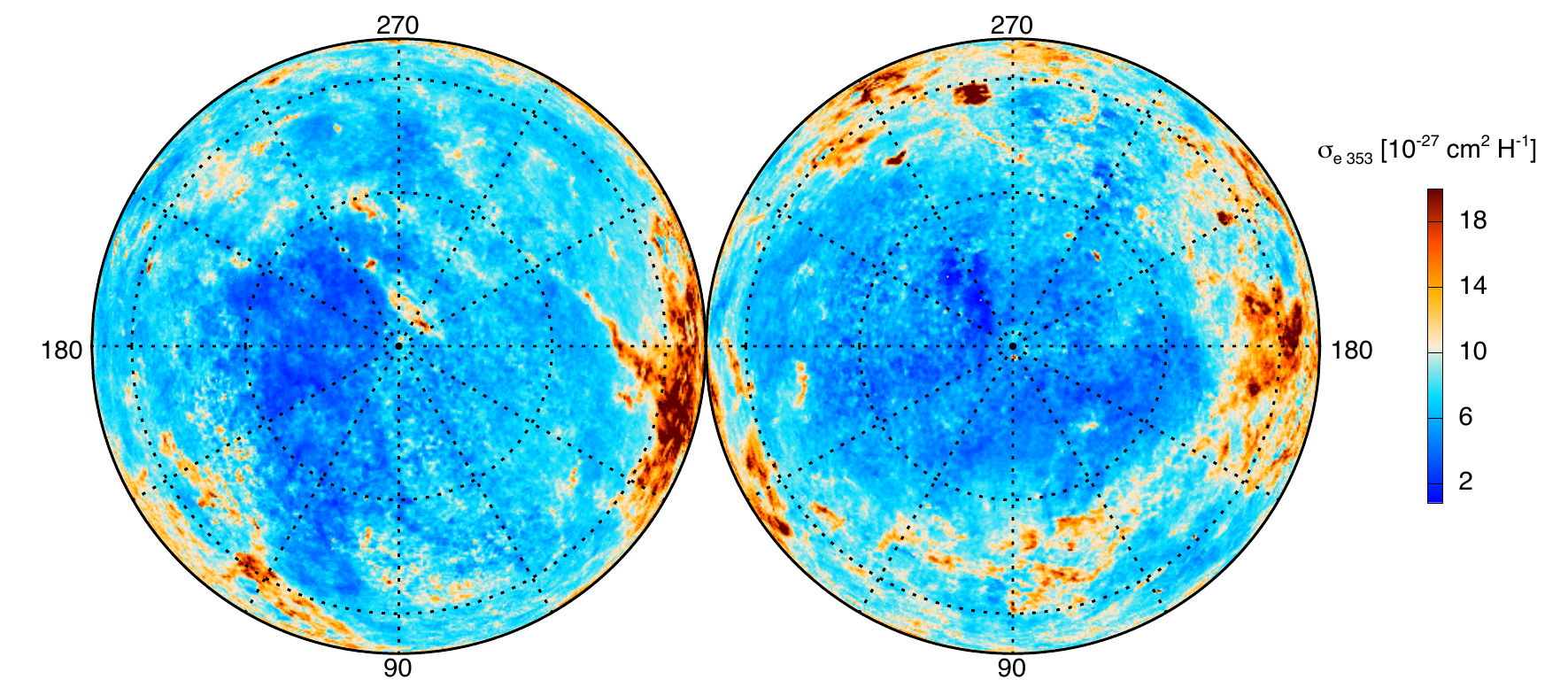}
\includegraphics[draft=false, angle=0]{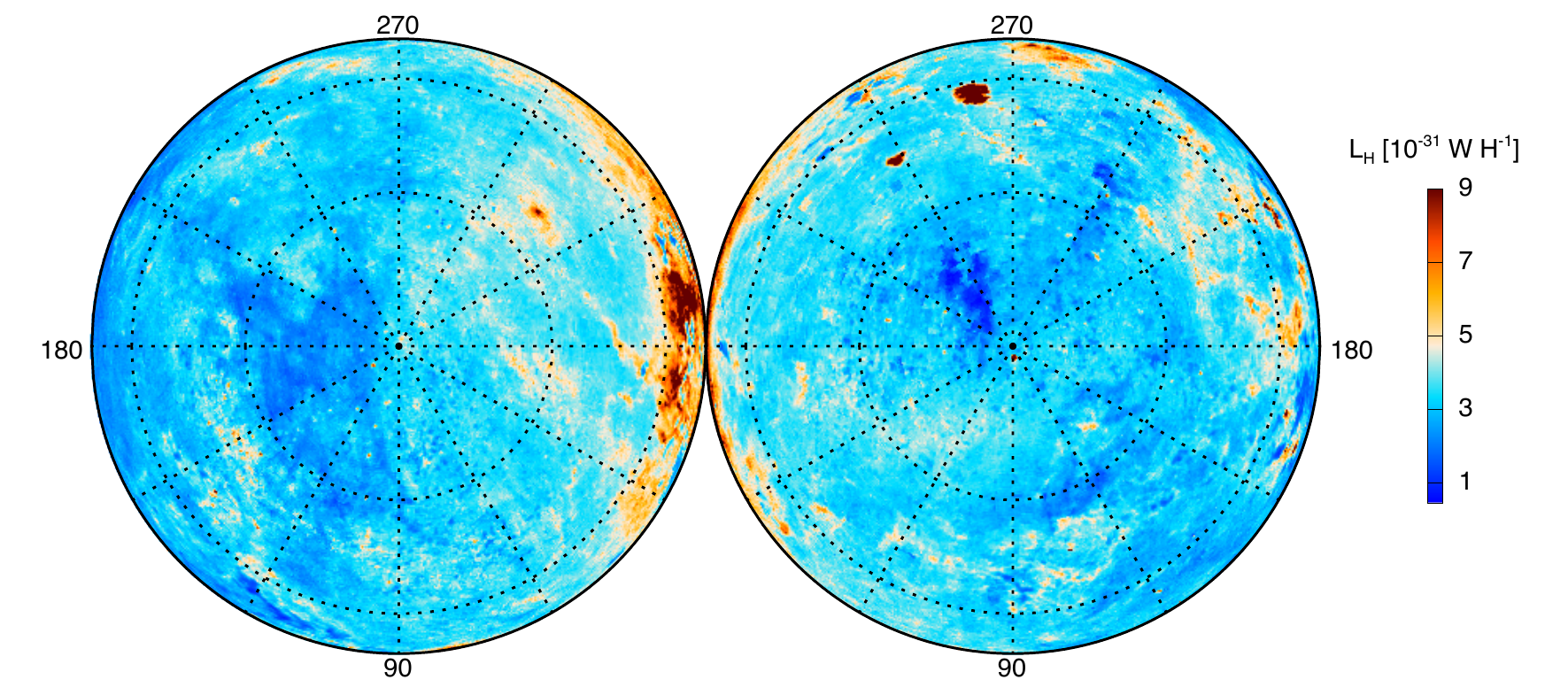}
\caption{\label{fig:fullres_polar_maps3} Polar view of \opacity\ (\textit{upper}) and \Ldust\ (\textit{lower}).}
\end{figure*}

\begin{table*}
  \caption{\label{tab:summary2} Summary of the mean and standard
deviation of the dust opacity and dust specific luminosity for
different masks.}
\begin{center}
\begin{tabular}{lcccc}\specialrule{\lightrulewidth}{0pt}{0pt} \specialrule{\lightrulewidth}{1.5pt}{\belowrulesep}
Mask & 
$\langle \sigma_{\mathrm{e}\,353} \rangle$ &
$\sigma(\sigma_{\mathrm{e}\,353})$ &
$\langle L_{\rm H} \rangle$ &
$\sigma(L_{\rm H})$ \\
& [cm$^2$\,H$^{-1}$] & [cm$^2$\,H$^{-1}$] & [W\,H$^{-1}$] & [W\,H$^{-1}$] \\ \midrule
Whole sky & $8.4\times 10^{-27}$ & $3.0\times 10^{-27}$ & $3.5\times 10^{-31}$ & $0.9\times 10^{-31}$ \\
G56 & $7.1\times 10^{-27}$ & $1.9\times 10^{-27}$ & $3.4\times 10^{-31}$ & $0.6\times 10^{-31}$ \\
$|b| > 15$\deg & $7.0\times 10^{-27}$ & $2.0\times 10^{-27}$ & $3.4\times 10^{-31}$ & $0.6\times 10^{-31}$ \\
G45 & $6.8\times 10^{-27}$ & $1.8\times 10^{-27}$ & $3.3\times 10^{-31}$ & $0.6\times 10^{-31}$ \\
G35 & $6.5\times 10^{-27}$ & $1.8\times 10^{-27}$ & $3.3\times 10^{-31}$ & $0.6\times 10^{-31}$ \\
South cap & $6.5\times 10^{-27}$ & $1.9\times 10^{-27}$ & $3.4\times 10^{-31}$ & $0.5\times 10^{-31}$ \\
Low $N_\ion{H}{i}$ & $6.6\times 10^{-27}$ & $1.7\times 10^{-27}$ & $3.5\times 10^{-31}$ & $0.6\times 10^{-31}$ \\
Lowest 1\,\% & $7.9\times 10^{-27}$ & $1.9\times 10^{-27}$ & $3.8\times 10^{-31}$ & $0.7\times 10^{-31}$ \\
\bottomrule[\lightrulewidth]
\end{tabular}
\end{center}
{\bf Note: } All quantities were computed using maps at 30\arcm\
resolution. The map of $N_{\rm H}$ is a combination of \hi\ (21\,cm LAB
data) and CO (\Planck) assuming $X_{\rm CO}=2\times
10^{20}$\,\XCOunits\ following \citet{bolatto2013}. This estimate of
$N_{\rm H}$ is a lower limit as it does not account for the ionized
gas and the molecular gas not detected via CO.  See
Table~\ref{tab:summary} for the definitions of each mask. 
\end{table*}

\section{Dust emission in relation to gas column density}
\label{sec:column_density}

In the previous section we have described the properties of the
parameters that define the dust SED.  Now we concentrate on the link
between the dust emission and the interstellar gas column density,
following on many detailed studies in environments
from the diffuse ISM \citep{boulanger1988,boulanger1996} to molecular clouds
\citep{pineda2008,goodman2009}.

Here the \emph{estimate} of gas column density, $N_{\rm H}$, accounts
for atomic and molecular gas:
\begin{equation}
\label{eq:ntot}
N_{\rm H} = N_\ion{H}{i} + 2\,X_{\rm CO}\,  W_{\rm CO}\, ,
\end{equation}
where the $N_\ion{H}{i}$ is from the LAB data assuming optically-thin
emission, $W_{\rm CO}$ is from the \Planck\ $^{12}$CO $J$=1$\rightarrow$0 map (type
3) \citep{planck2013-p03a}, and $X_{\rm CO}$ is not constant but is
typically $2\times10^{20}$\,\XCOunits\ \citep{bolatto2013}.  ``Dark''
neutral matter \citep{planck2011-7.0}
is by definition left out in this formulation,
though it is among the total that can be traced by $\gamma$-rays
\citep{grenier2005}.
Ionized gas is left out for lack of a proper template.

\subsection{Opacity and dust specific luminosity}

The optical depth ($\tau_{353}$ here) is often taken as a tracer of
$N_{\rm H}$ but this is only accurate if the opacity is constant
(Eq.~\ref{eq:opacity}).  This requirement can be assessed in the
all-sky map of the opacity $\sigma_{\mathrm{e}\,353} = \tau_{353}/N_{\rm H}$ in
Figs.~\ref{fig:allsky_tau_and_L_overNH} and
\ref{fig:fullres_polar_maps3}, smoothed to 30\arcm.  Although the
large dynamic range over the $\tau_{353}$ sky is greatly compressed,
so that a linear scale can be used, it is clear that there are changes
in opacity, even in the diffuse atomic ISM in the high-latitude sky
where $N_{\rm H}$ is well measured.  Related to these changes in
opacity are changes in the equilibrium dust temperature
\citep{planck2011-7.12,planck2013-XVII}, driving
complementary changes in the SED parameter $\tau_{353}$ through
Eq.~\ref{eq:mbb}.  This demonstrates how $\tau_{353}$ is compromised
as a tracer of column density.

We saw in Sect.~\ref{sec:luminosity} how \radiance\ compensates for
such effects, being smoother than $\tau_{353}$.  This is expected to
carry over into the dust specific luminosity
\begin{equation}
\label{eq:dsl}
L_{\rm H}= 4\pi{\cal{R}}/N_{\rm H}\, ,
\end{equation}
also shown as an all-sky map in
Figs.~\ref{fig:allsky_tau_and_L_overNH} and
\ref{fig:fullres_polar_maps3}.  At high latitudes this is indeed more
uniform.  This uniformity and the excursions to both higher and lower
values at higher column densities relating to the ambient ISRF are
taken up in Sect.~\ref{sec:radiation_field}.

\begin{figure}
\includegraphics[draft=false, angle=0]{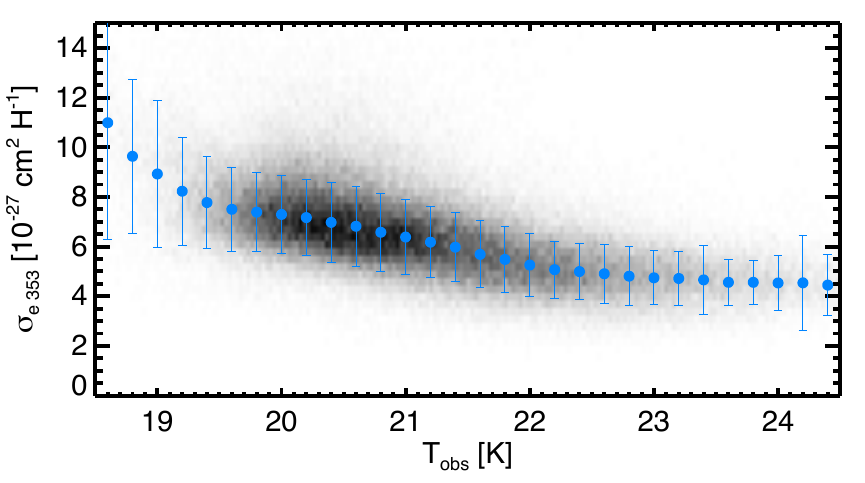}
\includegraphics[draft=false, angle=0]{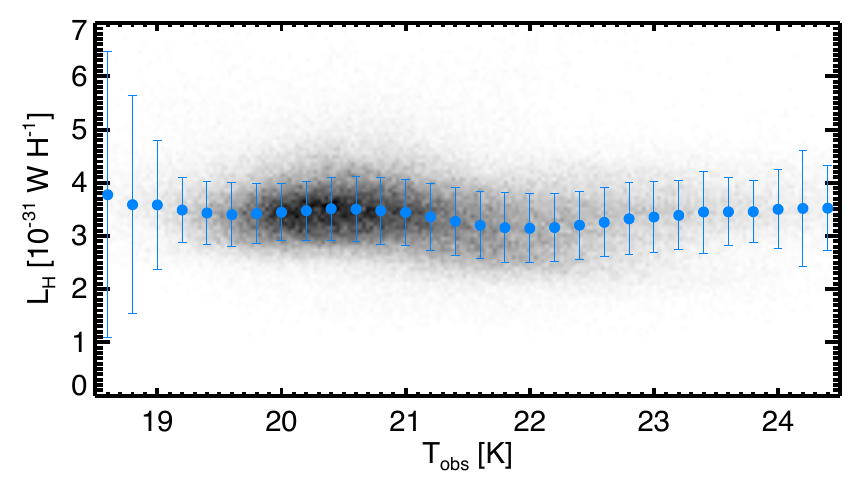}
\caption{\label{fig:tau_over_nh_vs_t} Variation of 
  \opacity\ (\textit{upper}) and \Ldust\ (\textit{lower}) as a function of $T_{\rm obs}$ in the diffuse
  ISM. Data used were smoothed to 30\arcm, the angular resolution of
  the \hi\ data. Only pixels where $1 \times 10^{20} < N_\ion{H}{i} <
  2.5 \times 10^{20}$\,cm$^{-2}$ were selected. The greyscale shows
  the point density in the two-dimensional histogram while the blue
  points indicate the average and standard deviation of \Ldust\ and
  \opacity\ in bins of $T_{\rm obs}$.  }
\end{figure}

Complementing the above, for low-column-density lines of sight with $1
\times 10^{20} < N_{\rm H} < 2.5 \times 10^{20}$\,cm$^{-2}$, the
dependence of \opacity\ on $T_{\rm obs}$, and by contrast the relative
lack of dependence of \Ldust\ on $T_{\rm obs}$, are evident in
Fig.~\ref{fig:tau_over_nh_vs_t}, lower.

The statistics of \opacity\ and \Ldust\ for the various masks are
presented in Table~\ref{tab:summary2} using $X_{\rm CO} = 2
\times10^{20}$\,\XCOunits\ where relevant.\footnote{
Use of a constant $X_{\rm CO}$ is certainly not realistic, 
given the large ranges in density and temperature covered. 
On the other hand, it is used here only to provide basic statistics 
of \opacity\ and \Ldust\ for the ``Whole sky" mask in Table~\ref{tab:summary2}.
The sky fraction with significant CO emission, greater than 0.15\,K\,km\,s$^{-1}$,
is only about 18\% and in all of the other masks considered here
CO does not contribute.}
Note how for these normalized quantities the systematic ranking seen in
Table~\ref{tab:summary} is not preserved.

\begin{figure}
\includegraphics[draft=false, angle=0]{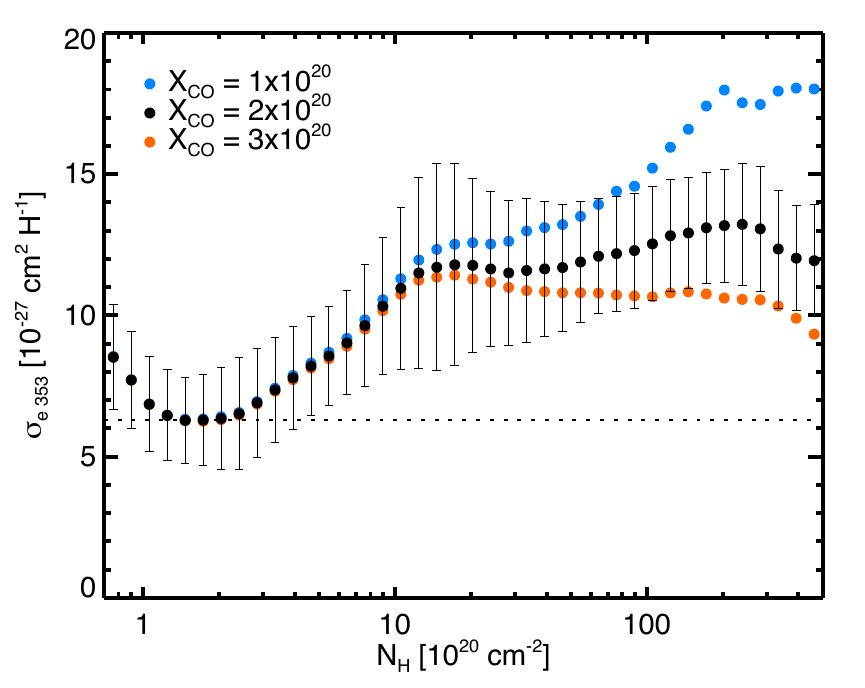}
\includegraphics[draft=false, angle=0]{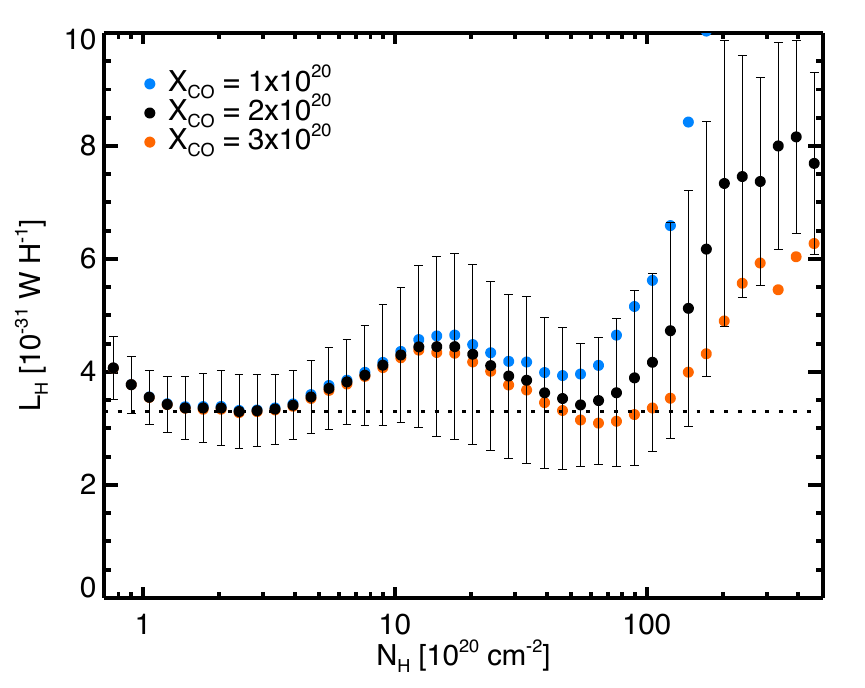}
\caption{\label{fig:tau353_over_nh_vs_nh} \opacity\ (\textit{upper}) and
  \Ldust\ (\textit{lower}) vs.\ $N_{\rm H}$ computed at 30\arcm\ resolution
  from Eq.~\ref{eq:ntot} with three values of $X_{\rm CO}$:
  $[1,2,3]\times10^{20}$\,\XCOunits. The mean values of \opacity\ and
  \Ldust\ were computed in bins of $N_{\rm H}$ equally spaced in log;
  the error bar is the standard deviation, shown only for $X_{\rm
    CO}=2\times10^{20}$\,\XCOunits. The horizontal dotted line indicates
  the average values found in the low $N_\ion{H}{i}$ mask. 
}
\end{figure}

To quantify the trends with column density,
Fig.~\ref{fig:tau353_over_nh_vs_nh} shows the mean and standard
deviation of \opacity\ and \Ldust\ in bins of $N_{\rm H}$ equally
spaced in log.  The results for $X_{\rm CO} = [1,2,3]
\times10^{20}$\,\XCOunits\ are shown.

\subsection{The strength of the ISRF}
\label{sec:radiation_field}

In thermal equilibrium, \radiance\ is equal to the amount of light
absorbed by dust (see Eq.~\ref{eq:absorb}).  
In normalized form, $L_{\rm H} \propto U\, \overline{\sigma_a}$.  Also from
Eq.~\ref{eq:L_analytical} for \radiance\ evaluated from emission,
$L_H$ depends on $T_{\rm obs}$, $\beta_{\rm obs}$, and $\sigma_{\mathrm{e}\,353}$.
Therefore, under the constraint of thermal equilibrium, the measured
\radiance\ and \Ldust\ provide insight into not only $U$ and
$\overline{\sigma_a}$ relating to absorption, but also their
relationship to the SED parameters for emission.

Under the hypothesis of a constant dust-to-gas ratio, constant dust
absorption cross section, and constant shape of the ISRF spectrum,
i.e., constant $\overline{\sigma_a}$, the all-sky map of \Ldust\
provides a way to trace the spatial structure of the radiation field
$U$ over the whole sky quite directly.  The large-scale structure of
this map is similar to the map of $C^+/N_\ion{H}{i}$, obtained from
lower-resolution data by \cite{bennett1994}, that also traces $U$.

\subsubsection{High latitudes}
\label{sec:hilat}

At high latitudes, best seen in the polar maps, \Ldust\ is fairly
uniform, much more so than the opacity, as quantified by the relative
fractional size of their standard deviations
(Table~\ref{tab:summary2}). This can also be seen over the low column-density
 range of Fig.~\ref{fig:tau353_over_nh_vs_nh}, lower, where \Ldust\ is
constant up to $N_\ion{H}{i} = 5.5 \times 10^{20}\,$cm$^{-2}$ which is a
threshold criterion in the $|b|>15$\deg\ mask.  Even in that mask the
standard deviation of \Ldust\ is less than 20\,\%. We also note again
that this column density is below that for which significant H$_2$ is
seen in the diffuse ISM
\citep{gillmon2006,wakker2006,rachford2002,rachford2009}.
Furthermore, there is unlikely to be local attenuation of the ISRF at
such low column densities.\footnote{Because the ISRF illumination is
  not just from along our line of sight, it is difficult to quantify
  the attenuation just from the observed column density.  The total
  line of sight extinction is $A_V = 0.053\, N_{\rm H}/(10^{20}\, {\rm
    cm}^{-2})$ in largely atomic regions \citep[see discussion and
  references in][]{martin2012}, and so roughly half of this amount to
  the centre of a structure.}  
Short of a conspiracy among the several
factors affecting \Ldust, this suggests that each of the factors is
fairly uniform in the diffuse atomic high-latitude ISM, up to column
densities of at least $5 \times 10^{20}\,$cm$^{-2}$.

There is a relatively flat trend of \Ldust\ with respect to $T_{\rm obs}$
in Fig.~\ref{fig:tau_over_nh_vs_t}, lower, which is for moderate column
densities $1 \times 10^{20} < N_\ion{H}{i} < 2.5 \times
10^{20}$\,cm$^{-2}$.  This uniformity is in contrast to that for
\opacity\ in the same figure which is anticorrelated with $T_{\rm obs}$
along a locus of constant \Ldust, a phenomenon also reported by 
\citet{planck2011-7.12} and \citet{planck2013-XVII}.
There is also a striking difference between the polar maps of \Ldust\
and of $T_{\rm obs}$.  This demonstrates that $T_{\rm obs}$ is not a
simple tracer of $U$ as is often assumed. In particular, it suggests
that the increase of $T_{\rm obs}$ observed toward the Galactic pole
is not a direct result of an increase of $U$. One interpretation, put
forward by \cite{martin2012}, is that grains in different regions of
the diffuse ISM retain the effects of different past histories of
evolution, e.g., through aggregation and fragmentation, even though
the density and timescale argue against present in situ evolution by
such processes \citep{planck2011-7.12}. Alternatively,
\citet{planck2013-XVII} review arguments that grain
evolution could be occurring in situ due to UV radiative processing or
exposure to cosmic rays.  In either case, $T_{\rm obs}$ would be a
response to and tracer of variations in dust properties (grain
structure, size distribution, material changes) rather than variations
in the strength of the ISRF.\footnote{
This result was shown 
by \citet{planck2011-7.12} to be robust against $\beta - T$ anticorrelation effects.
These authors reported even stronger variations of $T_{\rm obs}$ at constant \Ldust\ 
using a fit with a fixed $\beta_{\rm obs}$.
}
A corollary is that \radiance\ could be a better alternative to
$\tau_{353}$ as a tracer of $N_{\rm H}$, at least at high latitudes.

There is a remarkable region near the south Galactic pole with
abnormally low \Ldust, but it does not show up in the \radiance\ map.
\citet{planck2013-XVII} argue that it arises because of gas
in the Magellanic Stream (MS) that has Galactic velocities and so is
counted in $N_{\rm H}$ while at the same time the dust abundance and
dust emission in the low-metalicity MS is very low.
\citet{planck2011-7.12} have shown that high velocity clouds
(HVC) have relatively low emissivities, which suggests more generally
that anomalously low \Ldust\ is an interesting diagnostic of HVC-like
material that does not have a distinctive HVC velocity.  But it is not
an argument against using \radiance\ as a tracer of column density.

However, comparison of the polar maps of IVC
(Fig.~\ref{fig:fullres_polar_mapsm1}, lower) and \Ldust\ shows a
correlation of IVC column density with slightly lower \Ldust\ over
widespread regions.  Our interpretation follows
\citet{planck2011-7.12} who studied the emissivity of LVC
and IVC gas separately and concluded that IVC is Galactic gas that
often, though not always, has a lower \Ldust\ because dust has been
partially destroyed; the dust-to-gas ratio is lower in that component 
of gas along the line of sight. The amount of such IVC gas would be
underpredicted by \radiance.\footnote{Note that pixels with strong IVC
  were excluded from the low $N_\ion{H}{i}$ mask used to establish the
  zero points of the intensity maps.}

\subsubsection{Intermediate to low latitudes}

The all-sky map also shows that \Ldust\ is not constant, thus strong
evidence against using \radiance\ everywhere as a tracer of column
density.  For example, the increase of \Ldust\ in the inner Galaxy
instead implies an increase of the radiation field strength there, by
a factor about three compared to the local ISM. 
The all-sky \Ldust\ map also suggests that the ISRF is generally weaker in the
outer Galaxy, as expected.
Note also that an increase (decrease) in \Ldust\ can also be the result of 
an increase (decrease) in the dust-to-gas ratio accompanying a higher (lower) metallicity.

More localized regions of high \Ldust\ are present too: active star
formation sites like Cyg X where local sources significantly enhance
the IRSF illuminating the dust.

There are localized decreases in \Ldust\ as well, coincident with
recognizable intermediate-latitude molecular clouds.  Our
interpretation is that this is a result of a lower ISRF because of
attenuation, lowering the energy absorbed by dust within the clouds
and hence available to be emitted.  In these regions too, \radiance\
would be compromised as a quantitative linear tracer of column density
(see discussion below relating to the Taurus cloud in
Fig.~\ref{fig:maps_Taurus}).

All of these factors contribute to the complicated change of the mean
and standard deviation of \Ldust\ in
Fig.~\ref{fig:tau353_over_nh_vs_nh}, lower, at $N_\ion{H}{i} > 5.5 \times
10^{20}\,$cm$^{-2}$, the part of the sky that is the complement to the
$|b|>15$\deg\ mask.

\subsection{Dust opacity from the diffuse ISM to molecular clouds}\label{sec:doftism}

\begin{figure}
\includegraphics[draft=false, angle=0]{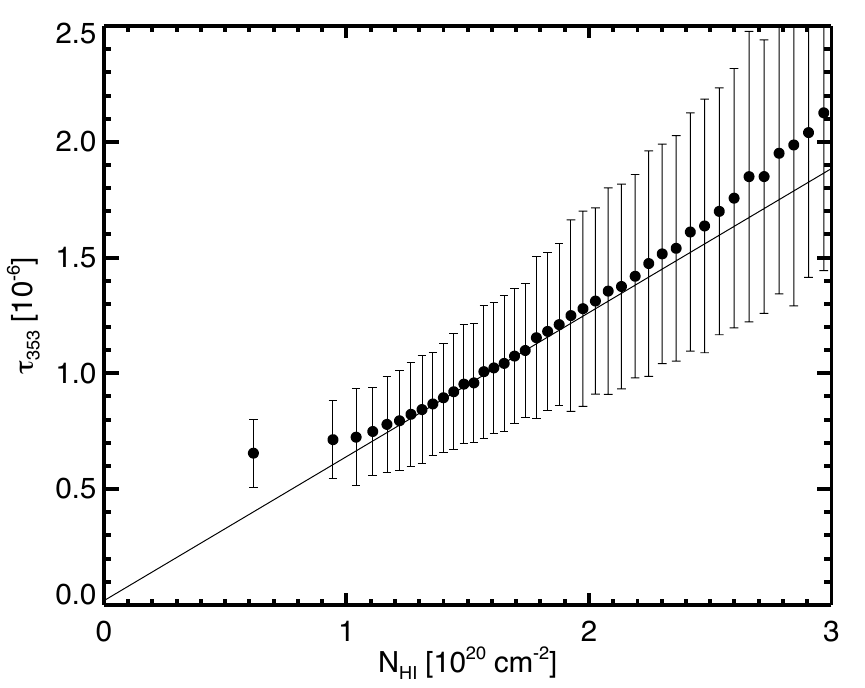}
\caption{\label{fig:zody_or_not}  $\tau_{353}$ as a function of
$N_\ion{H}{i}$ estimated at 30\arcm\ resolution.  Each point and its
associated bar is the mean and standard deviation of $\tau_{353}$
in bins of $N_\ion{H}{i}$.  The solid line is the linear regression fit
using pixels for which $1.2<N_{\rm
HI}<2.5\times10^{20}$\,cm$^{-2}$. Its parametrization is $\tau_{353} =
6.3 \pm 0.1 \times 10^{-27} \, N_\ion{H}{i} - 0.02 \times 10^{-6}$.}
\end{figure}

The maps of \opacity\ in the upper panels of
Figs.~\ref{fig:allsky_tau_and_L_overNH} and
\ref{fig:fullres_polar_maps3} reveal variations of the opacity over
the sky, variations that are spatially coherent.  In the polar plots,
the anticorrelation of \opacity\ with $T_{\rm obs}$ is apparent, the same
as summarized in Fig.~\ref{fig:tau_over_nh_vs_t} (upper).  Because of
the $\beta_{\rm obs} - T_{\rm obs}$ anticorrelation discussed in
Sect.~\ref{sec:betaT}, also seen clearly in the polar maps, there is a
correlation of \opacity\ and $\beta_{\rm obs}$ as well.  Thus at high
latitude $\tau_{353}$ is not a reliable measure of $N_\ion{H}{i}$.

The general increase of \opacity\ by almost a factor of three toward
higher column density can be followed on the all-sky map down to
intermediate latitudes, to known molecular clouds (e.g., Taurus,
Orion, $\rho$~Ophiuchi).  These tend to have lower $T_{\rm obs}$ and
higher $\beta_{\rm obs}$.\footnote{The Magellanic Clouds have an opacity
  almost five times that in the diffuse ISM, despite the low
  metallicity.  But $\beta_{\rm obs}$ is unusually low, pointing to a
  mixture of conditions within the beam and so an SED that is unlikely
  characterized by a single temperature.  $T_{\rm obs}$ is relatively
  high and dust in ionized gas could be contributing.}  

Fig.~\ref{fig:tau353_over_nh_vs_nh}, upper, shows the dependence of
\opacity\ on $N_\ion{H}{i}$.  There is a small range at low $N_{H}$ over
which \opacity\ is at a minimum and roughly constant.  But as shown in
Fig.~\ref{fig:zody_or_not}, where the slope of a fit of $\tau_{353}$
vs.\ $N_\ion{H}{i}$ over the same range of $N_\ion{H}{i}$ corresponds to the same
\opacity, the non-linear increase of $\tau_{353}$ (and \opacity) with
$N_\ion{H}{i}$ sets in at a rather low column density.  This in a range
where as discussed above \Ldust\ is constant and there is no
significant molecular hydrogen or \ion{H}{i} self-absorption.  Thus the
increase of \opacity\ is real and not a reflex of unaccounted dark
gas.  Again, this compromises $\tau_{353}$ as a measure of $N_\ion{H}{i}$ in the 
diffuse ISM.

The opacity continues to increase over the range $3 \times 10^{20} <
N_{\rm H} < 1 \times 10^{21}$\,cm$^{-2}$ reaching a plateau
thereafter, with a dependency on the choice of $X_{\rm CO}$ since the
gas is predominantly molecular there.  The choice of $X_{\rm CO}
=2.0\times10^{20}$\,\XCOunits\ recommended by \citet{bolatto2013},
results in a flat plateau at about twice the value in the diffuse ISM (dotted
line).

\subsection{Dust at the lowest column densities}
\label{sec:WIM}

At the lowest column densities ($N_{\rm H} <
1\times10^{20}$\,cm$^{-2}$) we note an increase of \opacity\ and
\Ldust\ (Fig.~\ref{fig:tau353_over_nh_vs_nh}). This effect is also
seen directly in the correlation of $I_\nu$ vs.\ $N_\ion{H}{i}$ in
Fig.~\ref{fig:offset} where all 857 and 3000\,GHz data points fall
above the correlation for $N_\ion{H}{i}<1.0\times10^{20}$\,cm$^{-2}$. It
is also the case at 545 and 353\,GHz and it thus propagates into the
map of $\tau_{353}$ and \radiance. We checked that this effect is
independent of the removal of the zodiacal emission. It is also present
in $E(B-V)/N_{\rm H}$ using the \Ebv\ map of \citet{schlegel1998}
which is based on \DIRBE.

Using correlation studies, \citet{planck2011-7.12} showed
that \hi\ is a reliable tracer of dust up to at least $N_{\rm
  HI}=2\times10^{20}$\,cm$^{-2}$, or  as discussed in
Sect.~\ref{sec:hilat} \radiance\ is a good tracer of $N_{\rm H}$  to
somewhat higher column densities.  This
suggests that the excess opacity at the lowest $N_\ion{H}{i}$ seen here
in this pixel by pixel analysis is the signature of dust associated
with the warm ionized medium (WIM), i.e., interstellar dust that is mixed with ionized
hydrogen, H$+$, that is not traced by \hi\ emission.

Assuming that dust in the WIM has a similar \Ldust\ as in the \hi, the
WIM gas column density needed to explain the rise at low $N_\ion{H}{i}$
is $N_{\rm H+}\sim 3\times 10^{19}$\,cm$^{-2}$.\footnote{This result
  seems compatible with \citet{lagache1999,lagache2000a} who showed
  that dust in the WIM has similar emissivity to dust in the WNM but a
  slightly higher temperature. These authors also concluded that about
  25\,\% of the dust emission in the diffuse ISM is associated with
  the WIM, uncorrelated with \hi, a value which corresponds well with
  what is seen here for $N_\ion{H}{i} <1.2 \times 10^{20}$\,cm$^{-2}$.}
However, a constant value of $N_{\rm H+}$ cannot explain the shape of
the rise of \Ldust. The rise is more compatible with $N_{\rm
  H+}+N_\ion{H}{i} \approx 1.1 \times 10^{20}$\,cm$^{-2}$ suggestive of
an increase of the ionization fraction of WNM toward lower
$N_\ion{H}{i}$. The apparent extra dust emission seen here would then
come from diffuse regions where \hi\ is partly ionized.

\subsection{Discussion}\label{sec:opacitynhdiscussion}

The comparison of dust emission and gas column density reveals an increase of 
dust opacity of a factor about two
from the diffuse ISM to molecular clouds (Fig.~\ref{fig:tau353_over_nh_vs_nh}, upper). 
In the translucent transition region 
($3\times 10^{20} < N_{\rm H} < 2 \times 10^{21}$\,cm$^{-2}$) H$_2$ might start 
rising in importance before CO and self-absorption could begin to affect the
21\,cm line emission. These effects are difficult to quantify from the
present data but they cannot be responsible for the systematic difference
in \opacity\ observed between the \hi\ dominated and CO dominated regimes.
The presence of ``dark gas," whether from H$_2$ not traced by CO or from 21\,cm self-absorption,
would simply flatten the rising
profile of \opacity\ vs.\ $N_{\rm H}$, reaching the plateau somewhat later. 
The fact that at $N_{\rm H} \sim 6\times 10^{21}$\,cm$^{-2}$, 
a column density where CO is thought to be a reliable tracer of $N_{\rm H}$, 
\Ldust\ dips to the diffuse ISM value while \opacity\ remains at the plateau value (Fig.~\ref{fig:tau353_over_nh_vs_nh})
is also consistent with an increased dust opacity in denser regions.

As discussed further in Sect.~\ref{sec:ejks}, 
this increase of \opacity\ in denser regions is also seen when $N_{\rm H}$ is estimated
using near-infrared colour excess or star counts
\citep{arce1999a,cambresy2001,stepnik2003,planck2011-7.13,martin2012,roy2013}.  
It is generally accompanied by a decrease of $T_{\rm obs}$, which is quite
challenging to explain just with radiative transfer effects
\citep{ysard2012}.  When the gas has become dense, an increased \opacity\ might
be attributed to an increase of dust emissivity related to dust
aggregation/coagulation
\citep{ossenkopf1994,ormel2011,kohler2012}.\footnote{With the increase
  of gas density, smaller grains stick on the surface of bigger ones,
  modifying their structure to a more open one, resulting in an
  increase of emissivity. Being more emissive, the grains cool more
  efficiently and are therefore colder.}  
However, we have seen
opacity changes in the diffuse high-latitude ISM as well that need
alternative interpretation if the evolution is in situ
(Sect.~\ref{sec:hilat}).

Because \radiance\ is less affected by the CIBA and because of its correlation
with $N_{\rm H}$ over a larger range in column density (see \Ldust\ in Fig.~\ref{fig:tau353_over_nh_vs_nh}, lower),
we conclude that \radiance\ is preferred over $\tau_{353}$ as a tracer of column density
in the high-latitude diffuse ISM, at least for $N_H < 5\times 10^{20}$\,cm$^{-2}$.
However, this preference does not hold in molecular clouds and star forming regions where 
\radiance\ traces not only the column density but also variations of 
the radiation field strength due
to attenuation and/or local sources of heating photons.  In such
regions, $\tau_{353}$ is the preferred tracer of column density,
to the extent that \opacity\ is constant there.\footnote{Using higher resolution \Herschel\ 
  data to probe opacity to
  high column densities, \citet{roy2013} found evidence for a
  non-linear increase of $\tau_{353}$ with $N_{\rm H}$. 
 }
This is supported empirically by the good correlation with the colour excess $E(J-K_s)$
discussed below in Sect.~\ref{sec:ejks}.  However, finding the absolute, rather than relative,
column density depends on proper
calibration of the opacity $\sigma_{\mathrm{e}\,\nu}$, which appears to vary with
column density and be larger in these regions.  Furthermore, caution
is advised because the opacity changes from diffuse to dense regions, which
might occur over the range of column densities encountered in the region being analysed.

\section{Dust emission in relation to extinction}
\label{sec:extinction}

A quantity often used to estimate interstellar column density is
visible or near-infrared extinction measured along lines of sight to
point sources: stars, globular clusters, galaxies, or quasars. It has
been established long ago that there exists a correlation between gas
and dust column densities, in particular through the comparison of
21\,cm emission and visible extinction \citep[e.g.,][]{lilley1955}. The
linear relationship between $N_\ion{H}{i}$ and \Ebv\ was established in
the 1970s \citep{savage1972,knapp1974a,ryter1975,bohlin1978}.\footnote{Key
  information on dust is derived from this relationship, for example
  that dust contains only 1\,\% of the mass of the ISM, and this
  relationship remains a important constraint for dust models
  \citep{draine2007a,compiegne2011}.} 
\citet{knapp1974a} advocated
using 21\,cm observations of $N_\ion{H}{i}$ as a proxy for extinction and
this correlation, especially as calibrated in the diffuse ISM using
measurements on extragalactic objects, has been key to correct
extragalactic observations for Galactic reddening.

We have seen in Sect.~\ref{sec:column_density} how the amount of dust
emission is, not surprisingly, also correlated with $N_{\rm H}$.
However, because dust is the agent in both extinction and emission, we
make a direct comparison of these observables rather than using
$N_{\rm H}$ as an intermediary.  An important example of this direct
approach is the proposal by \citet{schlegel1998} to use dust optical
depth obtained from FIR emission (\IRAS\ and \DIRBE), rather
than \hi, to estimate reddening ($E(B-V)_{\rm SFD}$), through a 
correlation calibrated on reddening measurements of galaxies.  
Such an approach is pursued in
Sect.~\ref{sec:ebv_qso}.  For higher column density lines of sight, we
compared dust emission to colour excess measurements based on 2MASS
stellar photometry.\footnote{The Two Micron All Sky Survey
  \citep{skrutskie2006} is a joint project of the University of
  Massachusetts and the Infrared Processing and Analysis
  Center/California Institute of Technology, funded by the National
  Aeronautics and Space Administration and the National Science
  Foundation.}

\subsection{Correlation with $E(B-V)$ from quasars}
\label{sec:ebv_qso}

\begin{figure}
\centering
\includegraphics[draft=false, angle=0]{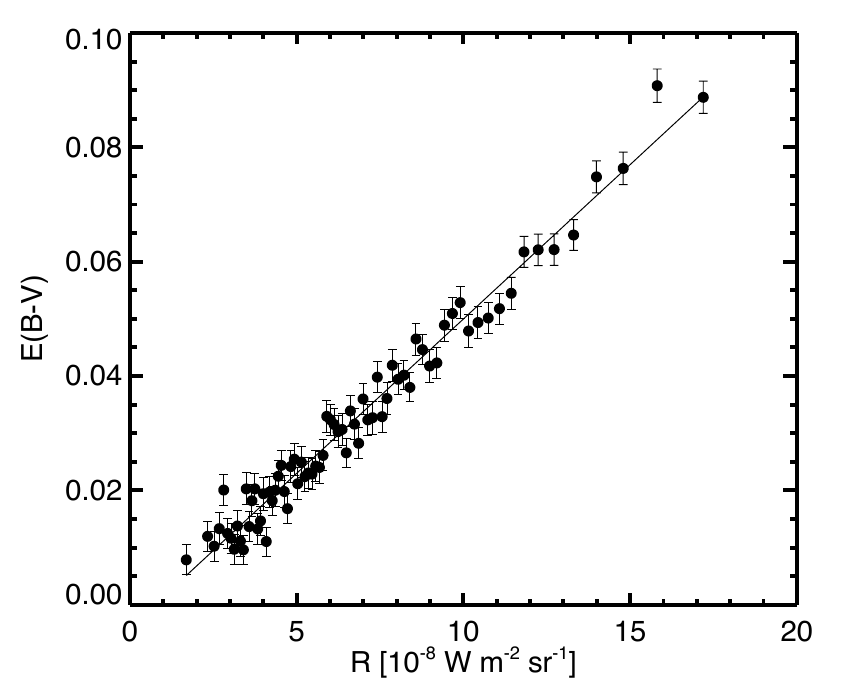}
\includegraphics[draft=false, angle=0]{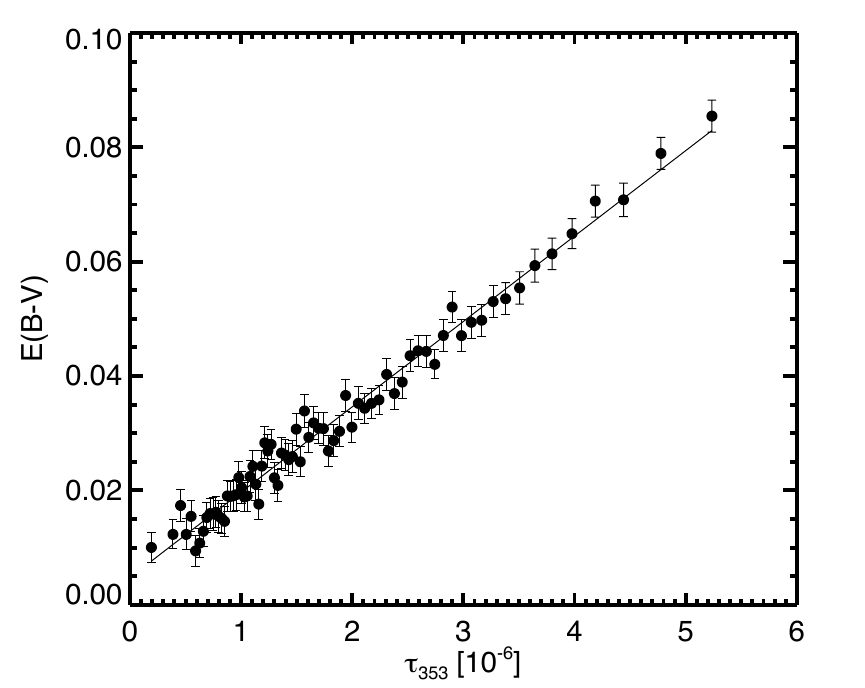}
\caption{\label{fig:ebv_vs_tau} \Ebv\ measured with quasars 
in the diffuse ISM at high Galactic latitude as a
function of \radiance\ (\textit{upper}) and $\tau_{353}$ (\textit{lower}). Each point is
the average of \Ebv\ values in a bin of $\tau_{353}$ or \radiance. The
bin size varies for each bin such that there is always the same number
of samples per bin ($N=1000$). The error bar is the standard deviation
of \Ebv\ values in the bin, divided by $\sqrt{N}$. The solid line is
the linear regression fit to the whole sample; $E(B-V)/{\cal{R}} =
(5.40 \pm 0.09)\times 10^5$ and $E(B-V)/\tau_{353} = (1.49 \pm
0.03)\times 10^4$. 
In each case, the intercept is small 
(-0.0040 for $\cal{R}$ and 0.0048 for $\tau_{353}$), in accordance
with the fact that the zero levels of \Ebv\ (i.e., the intrinsic colours) 
and of the \Planck\ and \IRIS\ data were
estimated using a correlation with $N_\ion{H}{i}$. 
}
\end{figure}

Here, based on the \Planck\ dust emission, we develop a map of \Ebv\
applicable to the diffuse ISM at high Galactic latitude.
In the years since the work of \citet{schlegel1998}, many models have
been put forward self-consistently describing dust emission and
extinction  \citep[e.g., ][]{draine2007a,compiegne2011} and
they could be used to convert emission to extinction.
However, to be independent of any assumption about dust properties, we
decided to remain with an empirical approach.  We estimate the
conversion factor to \Ebv\ using measurements of extinction of
extragalactic objects rather than stars to avoid potential biases
due to background dust emission.

In particular we estimated \Ebv\ using Sloan Digital Sky Survey (SDSS)
measurements of quasars. We used the final edition of the SDSS-II
quasar catalogue \citep{schneider2010} based on the
seventh SDSS data release \citep{abazajian2009}. The catalogue
contains 105\,783 objects spread over 8\,400\,deg$^2$ mostly on the
northern Galactic hemisphere. For each quasar, the observed magnitudes
in bands $u$, $g$, $r$, $i$, and $z$ are given together with their
uncertainties. All objects in this catalogue have highly reliable
redshift estimates.  We limited the sample to a subset of 53\,399
quasars at redshifts for which Ly$\alpha$ does not enter the SDSS
filters.
One benefit compared to the work of \citet{schlegel1998} is
the much larger number of objects.  Another is that many studies based
on SDSS data have shown that the shape of the extinction curve in the
diffuse ISM is compatible with that for stars from
\citet{fitzpatrick1999} with $R_V=3.1$
\citep{jones2011b,schlafly2011,mortsell2013}, so that we can take
advantage of all of the multi-colour measurements. The details of how
we estimate \Ebv\ for each quasar are given in
Appendix~\ref{sec:ebv}.  

The correlations of \Ebv\ with \Planck\ \radiance\ and $\tau_{353}$,
from which point sources have been removed
(Sect.~\ref{sec:luminosity}), are shown in
Fig.~\ref{fig:ebv_vs_tau}.\footnote{Although not an explicit selection
  criterion, the range of $\tau_{353}$ sampled by 
  the selected quasars corresponds the conditions in the low
  $N_\ion{H}{i}$ mask, Fig.~\ref{fig:hi_mask}, whose NDF for $\tau_{353}$ is shown in
  Fig.~\ref{fig:t_beta_degeneracy}.  The positions of the bins along
  the x-axes in Fig.~\ref{fig:ebv_vs_tau} reflect this NDF.}
Each is strongly correlated: $E(B-V)/{\cal{R}} = (5.40 \pm 0.09)\times
10^5$ and $E(B-V)/\tau_{353} = (1.49 \pm 0.03)\times 10^4$.  The
fractional uncertainty of the slope of the correlation with \radiance\
is about 20\,\% lower than that for $\tau_{353}$.  This is not
unexpected, both from our discussion of Fig.~\ref{fig:maps_tau_vs_L}
and, in Sect.~\ref{sec:hilat}, of the factors that influence
$\tau_{353}$ but not \radiance.  
This leads us to prefer the solution 
based on \radiance, $E(B-V)_{\cal{R}}$, 
to that based on $\tau_{353}$, $E(B-V)_{\tau353}$, for 
the low column density regions of the sky (see also Sect.~\ref{sec:opacitynhdiscussion}).

The product of $E(B-V)/{\cal{R}}$ and the diffuse ISM estimates of \Ldust\
gives the ratio $E(B-V)/N_{\rm H} =$ (1.42--1.46)$\times 10^{-22}$\,mag\,cm$^{2}$, 
a factor just 0.82--0.85 lower than that measured using background stars for 
lines of sight with considerably larger \Ebv\ \citep{bohlin1978,rachford2009}. 
Given all the potential for differences, this agreement is remarkable.
On the other hand, the same ratio found using $E(B-V)/\tau_{353}$ and the low 
value of \opacity\ in Figs.~\ref{fig:tau353_over_nh_vs_nh}, upper, and 
\ref{fig:zody_or_not} and in Table~\ref{tab:summary2} results in a factor 
0.55--0.60 lower. Using the \Ebv\ map derived from dust emission 
$\tau_{3000}$ by \citet{schlegel1998},  \citet{liszt2014} also found a lower 
ratio, by a factor 0.7, for lines of sight with $E(B-V) < 0.1$.
%

\subsection{Comparing to $E(J-K_s)$ from star colours in molecular clouds}\label{sec:ejks}

\begin{figure*}
\includegraphics[draft=false, angle=0]{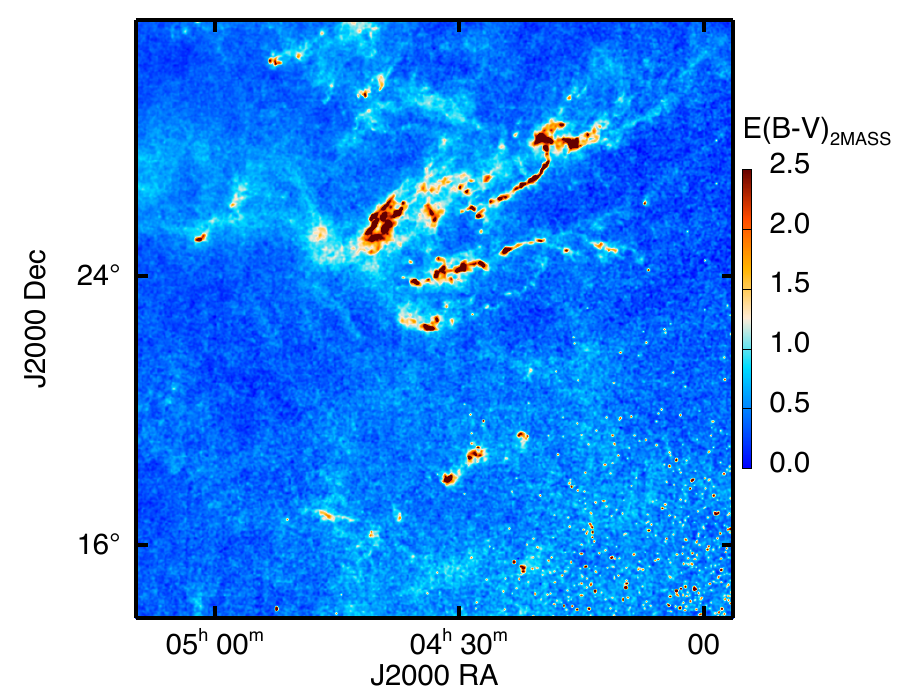}
\includegraphics[draft=false, angle=0]{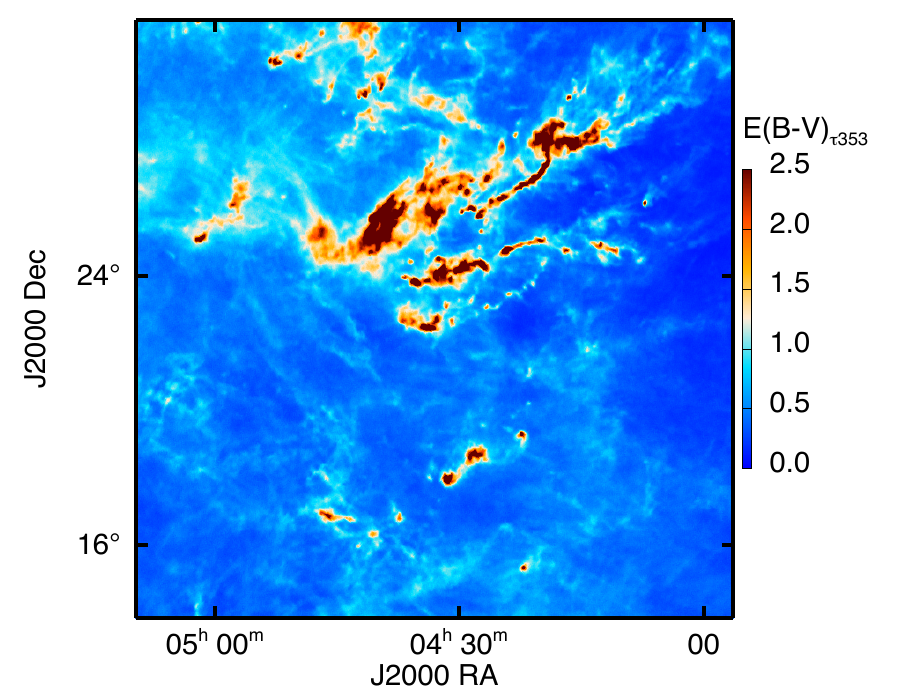}
\includegraphics[draft=false, angle=0]{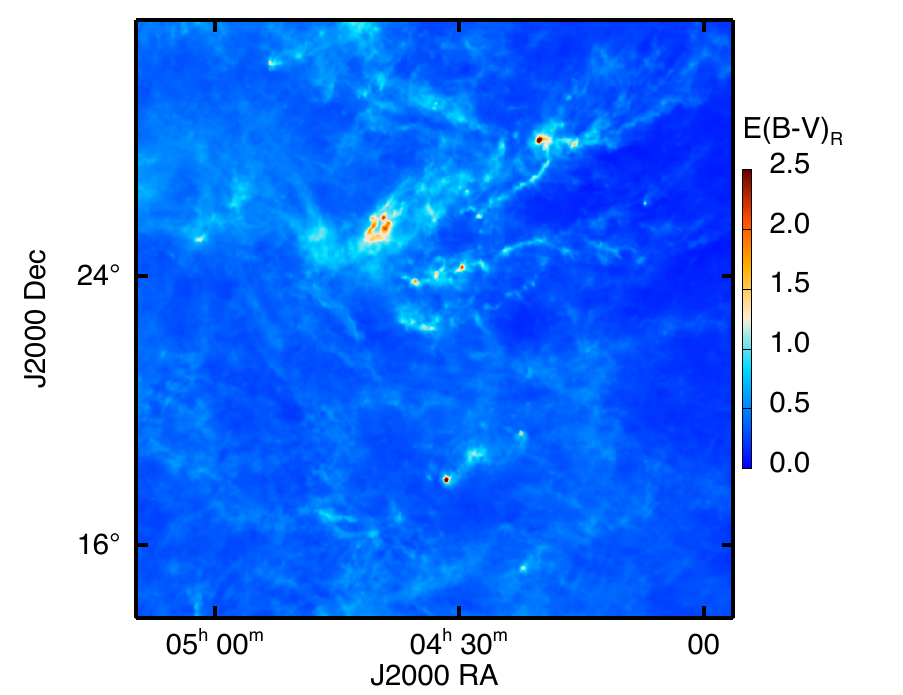}
\includegraphics[draft=false, angle=0]{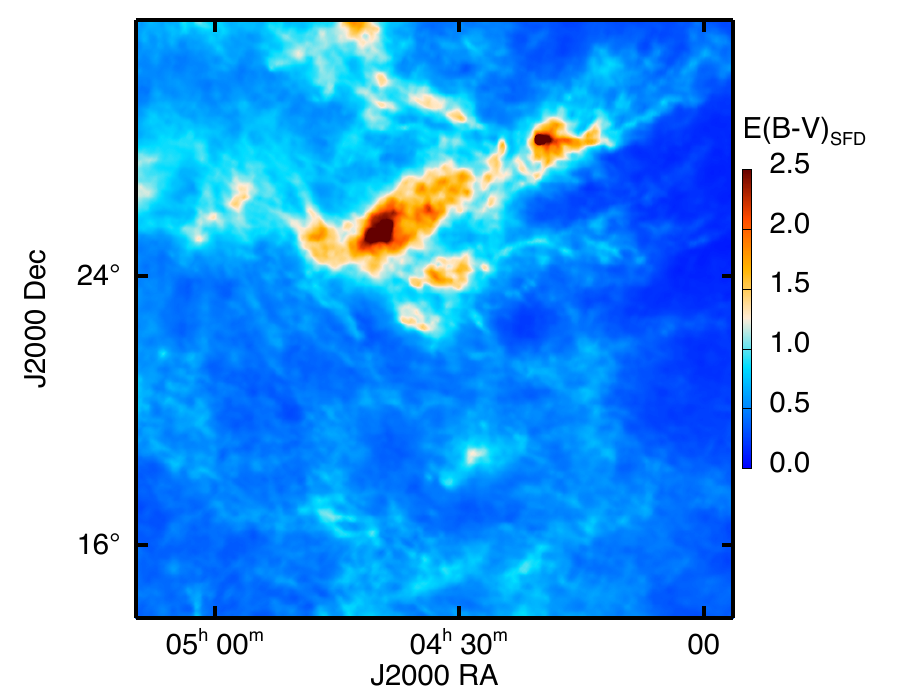}
\caption{\label{fig:maps_Taurus} Estimates of \Ebv\ in the Taurus
  molecular cloud. \textit{Clockwise from upper left}: from 2MASS
  data, from $\tau_{353}$, from \citet{schlegel1998}, and from \radiance.}
\end{figure*}

\begin{figure*}
\includegraphics[draft=false, angle=0]{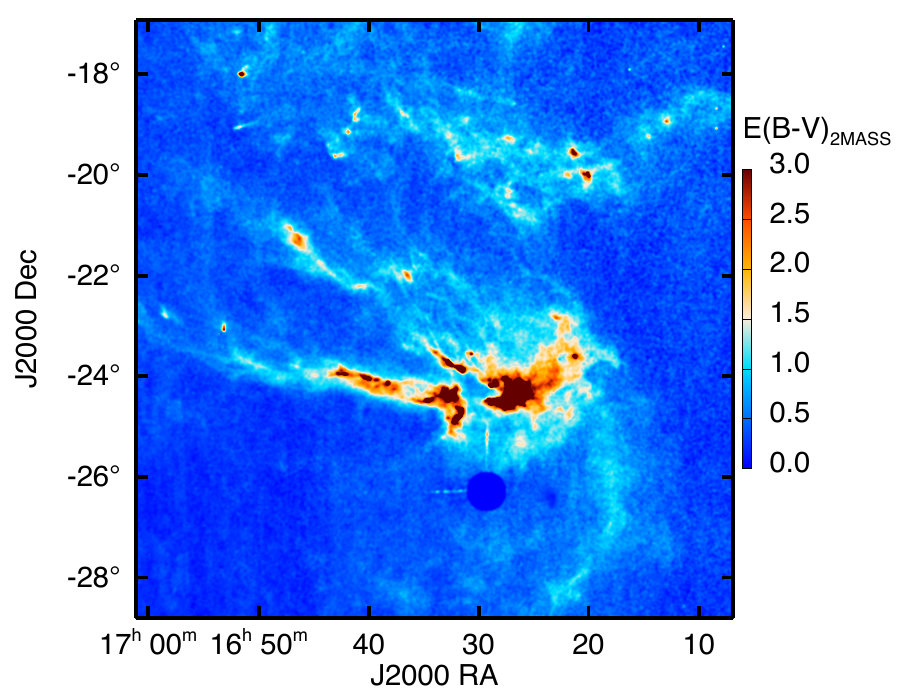}
\includegraphics[draft=false, angle=0]{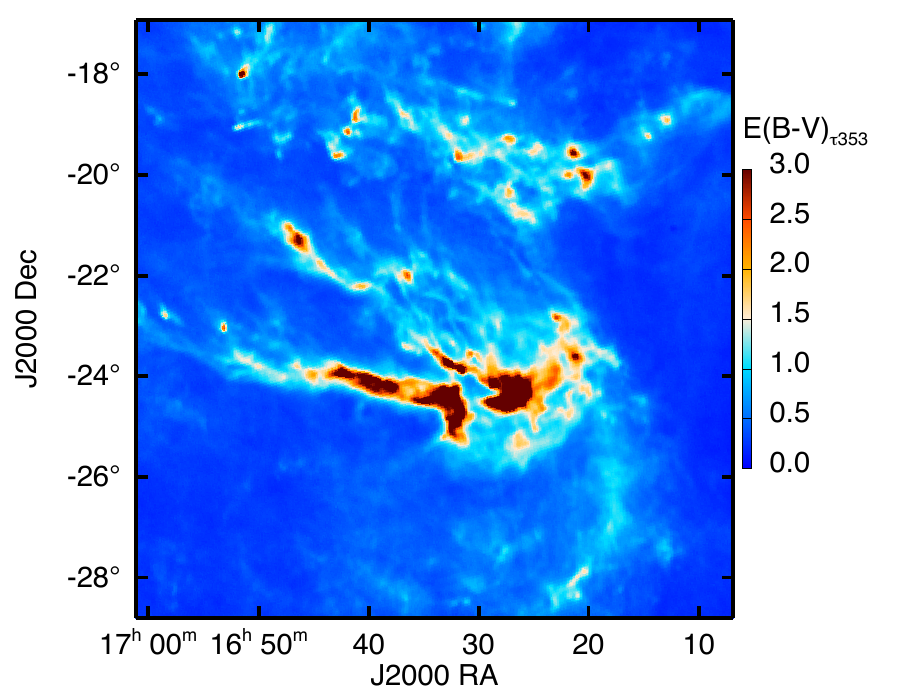}
\includegraphics[draft=false, angle=0]{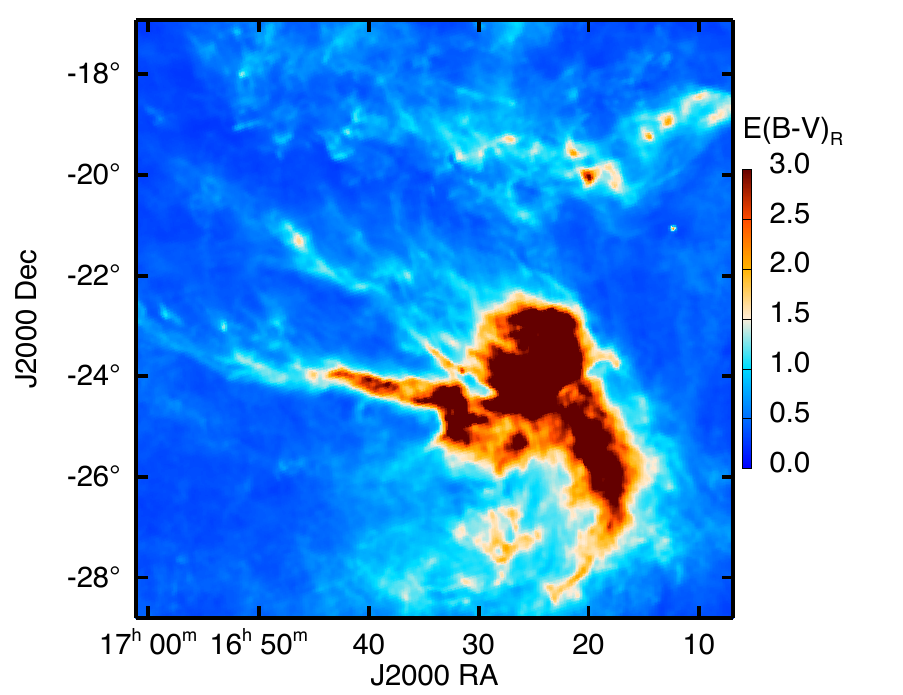}
\includegraphics[draft=false, angle=0]{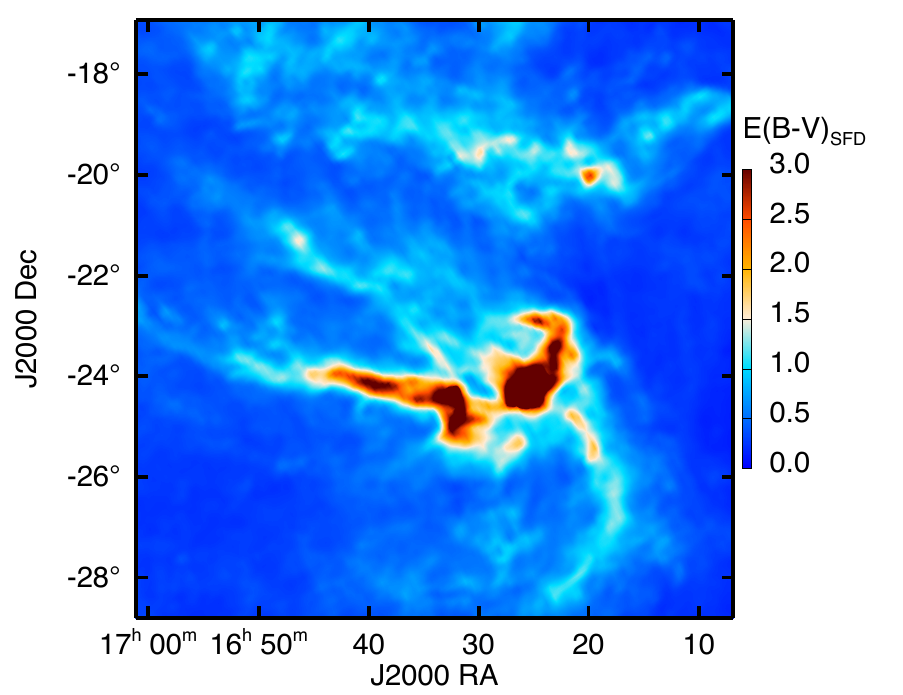}
\caption{\label{fig:maps_Ophiuchus} 
Estimates of \Ebv\ in the $\rho$~Ophiuchi
  molecular cloud. \textit{Clockwise from upper left}: from 2MASS
  data, from $\tau_{353}$, from \citet{schlegel1998}, and from \radiance.}
\end{figure*}

We have investigated how to estimate \Ebv\ for nearby molecular
clouds from \Planck\ dust emission.
A point of comparison for such regions is mapping of colour excesses
$E(J-H)$ and $E(H-K_s)$, or their sum $E(J-K_s)$, based on stellar
colours \citep[e.g.,][]{goodman2009}.
Here we use maps produced with the AvMAP technique 
and based on colour excesses from the 2MASS data base
\citep{schneider2011}. 
Similar maps can be obtained with the NICER and NICEST techniques 
\citep[e.g.,][]{lombardi2011}.
The effective angular resolution of the 2MASS extinction maps used here
is a few minutes of arc \citep[see, e.g.,][]{roy2013}, close to that of \Planck.

Although the optical extinction is rarely directly measured in the
same high column density regions \citep{martin2012,roy2013}, these
near-infrared colour excesses are usually expressed as $E(B-V)_{\rm
  2MASS}$ after conversion using an assumed shape of the extinction
curve, that for stars with $R_v=3.1$.  This
conversion might be inappropriate, and even variable across a field,
because of dust evolution affecting all of the colour excess ratios
for a given column of dust.  
Nevertheless, it is still very interesting to compare the spatial 
details of the dust column density revealed by dust extinction 
and by dust emission, because each is affected by different systematic effects.

For the Taurus and $\rho$~Ophiuchi molecular clouds,
Figs.~\ref{fig:maps_Taurus} and \ref{fig:maps_Ophiuchus} present a
comparison of four different estimates of \Ebv: $E(B-V)_{\rm 2MASS}$,
$E(B-V)_{\tau353}$, $E(B-V)_{\rm SFD}$, and $E(B-V)_{\cal{R}}$.  The
Pearson correlation coefficients of $E(B-V)_{\rm 2MASS}$ with the
three other maps are given in Table~\ref{tab:pearson}. The correlation
of $E(B-V)_{\tau353}$ map with $E(B-V)_{\rm 2MASS}$ is excellent
(Pearson coefficient of 0.86 for Taurus and 0.95 for
$\rho$~Ophiuchi); this agreement is remarkable given that very
different methods and data sets were used to build these two
maps.  However, notice how the brightest filamentary structures appear
with more contrast in $E(B-V)_{\tau353}$ than in $E(B-V)_{\rm 2MASS}$,
a point to which we shall return.

The \citet{schlegel1998} map $E(B-V)_{\rm SFD}$ was also produced from
a dust optical depth map (at 3000\,GHz in that case; see
Sect.~\ref{sec:sfd} for details) and so the scale is similar.
However, the correlation coefficient with $E(B-V)_{\rm 2MASS}$ is
lower and it is clear that a lot of spatial detail is absent. This
arises because they estimated $T_{\rm obs}$ using the lower resolution
\DIRBE\ data, thus missing the dust temperature decrease at small
scales that accompanies the increase of column density in molecular
clouds, and so their map of optical depth which underlies $E(B-V)_{\rm
  SFD}$ has lower contrast as well as lower resolution.
Being able to follow the small-scale variations of $T_{\rm obs}$
appears to be essential to gauge properly the full structural details
of the molecular clouds.  This was revealed by previous studies
\citep{cambresy2001,stepnik2003,planck2011-7.13}, in
particular using higher resolution \Herschel\ data
\citep{battersby2011,roy2013}, and is now confirmed and reinforced by
our \Planck\ analysis.
 
Although a lot of spatial detail is faithfully reproduced in
$E(B-V)_{\cal{R}}$, thanks to the \Planck\ resolution, the correlation
with $E(B-V)_{\rm 2MASS}$ is less good.  Furthermore, the scale is
off.  In these denser regions of the ISM, the radiation field
strength, and hence \radiance, varies locally due to attenuation
and/or local production of photons. 
The first effect (attenuation) is apparent in Taurus where the densest
and brightest filamentary structures appear with less contrast in
$E(B-V)_{\cal{R}}$ than in $E(B-V)_{\rm 2MASS}$, and these regions
are, consistently, also colder.
The case of $\rho$~Ophiuchi is different; because of active star
formation, and thus local sources of heating photons, it is a
photon-dominated region.  The spatial structure of \radiance\ is
therefore visually different than that of $\tau_{353}$ or $E(B-V)_{\rm
  2MASS}$, due to the spatial variation of the radiation field.

Opposite to what was found in the diffuse ISM, $\tau_{353}$ appears
preferable to \radiance\ as a tracer of column density, in this case
\Ebv.  But as discussed next it is a complex situation warranting
caution.  A high correlation coefficient is an important criterion,
but the scale and dynamic range are also important.

\begin{figure*}
\centering
\includegraphics[draft=false, angle=0]{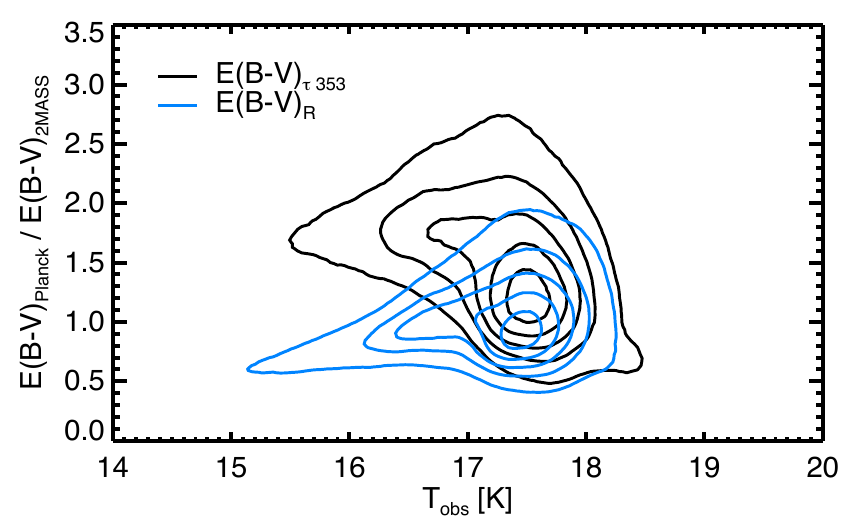}
\includegraphics[draft=false, angle=0]{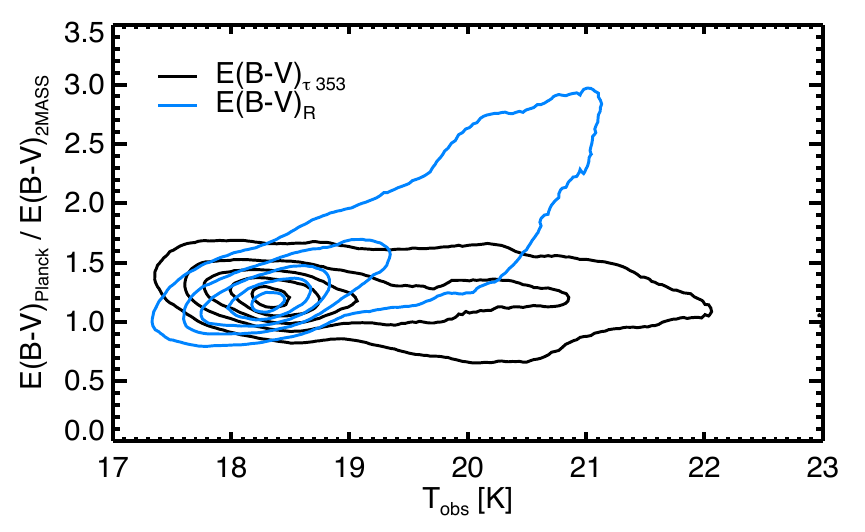}
\caption{\label{fig:Ebv_ratio_vs_Tobs} Two-dimensional histogram of
  $E(B-V)_{\rm \tau 353}/E(B-V)_{\rm 2MASS}$ (black contours) and
  $E(B-V)_{\cal{R}}/E(B-V)_{\rm 2MASS}$ (blue contours) as a function of $T_{\rm obs}$, 
  for the Taurus (\textit{left}) and $\rho$~Ophiuchi (\textit{right}) molecular clouds. 
The maps used are those shown in Figs.~\ref{fig:maps_Taurus} and
\ref{fig:maps_Ophiuchus}.}
\end{figure*}

These effects can be appreciated by the quantification in
Fig.~\ref{fig:Ebv_ratio_vs_Tobs} where the ratios
$E(B-V)_{\tau353}/E(B-V)_{\rm 2MASS}$ and
$E(B-V)_{\cal{R}}/E(B-V)_{\rm 2MASS}$ are plotted as a function of
$T_{\rm obs}$. 

In Taurus, $E(B-V)_{\tau353}$ agrees with $E(B-V)_{\rm
  2MASS}$ over most of the map, although it is systematically high by
typically 25\,\%. 
\citet{arce1999a} reported a similar result  
in their study of $E(B-V)_{\rm SFD}$ in Taurus. The exception here is in the coldest parts
of the cloud where $E(B-V)_{\tau353}/E(B-V)_{\rm 2MASS}$ increases even more.
The systematic departure and spatial variations of $E(B-V)_{\tau353}/E(B-V)_{\rm 2MASS}$
appear to be the result of an increase in \opacity\ even in the relative 
diffuse parts of the map and even more in the coldest (densest) regions.
The opacity changes at higher column densities are argued to be related 
to dust evolution \citep{planck2011-7.13}, and unless independently
characterized these opacity changes compromise the interpretation of
$E(B-V)_{\tau353}$ as a quantitative measure of dust column density.
This is a general concern for all column densities derived from
FIR and submillimetre optical depth.

In Taurus, $E(B-V)_{\cal{R}}$ also agrees with $E(B-V)_{\rm 2MASS}$
over large parts of the map, although it is slightly low systematically
(Fig.~\ref{fig:Ebv_ratio_vs_Tobs}).  In the densest regions, which are
cold because of attenuation of the ISRF, \Ldust\ is depressed even
further and $E(B-V)_{\cal{R}}/E(B-V)_{\rm 2MASS}$ decreases.  This
greatly reduces the contrast across the map of $E(B-V)_{\cal{R}}$.

In $\rho$~Ophiuchi, there is a similar scale difference in the
typical $E(B-V)_{\tau353}/E(B-V)_{\rm 2MASS}$ and an upturn toward
lower $T_{\rm obs}$.  Unlike in Taurus, \Ldust\ is not generally
depressed.  There is a correlation of
$E(B-V)_{\cal{R}}/E(B-V)_{\rm 2MASS}$ with $T_{\rm obs}$ as expected
if $T_{\rm obs}$ reflects changes in the strength of radiation.
Without an independent measure of changes in the ISRF, \radiance\ is
not a reliable quantitative tracer of the dust column density.

\begin{table}
\centering
\caption{\label{tab:pearson} Pearson correlation coefficient with the 
\Ebv\ map obtained with colour excess using 2MASS data. }
\begin{tabular}{lccc}\specialrule{\lightrulewidth}{0pt}{0pt} \specialrule{\lightrulewidth}{1.5pt}{\belowrulesep}
Cloud & $E(B-V)_{\tau 353}$ & $E(B-V)_{\cal{R}}$ & $E(B-V)_{\rm SFD}$ \\ \midrule
Taurus & 0.86 & 0.75 & 0.67\\ 
$\rho$~Ophiuchi & 0.95 & 0.70 & 0.79 \\
\bottomrule[\lightrulewidth]
\end{tabular}
\end{table}

\subsection{Discussion}\label{sec:opacityebvdiscussion}

As in the comparison of column density measures from dust emission with gas column density, 
the comparison with dust extinction leads us to the conclusion
that \radiance\ is a slightly better tracer of $E(B-V)$ for
diffuse low column density lines of sight.
This tracer, of particular interest for extragalactic studies,
is the product that we call $E(B-V)_{\rm xgal}$ in the \Planck\ Legacy Archive
(PLA; Appendix~\ref{revisedPLA}).  In addition to its usefulness in
estimating Galactic \Ebv\ for extragalactic studies, it can also be
used to study the structure of the diffuse ISM in regions where the
CIBA is dominating the fluctuations at small scales. 
We stress 
that $E(B-V)_{\rm xgal}$ should not be used for $E(B-V)>0.3$,
where attenuation effects on \radiance\ become important.
In particular, this counter-indication applies
in molecular clouds and star forming
regions where \radiance\ traces not only the dust column density but also
changes in the ISRF arising from attenuation and/or
local sources of radiation.  
In line with the discussion in Sect.~\ref{sec:opacitynhdiscussion},
in such regions $E(B-V)_{\rm xgal}$ should not be used.

However, in such regions $\tau_{353}$ is well correlated with
$E(B-V)_{\rm 2MASS}$, suggesting an alternative tracer.  Again there
is an issue with absolute amount, as in the conversion of $\tau_{353}$
to $N_{\rm H}$, relating to changes in \opacity.  Adopting the scaling
factor estimated using the correlation with quasars for more
diffuse lines of sight ($E(B-V)/\tau_{353}=1.49\times 10^4$ -- see
Fig.~\ref{fig:ebv_vs_tau}) along with the $\tau_{353}$ maps could
systematically overestimate \Ebv, and this could be exacerbated in the
most dense regions where \opacity\ increases further.

\section{\Planck\ dust products and comparisons with forerunners}
\label{sec:products}

\subsection{Description of products}
\label{sec:plaproducts}

As described more fully in Appendix~\ref{revisedPLA}, the following
maps are available in the PLA: the three MBB
parameters $T_{\rm obs}$, $\beta_{\rm obs}$, and $E(B-V)_{\rm xgal}$, 
together with the associated uncertainty maps.
The map of dust integrated intensity, \radiance, can be obtained
readily from the three MBB parameter maps using the analytical
expression of Eq.~\ref{eq:L_analytical}.

$E(B-V)_{\rm xgal}$, a scaled version of \radiance, was obtained from
MBB parameters of a fit to data from which point sources have been
removed (see Sect.~\ref{sec:luminosity} and
Appendix~\ref{sec:pointsources}).  The map of this \radiance\ can
therefore be obtained by dividing the $E(B-V)_{\rm xgal}$ map by the
conversion factor from Sect.~\ref{sec:ebv_qso}
(Fig.~\ref{fig:ebv_vs_tau}, upper).

The main limitations on the \Planck\ dust products from the data
themselves are related to the \IRAS\ data. There is some residual
striping in the 3000\,GHz \IRAS\ data that propagates mainly into the
map of $T_{\rm obs}$. There is also about 4\,\% of the sky that was
not observed by \IRAS\ \citep{beichman1988}. This area was filled with
\DIRBE\ data \citep{miville-deschenes2005a}. Finally, we stress that
the dust model is based on data that unavoidably include the CIBA.

\begin{figure*}
\centering
\includegraphics[draft=false, angle=0]{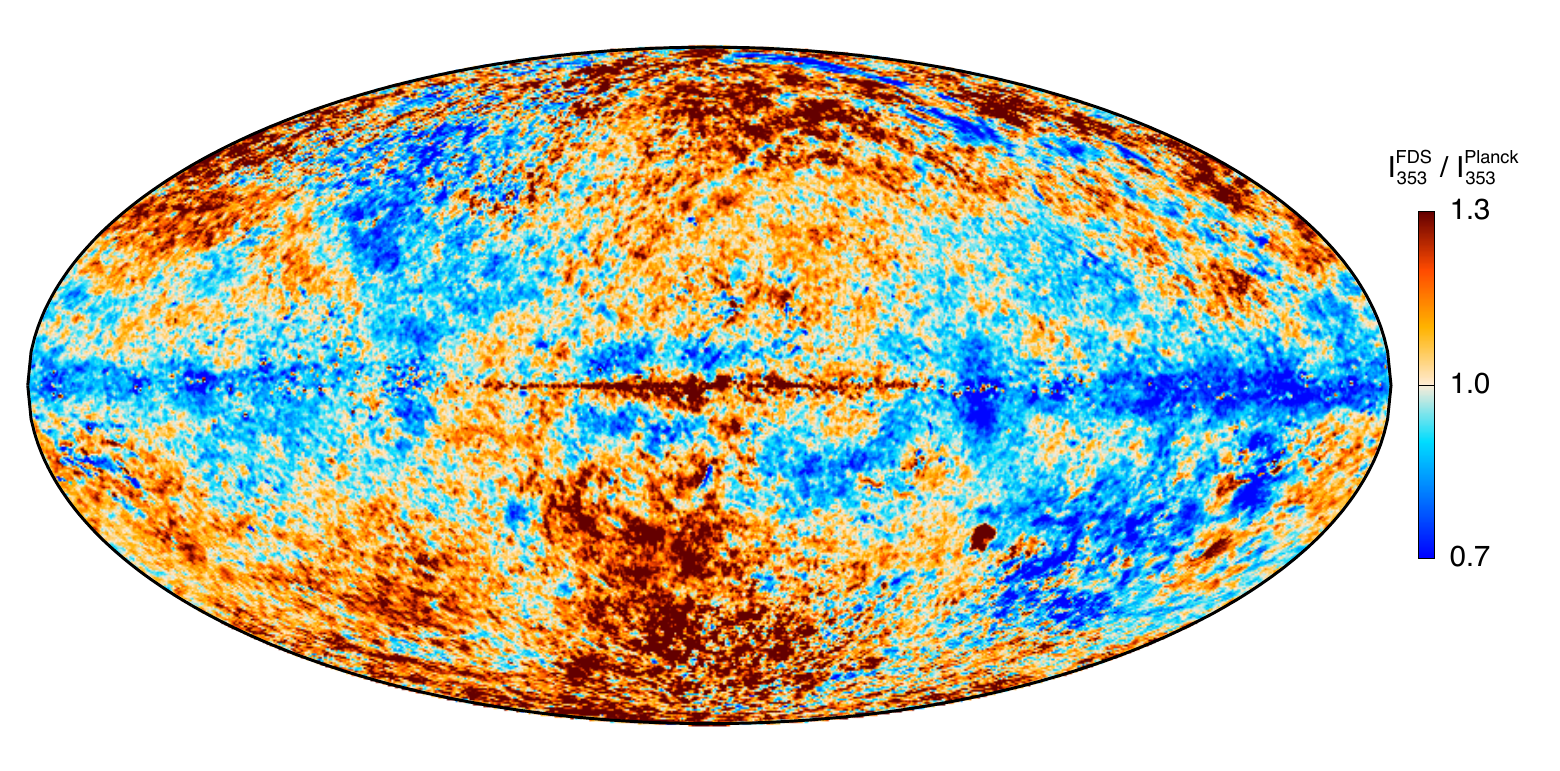}
\caption{\label{fig:Planck_over_FDS} Ratio of the dust specific
  intensity at 353\,GHz from the \citet{finkbeiner1999} and \Planck\
  models, $I_{353}^{\rm FDS}/I_{353}^{\rm Planck}$, using maps
  smoothed to 30\arcm.}
\end{figure*}

\subsection{Modelling dust emission: comparison of \Planck\ with Finkbeiner et al. (1999)}

\label{sec:fds}

One of the important applications of these parameter maps will be to
combine them using Eq.~\ref{eq:mbb} to model the SED of the dust
emission in the submillimetre range. 

A benchmark for comparison is the parametric SED model developed by
\citet{finkbeiner1999}.  Motivated by data from the \FIRAS\ experiment, 
they modelled the dust emission as the sum of two MBBs.  
Each MBB component is in principle characterized by three
parameters.  However, by adopting the view that some parameters (or
related ratios) are global and by using the constraints provided by
thermal coupling to the same radiation field at a given sky position
(like Eqs.~\ref{eq:L_analytical} and\ref{eq:absorb}) so that the two
temperatures are coupled, their model is simplified to only two
degrees of freedom instead of six.  Thus over the whole sky, fitting
\IRAS\ and lower resolution \DIRBE\ data, the model can be summarized
by two templates corresponding to the two degrees of freedom: the
total dust optical depth at $100\,\mu$m at a resolution of 6\farcm1
and a dust temperature map (for either component) at a resolution of
several degrees (it is almost constant at high Galactic latitude).

The dust model can be improved significantly by exploiting the
\Planck\ data. The exploration of the parametrization of the dust SED
done previously at 7\deg\ resolution with \FIRAS\ data can be done at
5\arcm\ with much better sensitivity. 
With the recalibration of the \Planck\ 545 and 857\,GHz data we have
shown that a single MBB is a good representation of the dust SED over
the 353--3000\,GHz frequency range, well within the relatively large
calibration uncertainties of the data (about 10\,\% at 545, 857, and
3000\,GHz) (Sect.~\ref{sec:fullskymaps},
Fig.~\ref{fig:data_model_noise}). 
\citet{planck2013-XVII} reached the same conclusion. Even
though the \Planck\ dust model assumes that the dust emission can be
modelled by a single MBB from 353 to 3000\,GHz, it has one extra
degree of freedom compared to \citet{finkbeiner1999} because
$\tau_{353}$, $T_{\rm obs}$, and $\beta_{\rm obs}$ are estimated at
each sky position. Finally, the parameter maps are at higher
resolution (5\arcm, 5\arcm, and 30\arcm, respectively) and are less
noisy, providing a very tight description of the data over that
frequency range.  With frequency coverage spanning the SED, we can
also measure the radiance \radiance.

Fig.~\ref{fig:Planck_over_FDS} shows the ratio of the predicted
brightness at 353\,GHz from the \citet{finkbeiner1999} dust model
(model 7) and that from the \Planck\ model, both at 30\arcm\
resolution.  

Although the global ratio of the two maps is compatible with one,
there are variations at all scales much larger than the uncertainties
of the \Planck\ model. Local variations larger than 30\,\% are seen
all over the sky, especially in the Galactic plane; the outer Galaxy
is significantly underpredicted in the \citet{finkbeiner1999} model
while the inner Galaxy is too bright. Because the \Planck\ dust model
is a particularly tight representation of the \Planck\ 353\,GHz data
(Fig.~\ref{fig:data_model_noise}), the same discrepancies are seen by
comparing the \citet{finkbeiner1999} model, integrated in the \Planck\
bandpass, directly with the 353\,GHz \Planck\ data.

The \Planck\ dust model produces an accurate 353\,GHz map almost free
of instrumental noise (but recall that the model includes the effects
of the CIBA).  That model map along with $T_{\rm obs}$ can be the
basis for extrapolation to lower frequencies, assuming that the
appropriate $\beta_{\rm obs, mm}$ can be identified.

\subsubsection{Frequency range of application}
\label{sec:freqrange}

We recall that our fit was done using data from 353 to
3000\,GHz. Extrapolating the model outside this range is not
recommended.  At higher frequencies the dust emission is known to be
in excess with respect to the big grain MBB, due to the emission from
smaller, stochastically heated, grains \citep[see,
e.g.,][]{draine2007a,compiegne2011}.

At frequencies below 353\,GHz the dust SED seems to be flatter
than that found for the frequency range here, i.e., from the
tests that we have made, extrapolation of the dust model underpredicts
the unmodelled \Planck\ dust emission at lower frequencies.
This is in accord with the results of
\citet{planck2013-XVII} for the south Galactic pole area
where $\beta_{\rm obs, mm}=1.54\pm0.03$ between 100 and 353\,GHz and
$\beta_{\rm obs, FIR}=1.65\pm0.10$ at higher frequencies. The spectral
index of dust between 100 and 353\,GHz estimated over the whole sky
using the {\tt Commander-Ruler} code \citep{planck2013-p06} has a mean
value of 1.49, thus also significantly flatter, and interesting
variations (their Fig.~16) that are similar though not identical to
what we see in our map of $\beta_{\rm obs}$
(Fig.~\ref{fig:fullres_maps}, lower).  While further discussion is
beyond the scope of this paper, it is clear that extrapolating the
current model to frequencies lower than 353\,GHz needs to be
approached with caution.

\subsection{Extinction: comparison of \Planck\  with Schlegel et al. (1998)}

\label{sec:sfd}

One of the expected uses of the \Planck\ dust products presented here
is to estimate reddening for extragalactic studies. Here we evaluate
how the \Planck\ $E(B-V)_{\rm xgal}$ map compares with the widely-used
$E(B-V)_{\rm SFD}$ map from \citet{schlegel1998}.

\begin{figure}
\centering
\includegraphics[draft=false, angle=0]{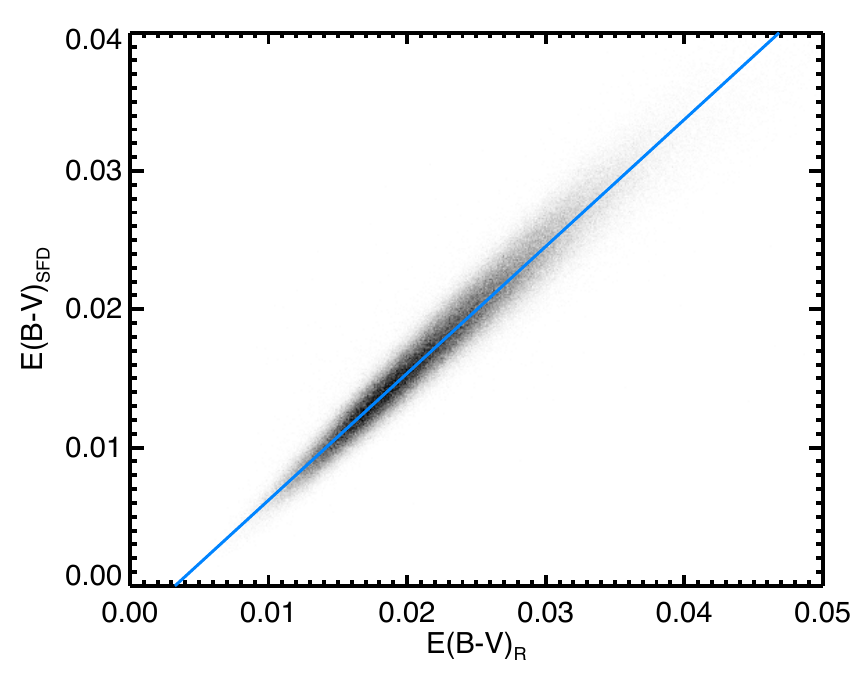}
\caption{\label{fig:Ebv_Planck_vs_SFD} Comparison of $E(B-V)_{\rm
    xgal}$ from \Planck\ (obtained from \radiance, smoothed to
  6\farcm1) and $E(B-V)_{\rm SFD}$ from \citet{schlegel1998} for the
  low $N_\ion{H}{i}$ mask (Fig~\ref{fig:hi_mask}). The greyscale gives
  the density of points and the solid line is the linear regression:
  $E(B-V)_{\rm SFD} = 0.92 \, E(B-V)_{\rm xgal}-0.003$. }
\end{figure}

The \Ebv\ map of \citet{schlegel1998} is proportional to their map of
$\tau_{3000}$, obtained assuming a constant $\beta_{\rm obs}$ and with
$T_{\rm obs}$ estimated with low resolution, low sensitivity \DIRBE\
data. The proportionality factor $E(B-V)/\tau_{3000}$ was estimated by
correlating $\tau_{3000}$ with colour excess measurements on 389
galaxies, assuming $R_V=3.1$ and the extinction curve from
\citet{cardelli1989} and \citet{o_donnell1994}. This factor
$E(B-V)/\tau_{3000}$ has been checked by different probes of extinction
and for much larger samples.
These studies all used SDSS data and showed that, globally, for $E(B-V)
< 0.5$, the regime of interest for extragalactic studies, $E(B-V)_{\rm
  SFD}$ is precise to 15\,\%, though not a fully consistent
picture. Using reddening of quasars, \citet{mortsell2013} concluded
that $E(B-V)_{\rm SFD}$ underestimates \Ebv\ by 20\,\% at low \Ebv\
values. Using reddening measurements of elliptical galaxies
\citet{peek2010} found local variations of $E(B-V)_{\rm
  SFD}/E(B-V)_{\rm elliptical}$ but no systematic bias of the
$E(B-V)_{\rm SFD}$ map.  Using reddening of stars,
\citet{schlafly2010} and \citet{schlafly2011} concluded that
$E(B-V)_{\rm SFD}$ overestimates \Ebv\ by 14\,\% but with spatial
variations of the normalization of the order of 10\,\% which might be
attributed to biases in the dust temperature map of
\citet{schlegel1998}. \citet{jones2011b} used SDSS colours of M dwarfs
to estimate the Galactic extinction properties. In their comparison to
\citet{schlegel1998} they often find \Ebv\ values lower than
$E(B-V)_{\rm SFD}$ at lower Galactic latitudes but mention that this
could be due to the fact that extinction toward stars does not trace
the full line-of-sight dust column density, an effect that could be
present in the analysis of \citet{schlafly2010,schlafly2011}. At high
Galactic latitudes \citet{jones2011b} report a number of lines of
sight where $E(B-V)_{\rm SFD}$ seems to underestimate \Ebv.

Based on near-infrared data in brighter areas ($E(B-V)>1.5$),
\citet{arce1999a} and \cite{cambresy2001} found that $E(B-V)_{\rm SFD}$
overestimates \Ebv\ systematically by more than 30\,\%. One possible
interpretation is an increase of the dust emission
efficiency, $\sigma_a$, relative to the dust absorption cross section,
$\overline{\sigma_a}$, in the dense medium, potentially caused by
changes in the grain structure, similarly to what is seen with
$\tau_{353}/N_{\rm H}$ in Fig.~\ref{fig:tau353_over_nh_vs_nh}, upper.  All of
these studies reveal variations of the ratio $E(B-V)_{\rm
  SFD}/E(B-V)_{\rm SDSS}$ that depend on position on the sky, on
methodology (stars, galaxies, quasars), and on the range of \Ebv. There
seem to be systematic trends where $E(B-V)_{\rm SFD}$ overestimates
reddening in dense regions and underestimates reddening in diffuse
areas at high Galactic latitudes.

The \Planck\ $E(B-V)_{\rm xgal}$ and $E(B-V)_{\rm SFD}$ are compared
in Fig.~\ref{fig:Ebv_Planck_vs_SFD}, for the low $N_\ion{H}{i}$ mask
(Fig.~\ref{fig:hi_mask}) corresponding to $N_\ion{H}{i}< 2\times
10^{20}$\,cm$^{-2}$ and low IVC emission. The comparison is done after
smoothing $E(B-V)_{\cal{R}}$ to 6\farcm1, the resolution of
$E(B-V)_{\rm SFD}$. The correlation between the two maps is excellent
with $E(B-V)_{\rm SFD} = 0.92 \, E(B-V)_{\cal{R}} -0.003$. The
dispersion around the correlation is 7\,\%. In relative terms $E(B-V)_{\rm SFD}$ 
underestimates $E(B-V)_{\rm xgal}$  by 8\,\% but in absolute terms the underestimate 
is more as there is also a negative offset of 0.003~mag. 

Our result depends on our choice of using \Ebv\ of quasars to
calibrate \radiance.  Given the variety of methodologies used in the
previous studies that have examined the calibration of the high
latitude extinction
\citep{schlafly2010,schlafly2011,peek2010,jones2011b,mortsell2013},
each having their own biases, the agreement is excellent.

\section{Conclusion}

\label{sec:conclusion}

We have presented an all-sky model of dust emission based on \Planck\
and \IRAS\ data at 5\arcm\ resolution, covering the frequency range from 353 to
3000\,GHz. We fit the data at each pixel of the {\HLP}\ $N_{\rm
  side}=2048$ grid assuming a modified blackbody model. The
parameters were estimated using a $\chi^2$ minimization, in two steps.
First the data were smoothed to 30\arcm\ and $\tau_{353}$, $T_{\rm
  obs}$, and $\beta_{\rm obs}$ were estimated in a three-parameter
fit. Then, using the spectral index $\beta_{\rm obs}$ at 30\arcm\ as a
fixed parameter, $\tau_{353}$ and $T_{\rm obs}$ were estimated at
5\arcm. We showed that this method minimizes the effect of noise on
the determination of the parameters (especially on the $T_{\rm
  obs}$ -- $\beta_{\rm obs}$ degeneracy in faint parts of the sky).

Over the whole sky, the mean and standard deviation of $T_{\rm obs}$
and $\beta_{\rm obs}$ are [$19.7$, $1.4$]\,K and [$1.62$, $0.10$],
respectively. The uncertainties of each parameter are about
3--6\,\%. We showed using Monte Carlo simulations that the
uncertainties are dominated by the CIBA, which are highly
correlated within the \Planck\ bands but only at a 30\,\% level
between \Planck\ and \IRAS\ 3000\,GHz. The CIBA is the dominant source
of uncertainty after smoothing the data to 30\arcm\ due to its
non-white power spectrum ($C_\ell \propto l^{-1}$).

This \Planck\ dust model reproduces the \Planck\
data well within the noise level at all frequencies.  Comparison of the
\citet{finkbeiner1999} dust model with the new data and model shows
only broad agreement, with variations of the order of 30\,\% at all
scales.

We found an increase of the dust opacity, $\tau_{353}/N_{\rm H}$, by a
factor of two from the diffuse to the higher column density (denser) ISM. 
Empirically, this increase is associated with a
decrease in $T_{\rm obs}$; because grains are in equilibrium with the
interstellar radiation field, we interpret this as a response to the
increased dust emissivity.  We also noted an excess of dust emission
and opacity at \hi\ column densities lower than $10^{20}$\,cm$^{-2}$
that might be attributed to dust in the WIM.

The combination of \Planck\ and \IRAS\ data allowed us to model the
dust emission from the Rayleigh-Jeans regime over the peak of the SED
to the Wien side and therefore to estimate the dust integrated
specific intensity or radiance \radiance\ for each line of sight.  We
also presented a map of the specific dust luminosity \Ldust\ by
normalizing \radiance\ with respect to \hi\ (Eq.~\ref{eq:dsl}).  Given
thermal equilibrium emission, this is a direct tracer of $U$, the
average strength of the interstellar radiation field (weighted by dust
absorption opacity) on each line of sight. This \Ldust\ map reveals an
increase of $U$ in the inner Galaxy, in active star forming regions,
and in the Magellanic Clouds. The map of \Ldust\ was shown to be
notably different from that of $T_{\rm obs}$, indicating that $T_{\rm obs}$ is
not a simple tracer of $U$ as often assumed. 
This is especially true at high Galactic latitudes where it was found
that \Ldust\ is fairly uniform and $T_{\rm obs}$ depends (inversely) on
the opacity $\tau_{353}/N_\ion{H}{i}$, confirming early \Planck\ results
\citep{planck2011-7.12}. This reveals that $\tau_{353}$ is
not the most reliable estimator of column density in the diffuse ISM.
The spatial variations of $T_{\rm obs}$ observed in the high-latitude
sky appear to be a response to variations of the dust emission opacity
resulting in grains of different equilibrium temperature even when
exposed to the same $U$.  The analysis at high Galactic latitude is
consistent with $U$ being fairly uniform, so that \radiance\ is a good
estimator of column density and can be used to estimate \Ebv\ there.

On the other hand, in molecular clouds we showed that variations of
$T_{\rm obs}$ are dominated by the effect of attenuation of the
interstellar radiation and/or local sources of heating photons. In
this type of environment, where the amplitude of the CIBA is
negligible, $\tau_{353}$ is a better estimator of column density than
\radiance, but the scale depends on the adopted opacity.  Compared to
the lower resolution work of \citet{schlegel1998}, the MBB analysis of
\Planck\ data in this paper provides estimates of $T_{\rm obs}$ at
5\arcm\ resolution and thus an improved higher-resolution estimate of
$\tau_{353}$, especially in high-contrast molecular regions where the
dust temperature and column density vary markedly at small scales.

The \Planck\ dust model was used to produce a map to measure Galactic
dust reddening for extragalactic studies at high Galactic latitude,
$E(B-V)_{\rm xgal}$. This map was based on the radiance \radiance\ and
calibrated by comparison with SDSS reddening measurements of
quasars. The correlation of $E(B-V)_{\rm xgal}$ with the \Ebv\ map of
\citet{schlegel1998} is very tight for $N_\ion{H}{i} < 2\times
10^{20}$\,cm$^{-2}$, but has a slope significantly different than one,
in the sense that the \Ebv\ map of \citet{schlegel1998} underestimates
\Ebv\ by $8$\,\% in the diffuse ISM. We stress that $E(B-V)_{\rm xgal}$ is
reserved for extragalactic studies; it should not be used to estimate
reddening in lines of sight where $E(B-V) > 0.3$, i.e., where
attenuation effects on \radiance\ become important. There we recommend
the map of $\tau_{353}$ multiplied by the $E(B-V)/\tau_{353}$ ratio
also calibrated using quasars.  However, systematic decreases of scale
can arise from region to region, and even locally within a region,
because of the increases in the opacity $\sigma_{\mathrm{e}\,353}$ that,
empirically, accompany increase in (column) density.

The \Planck\ dust products ($\tau_{353}$, $T_{\rm obs}$, $\beta_{\rm
  obs}$, \radiance\ and $E(B-V)_{\rm xgal}$) are available on the PLA 
(see Appendix~\ref{revisedPLA}).

\begin{acknowledgements}

The development of \Planck\ has been supported by: ESA; CNES and
CNRS/INSU-IN2P3-INP (France); ASI, CNR, and INAF (Italy); NASA and DoE
(USA); STFC and UKSA (UK); CSIC, MICINN, JA and RES (Spain); Tekes,
AoF and CSC (Finland); DLR and MPG (Germany); CSA (Canada); DTU Space
(Denmark); SER/SSO (Switzerland); RCN (Norway); SFI (Ireland);
FCT/MCTES (Portugal); and PRACE (EU). A description of the Planck
Collaboration and a list of its members, including the technical or
scientific activities in which they have been involved, can be found
at
\url{http://www.sciops.esa.int/index.php?project=planck&page=Planck_Collaboration}.
The research leading to these results has received funding from the 
European Research Council under the European Union's Seventh Framework 
Programme (FP7/2007-2013) / ERC grant agreement N$^o$ 267934.
Funding for the SDSS and SDSS-II has been provided by the Alfred
P. Sloan Foundation, the Participating Institutions, the National
Science Foundation, the U.S. Department of Energy, the National
Aeronautics and Space Administration, the Japanese Monbuk agakusho,
the Max Planck Society, and the Higher Education Funding Council for
England. The SDSS Web Site is \url{http://www.sdss.org/}.
The SDSS is managed by the Astrophysical Research Consortium for the
Participating Institutions. The Participating Institutions are the
American Museum of Natural History, Astrophysical Institute Potsdam,
University of Basel, University of Cambridge, Case Western Reserve
University, University of Chicago, Drexel University, Fermilab, the
Institute for Advanced Study, the Japan Participation Group, Johns
Hopkins University, the Joint Institute for Nuclear Astrophysics, the
Kavli Institute for Particle Astrophysics and Cosmology, the Korean
Scientist Group, the Chinese Academy of Sciences (LAMOST), Los Alamos
National Laboratory, the Max-Planck-Institute for Astronomy (MPIA),
the Max-Planck-Institute for Astrophysics (MPA), New Mexico State
University, Ohio State University, University of Pittsburgh,
University of Portsmouth, Princeton University, the United States
Naval Observatory, and the University of Washington.
Some of the results in this paper have been derived using the {\HLP}\
package.

\end{acknowledgements}


\allresultspapers

\bibliographystyle{aa}
\bibliography{draft_v48,Planck_bib}

\begin{appendix}

\section{Zodiacal emission}

\subsection{Zodiacal emission correction at 100\,$\mu$m}\label{sec:rmzody}

\begin{figure}
\centering
\includegraphics[draft=false, angle=0]{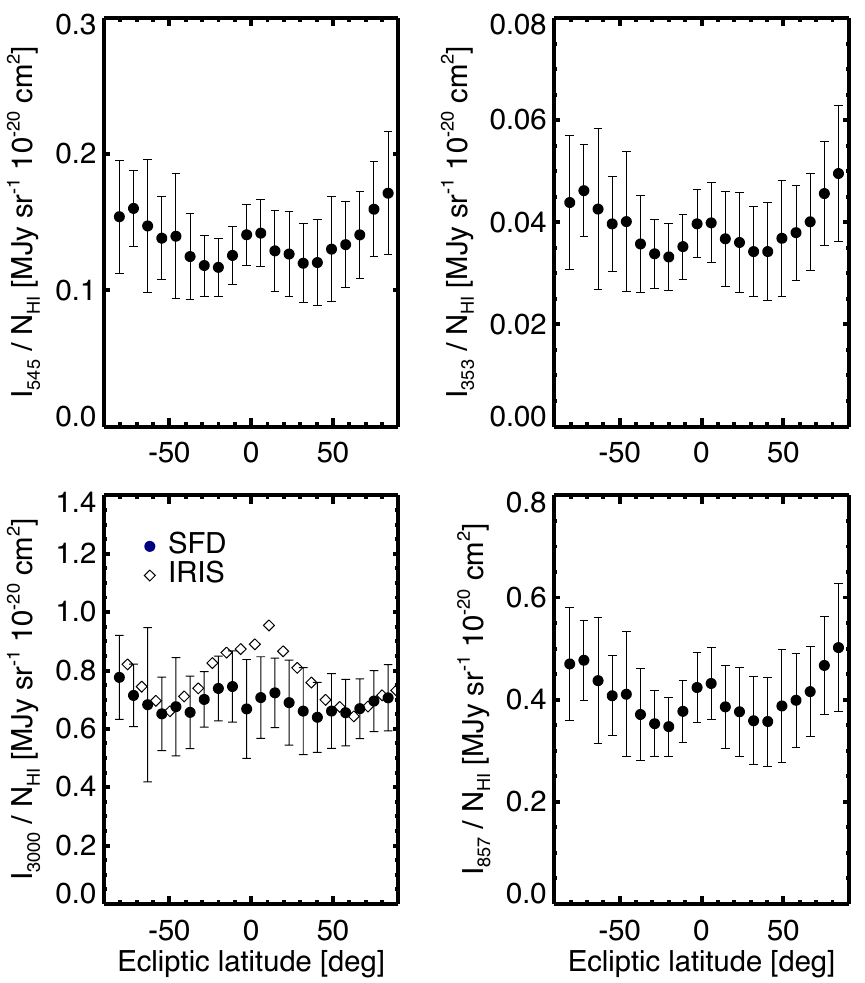}
\caption{\label{fig:ecliptic} Dust emissivity $I_\nu/N_\ion{H}{i}$ as a function of
ecliptic latitude for 3000\,GHz (\textit{lower left}), 857\,GHz (\textit{lower right}), 545\,GHz (\textit{upper left}) and 353\,GHz
(\textit{upper right}). Each point gives the average and standard deviation
(error bar) of all pixels in the corresponding bin in ecliptic
latitude (\IRIS\ error bars omitted for clarity). These plots were obtained using data smoothed to 1\deg\ and
selecting pixels with $N_\ion{H}{i} < 3\times 10^{20}$\,cm$^{-2}$.}
\end{figure}

Zodiacal emission (ZE) is a component that is difficult to remove from the data as
it is changing with sky position as well as time of observation. A
detailed model has been built to correct the \Planck\ maps for ZE
\citep{planck2013-pip88}. For the \IRAS\ 100\,$\mu$m map, as
mentioned in Sect.~\ref{sec:100um}, the \IRIS\ and SFD maps were not
corrected for ZE in the same way. The impact of the different ZE
correction for these two maps can be appreciated from
Fig.~\ref{fig:ecliptic} where the dust emissivity $I_\nu/N_\ion{H}{i}$
is plotted 
for the four frequencies, using data smoothed to 1\deg\ and selecting
only pixels with $N_\ion{H}{i}<3\times10^{20}$\,cm$^{-2}$. In such a low
column density regime, the correlation between dust emission and \hi\
column density was shown to be tight (Fig.~\ref{fig:offset}).
Therefore, any systematic
variation of $I_\nu/N_\ion{H}{i}$, especially at low angular
resolution, can reveal emission from components other than \hi, for example
the WIM or residual ZE.  
To assess the latter, in Fig.~\ref{fig:ecliptic} the emissivity is
plotted as a function of ecliptic latitude. For the \Planck\
frequencies the emissivity is almost constant with ecliptic latitude,
validating the \Planck\ ZE removal. The SFD data also show a constant
emissivity, the exception being the \IRIS\ map that shows a systematic
increase toward the ecliptic plane, indicative of residual ZE. This is
why we implemented the procedure described in Sect.~\ref{sec:100um}.

\subsection{Impact of zodiacal emission correction on dust parameters}

\label{sec:PIP82}

\begin{table*}
\caption{\label{tab:compare_PIP82} Data configuration exploration and comparison with result from \citet{planck2013-XVII}. }
\begin{center}
\begin{tabular}{lcccccc}\specialrule{\lightrulewidth}{0pt}{0pt} \specialrule{\lightrulewidth}{1.5pt}{\belowrulesep}
 & $\langle T_{\rm obs} \rangle$ & $\sigma(T_{\rm obs})$ & $\langle \tau_{353}/N_\ion{H}{i} \rangle$ & $\sigma(\tau_{353}/N_\ion{H}{i})$ \\ 
Configuration & [K] & [K] & [cm$^2$\,H$^{-1}$] & [cm$^2$\,H$^{-1}$] \\ \midrule
\Planck\ ZE rm, IRIS & 20.6 & 1.0 & $6.8\times 10^{-27}$ & $1.8\times 10^{-27}$ \\
\Planck\ ZE rm, SFD & 20.2 & 0.8 & $7.0\times 10^{-27}$ & $1.7\times 10^{-27}$ \\
\Planck, IRIS & 20.0 & 0.7 & $8.0\times 10^{-27}$ & $1.6\times 10^{-27}$ \\
\Planck, SFD & 19.6 & 0.5 & $8.3\times 10^{-27}$ & $1.6\times 10^{-27}$ \\
\citet{planck2013-XVII} & 19.7 & 0.9 & $7.3\times 10^{-27}$ & $2.2\times 10^{-27}$ \\ \bottomrule[\lightrulewidth]
\end{tabular}
\end{center}
{\bf Note:} The average and standard deviation of $T_{\rm obs}$ and $\tau_{353}/N_\ion{H}{i}$ were computed on the south Galactic pole area studied in \citet{planck2013-XVII}. Data smoothed to 60\arcm\ ($N_{\rm side}=128$) were used to compute the dust parameters. In this comparison a fixed $\beta_{\rm obs}=1.65$ was assumed in the fit.
\end{table*}

To evaluate the impact of the ZE on our analysis, we also compared our
results on fitting parameters with those of
\citet{planck2013-XVII} for the same masked region that they
studied, an area of 7500\,deg$^2$ toward the south Galactic
cap. This check is useful because there is a fundamental difference
between the two analyses. Here the dust SED is modelled using the
observed specific intensity for each pixel independently; results are
therefore sensitive to uncertainties in the zero levels of the maps.
On the other hand, \citet{planck2013-XVII} 
correlated the dust maps with \hi\ within regions 15\deg\ in
diameter; they showed that their results are insensitive to the zero
level of the maps and to the ZE that is very uniform on 15\deg\
scales.

To further evaluate the impact of the ZE correction on our analysis, we have explored different data configurations and compared our results and those of  \citet{planck2013-XVII} in the south Galactic pole area. 

Given the lower-resolution results of \citet{planck2013-XVII}, we used data smoothed to 60\arcm\ on an $N_{\rm side}=128$ grid. The comparison was done using different combinations of maps: \IRIS\ or SFD at 100\,$\mu$m (at such low resolution, SFD is equivalent to the combined \IRIS+SFD map, see Sect.~\ref{sec:100um}) and maps with and without ZE removed for \Planck. To be compatible with \citet{planck2013-XVII}, the fit was done using a fixed $\beta_{\rm obs}=1.65$. The results we obtained on $T_{\rm obs}$ and $\tau_{353}/N_\ion{H}{i}$ for this south Galactic cap region are compiled in Table~\ref{tab:compare_PIP82}. The differences in $T_{\rm obs}$ and $\tau_{353}/N_\ion{H}{i}$ between data configurations are limited, within the standard deviation observed over the region. They are especially small for $T_{\rm obs}$ that shows variations of less than 5\,\% between data configurations. The largest effect is from the removal of the ZE in the \Planck\ data that reduces $\tau_{353}$ by 15\,\%. The impact of the choice of \IRIS\ or SFD on $\tau_{353}$ is more limited; fitting the data with SFD produces a $\tau_{353}$ about 3\,\% higher than with \IRIS. 
Nevertheless and even though there is a general good spatial correlation between the maps of $T_{\rm obs}$ and $\tau_{353}/N_\ion{H}{i}$ obtained with the two methods, care should be taken in comparing them in greater detail. Contrary to the results obtained using a fit of the observed specific intensity, the results of \citet{planck2013-XVII} are not sensitive to dust emission associated with the WIM that is not spatially correlated with \hi\ and to the CIBA. In addition, \citet{planck2013-XVII} showed that in the southern Galactic cap area there are \hi\ clouds at local velocities that do not have associated dust emission. These effects produce spatial variations of $T_{\rm obs}$ and $\tau_{353}/N_\ion{H}{i}$ computed with the two methods. Even with these caveats, there is a good agreement between the two analyses.
In particular we note that the data configuration combining \Planck\ (ZE removed) together with the SFD map at large scales (the equivalent of the 100\,$\mu$m map built in Eq.~\ref{eq:100micron}) has $\langle \tau_{353}/N_\ion{H}{i} \rangle$, the closest to the values found by \citet{planck2013-XVII}.


\section{The $\chi^2$ fit}\label{sec:mpfit}

Adopting a reference frequency $\nu_0=353$\,GHz, the parameters
$\tau_{\nu_0}$, $T_{\rm obs}$ and $\beta_{\rm obs}$ (see
Eq.~\ref{eq:mbb}) were found at each sky position $p$ by fitting the
\Planck\ and \IRAS\ data $I_\nu(p)$. The {\tt MPFIT} $\chi^2$
minimization routine was used \citep{markwardt2009}. Colour
corrections due to finite bandpasses (\Planck\ and \IRAS) were taken
into account explicitly at each iteration. To speed up the
convergence of the fit, initial estimates of the parameters were
provided: $\beta_{\rm obs}$ was initialized to 1.65, $T_{\rm obs}$ was
initialized to the colour temperature $T_{3000-857}$ obtained from the
ratio of $I_{3000}/I_{857}$ and assuming $\beta_{\rm obs}=1.65$, and
$\tau_{353}$ was initialized to $I_{3000} / [ B_{3000}(T_{3000-857})
\times (3000/353)^{1.65}]$ for frequencies in GHz.

The fit takes into account the calibration uncertainty of the data
($c_\nu \, I_\nu(p)$ -- see Table~\ref{tab:data}), the uncertainty of
the CMB removal estimated to be $c_\nu \,I_\mathrm{CMB}(p)$ where $I_\mathrm{CMB}$ is the
{\tt SMICA} map \citep{planck2013-p06}, the uncertainty of the offset
($\delta O_\nu$ -- see Table~\ref{tab:data}), and the instrumental
noise $n_\nu(p)$. For both \IRAS\ and \Planck, the instrumental noise
is modulated inversely by the square root of the coverage map (number
of times a sky pixel $p$ has been observed).
All sources of uncertainty are added in quadrature:
\begin{equation}
\label{eq:noise}
N_\nu(p) = \sqrt{ c_\nu^2 \, I_\nu(p)^2 + c_\nu^2 \, I_\mathrm{CMB}(p)^2 + (\delta O_\nu)^2 + n_\nu(p)^2}\,.
\end{equation}
%

\section{Impact of the CIBA on the dust parameters}\label{sec:MC_CIBA}

This section describes the Monte Carlo simulations done to quantify the impact of instrumental noise and the CIBA on the parameters of the MBB fit.
Specifically, we simulated a single SED as the sum of dust emission, the CIBA, and instrumental noise, for data smoothed to different angular resolution. 

The dust emission was modelled using typical values of dust parameters and taking into account the \Planck\ and \IRAS\ bandpasses.
The noise level used at each frequency is the median value found in the low $N_\ion{H}{i}$ mask (Fig~\ref{fig:hi_mask}), properly scaled as a function of the angular resolution considered.
The CIBA levels in the \Planck\ channels follow the parametrization of \citet{planck2013-XVII} ($T_{\rm CIB} = 18.3$\,K and $\beta_{\rm CIB} = 1.0$) normalized at 857\,GHz to the value given by \citet{planck2011-6.6}. The CIBA level at 3000\,GHz is that of \citet{penin2012}.
The scaling of the CIBA level with resolution was done assuming a power spectrum $C_\ell \propto \ell^{-1}$. The noise and CIBA levels for each frequency and each resolution are given in Table~\ref{tab:CIB_levels}.

\begin{table*}
\caption{\label{tab:CIB_levels} Noise and CIBA levels used for the Monte Carlo simulations. }
\begin{center}
\begin{tabular}{c cc cc cc cc}\specialrule{\lightrulewidth}{0pt}{0pt} \specialrule{\lightrulewidth}{1.5pt}{\belowrulesep} 
  & \multicolumn{2}{c}{353 GHz} & \multicolumn{2}{c}{545 GHz} & \multicolumn{2}{c}{857 GHz} & \multicolumn{2}{c}{3000 GHz}\\ 
\cmidrule(rl){2-3} \cmidrule(rl){4-5} \cmidrule(rl){6-7} \cmidrule(rl){8-9} 
 $\theta$ & $n$ & $\sigma_{\rm CIBA}$ & $n$ & $\sigma_{\rm CIBA}$ & $n$ & $\sigma_{\rm CIBA}$ & $n$ & $\sigma_{\rm CIBA}$\\
 \,[arcmin] & [MJy\,sr$^{-1}$] & [MJy\,sr$^{-1}$] & [MJy\,sr$^{-1}$] & [MJy\,sr$^{-1}$] & [MJy\,sr$^{-1}$] & [MJy\,sr$^{-1}$] & [MJy\,sr$^{-1}$] & [MJy\,sr$^{-1}$]\\ 
\midrule
$\:\:\:\:5$ & $3.19\times 10^{-2}$ & $1.60\times 10^{-2}$ & $3.44\times 10^{-2}$ & $4.36\times 10^{-2}$ & $3.41\times 10^{-2}$ & $1.00\times 10^{-1}$ & $6.20\times 10^{-2}$ & $1.00\times 10^{-1}$\\
$\:\:15$ & $3.63\times 10^{-3}$ & $9.23\times 10^{-3}$ & $3.91\times 10^{-3}$ & $2.52\times 10^{-2}$ & $3.86\times 10^{-3}$ & $5.77\times 10^{-2}$ & $6.93\times 10^{-3}$ & $5.77\times 10^{-2}$\\
$\:\:30$ & $1.74\times 10^{-3}$ & $6.52\times 10^{-3}$ & $1.88\times 10^{-3}$ & $1.78\times 10^{-2}$ & $1.86\times 10^{-3}$ & $4.08\times 10^{-2}$ & $3.36\times 10^{-3}$ & $4.08\times 10^{-2}$\\
$\:\:60$ & $8.61\times 10^{-4}$ & $4.61\times 10^{-3}$ & $9.29\times 10^{-4}$ & $1.26\times 10^{-2}$ & $9.20\times 10^{-4}$ & $2.89\times 10^{-2}$ & $1.67\times 10^{-3}$ & $2.89\times 10^{-2}$\\
$120$ & $4.29\times 10^{-4}$ & $3.26\times 10^{-3}$ & $4.63\times 10^{-4}$ & $8.89\times 10^{-3}$ & $4.59\times 10^{-4}$ & $2.04\times 10^{-2}$ & $8.32\times 10^{-4}$ & $2.04\times 10^{-2}$\\ 
\bottomrule[\lightrulewidth]
\end{tabular}
\end{center}
{\bf Note:} The levels of noise, $n$, and the CIBA, $\sigma_{\rm CIBA}$, are given for each channel and for data smoothed to different angular resolution, $\theta$.
The noise level is the median value of the noise in the low $N_{\ion{H}{i}}$
mask (see Fig.~\ref{fig:hi_mask}).
The CIBA standard deviations at \Planck\ frequencies are derived from the parametrization of \citet{planck2013-XVII} ($T_{\rm CIBA} = 18.3$\,K and $\beta_{\rm CIBA} = 1.0$) normalized at 857\,GHz with the value given by \citet{planck2011-6.6}. The value at 3000\,GHz is from \citet{penin2012}. The scaling of $\sigma_{\rm CIBA}$ from the original measurements, made at 5\farcm4 by \citet{planck2011-6.6} and 4\farcm3 by \citet{penin2012}, to 5\arcm\ and 30\arcm\ was done assuming a CIBA power spectrum of $C_\ell \propto \ell^{-1}$.
\end{table*}

Because galaxies at different redshifts have their peak emission at different frequencies, the structure of the CIB on the sky is only partly correlated between frequencies.
Looking at \citet{planck2011-6.6} and \citet{planck2011-7.12}, who showed the residual brightness fluctuations in selected faint patches of the sky after removal of the interstellar dust emission traced by 21\,cm emission, one can appreciate visually the level of spatial correlation of the CIBA between frequencies. The fluctuations are strongly correlated in the 857--353\,GHz range, slightly less at 3000\,GHz. The level of frequency (de)coherence of the CIBA was quantified by \citet{planck2013-pip56} for $150 < \ell < 1000$. In the high-frequency channels of \Planck\ they report a strong correlation: 0.95 between 857 and 545\,GHz, 0.91 between 857 and 353\,GHz, and 0.98 between 545 and 353\,GHz. It is weaker between 3000\,GHz and the \Planck\ frequencies: 0.36, 0.31, and 0.29 at 857, 545 and 353\,GHz respectively. \citet{planck2013-XVII} modelled the part of the CIBA correlated with the ones at 857\,GHz based on the results of \citet{planck2013-pip56}. They found that it can be well fitted by a MBB function with $T_{\rm CIB} = 18.3$\,K and $\beta_{\rm CIB} = 1.0$. This parametrization is compatible with the one found in a similar study by \citet{planck2011-7.12}. 

To model the CIBA, $c_\nu$, taking into account the inter-frequency correlation, we have proceeded the following way. We made the assumption that $c_\nu$ is fully correlated in the \Planck\ frequencies. Therefore, the CIBA in the \Planck\ bands is simply modelled as a single random draw scaled by the CIBA levels:
\begin{equation}
c_\nu = A^{\rm cor} \, \sigma_{\rm CIBA\,\nu}\,,
\end{equation}
where $A^{\rm cor}$ is drawn from a normal distribution with unit standard deviation and zero mean, and $\sigma_{\rm CIBA\,\nu}$ is the CIBA level at each \Planck\ frequency given in Table~\ref{tab:CIB_levels}. 

At 3000\,GHz the CIBA is not strongly correlated with the CIBA at \Planck\ frequencies so it was divided into one part correlated with 857\,GHz and another part uncorrelated:
\begin{equation}
c_{3000} = A^{\rm cor} \, \sigma^{\rm cor}_{\rm CIBA\,3000} + A^{\rm uncor} \, \sigma^{\rm uncor}_{\rm CIBA\,3000}\,,
\end{equation}
where $A^{\rm uncor}$ represent a second random draw.
The correlated part, $\sigma^{\rm cor}_{\rm CIBA}$, is estimated using the [$T_{\rm CIBA} = 18.3$\,K, $\beta_{\rm CIBA} = 1.0$] parametrization, normalized with $c_{857}$. 
The uncorrelated part, $\sigma^{\rm uncor}_{\rm CIBA}$, is simply the quadratic complement to $\sigma_{\rm CIBA\,3000}$. 
At 5\arcm\ these contributions are $\sigma^{\rm cor}_{\rm CIBA\,3000}=0.049$ MJy\,sr$^{-1}$ and $\sigma^{\rm uncor}_{\rm CIBA\,3000}=0.087$ MJy\,sr$^{-1}$ highlighting the fact that the uncorrelated part is the dominant source of CIB fluctuations at 3000\,GHz.

Noise was assumed to be uncorrelated in frequency, with levels given in Table~\ref{tab:CIB_levels}.

Depending on the situation, this procedure was executed with different dust parameters ($\tau_{353}$, $T_{\rm obs}$, and $\beta_{\rm obs}$) and at different data resolutions. The fit of the simulated SED was done using the method that was used with the real data, including colour corrections. 
Simulations were done with and without noise to quantify the specific effect of noise and the CIBA on the recovered parameters.

\section{Production of maps with point sources removed}\label{sec:pointsources}

\begin{figure}
\centering
\includegraphics[draft=false, angle=0]{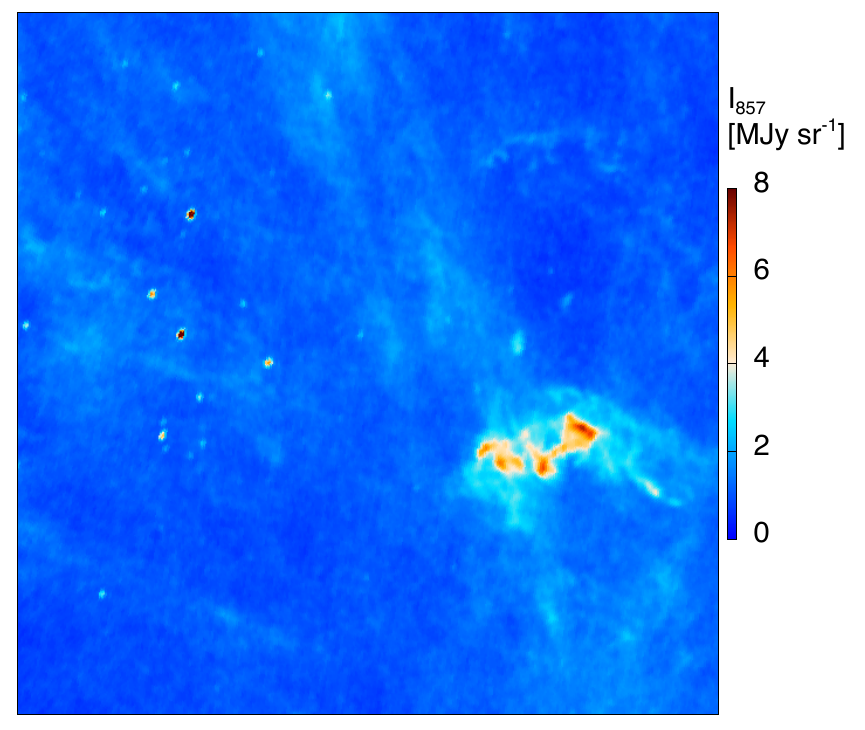}
\includegraphics[draft=false, angle=0]{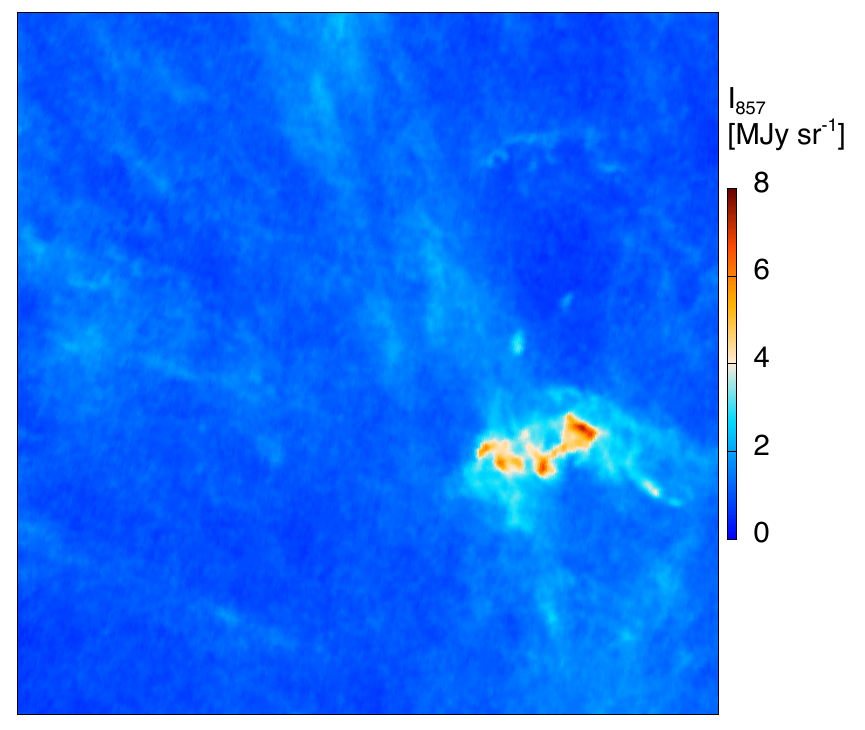}
\includegraphics[draft=false, angle=0]{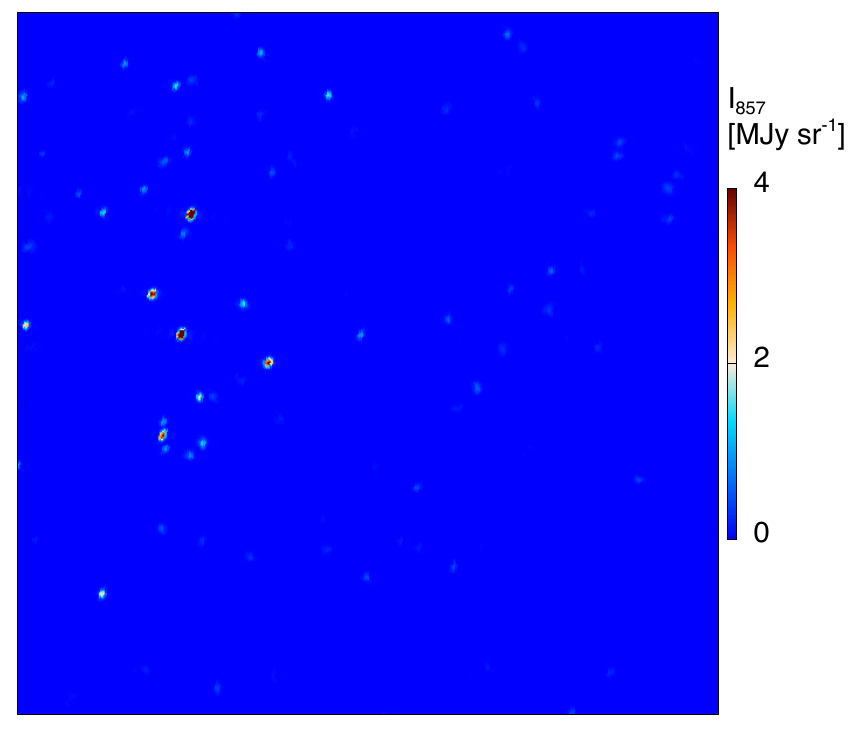}
\caption{\label{fig:rmsources} Example of the point source removal in a $10^\circ \times 10^\circ$ region centred on $l=260$\deg, $b=75$\deg; original map at 857\,GHz (\textit{upper}), point source removed map (\textit{middle}), and difference (\textit{lower}).}
\end{figure}

The $E(B-V)_{\rm xgal}$ product, to be used for estimating Galactic reddening for
extragalactic studies, is obtained with a fit of the dust model on a version
of the \Planck\ and \IRAS\ data from which point sources have been removed. 
This section describes how the removal was done. 
Most of the software used in this process is part of the {\em Planck Sky Model} software package ({\tt PSM}). A description of the {\tt PSM} and how to download it can be found in \citet{delabrouille2013}.

\subsection{Making the mask containing the radio and IR sources}
 
A source mask is made as a HEALPix map from the union of discs centred on a selected set of known FIR and radio sources. The FIR sources are compiled from the \IRAS\ Point Source Catalog \citep[PSC, ][]{beichman1988} and the Faint Source Catalog \citep[FSC, ][]{moshir1992}. 
Radio sources are derived from the \WMAP\ catalog of point sources detected in the 9-year sky maps \citep{bennett2013}, supplemented with a selection of extragalactic point sources detected by \Planck\ at 100\,GHz \citep[ERCSC, ][]{planck2011-1.10} that are not in the \WMAP\ catalog.
Pixels within a 15\arcm\ radius of each source are masked.

\subsection{Grouping adjacent pixels for each source}

While most sources are isolated, in rare cases masks around two or more sources overlap. This results in larger, non-circular holes in the mask. As a first step, we divide the mask into a set of small connected holes (most but not all of which correspond to the masking of one single source). This is performed by the {\tt PSM} routine {\tt GROUP\_ADJACENT\_PIXELS}.  

\subsection{Interpolating across the source}

Finally, we estimate the diffuse brightness in each of the small connected holes containing point sources using the {\tt PSM} routine {\tt FILL\_SMALLGAP}. The interpolation is done using a minimum curvature surface ({\tt IDL} routine {\tt MIN\_CURV\_SURF}). This step produces the final, de-sourced maps.

\section{Estimating \Ebv\ from colour excess measurements}\label{sec:ebv}

This section describes how \Ebv\ was estimated from the multi-colour SDSS measurements of quasars. The colour excess due to dust extinction is defined as
\begin{eqnarray}
\label{eq:Exy}
E(X-Y) & = & A_X - A_Y\\
& = & (m_X-m_{X0}) - (m_Y - m_{Y0})\\
\label{eq:Exy_3}
& = & -2.5 \log\left( \frac{F_X}{F_{X0}} \frac{F_{Y0}}{F_Y}\right),
\end{eqnarray}
where, for a given band $i$, $A_i$ is the extinction, $m_i$ the observed magnitude, $m_{i0}$ the absolute magnitude, $F_i$ the observed flux, and $F_{i0}$ the intrinsic flux of the source. 

The quantity $F_{i0}$ corresponds to the source flux density $f_\lambda$ convolved with the filter transmission $T_i(\lambda)$
\begin{equation}
\label{eq:Fi0}
F_{i0} = \int_i T_i(\lambda) \, f_\lambda \, d\lambda\,,
\end{equation}
while $F_i$ is the same quantity but affected by extinction;
\begin{equation}
\label{eq:Fi}
F_i = \int_i T_i(\lambda) \, f_\lambda \, 10^{-0.4 \, E(B-V) \, C_\lambda} \, d\lambda\,,
\end{equation}
where $C_\lambda$ is the extinction curve for a given value of $R_V$ and normalized to $E(B-V)$:
\begin{equation}
C_\lambda = A_\lambda / E(B-V)\,.
\end{equation} 
In what follows, we use $C_\lambda$ of \citet{fitzpatrick1999} and assume $R_v=3.1$.

Equations~\ref{eq:Exy} to \ref{eq:Fi} relate the observed colour excess of a source in bands $X$ and $Y$ to its intrinsic spectrum $f_\lambda$, to the transmission of the filters $T_i$ and to the extinction of dust along the line of sight. As mentioned by \citet{fitzpatrick2005}, if the intrinsic spectrum of the source and the filter transmissions are known, the normalization of the extinction curve (i.e., \Ebv) can be estimated directly from the observed colour excess $E(X-Y)$. 

An often overlooked fundamental fact follows from this description. Because the observed magnitudes $m_i$ are obtained with broad band filters, the measured value of $E(X-Y)$ for $X=B$ and $Y=V$ will in general be different from the normalization of the extinction curve \Ebv\ in Eq.~\ref{eq:Fi}. The observed colour excess is indeed a convolution of the source spectrum, the filter transmission and the extinction curve, and so there are bandpass corrections. We stress that the quantity \Ebv\ we are estimating here is independent of the spectrum of the background source. It is the value that is used to scale the extinction curve.

\begin{figure}
\includegraphics[draft=false, angle=0]{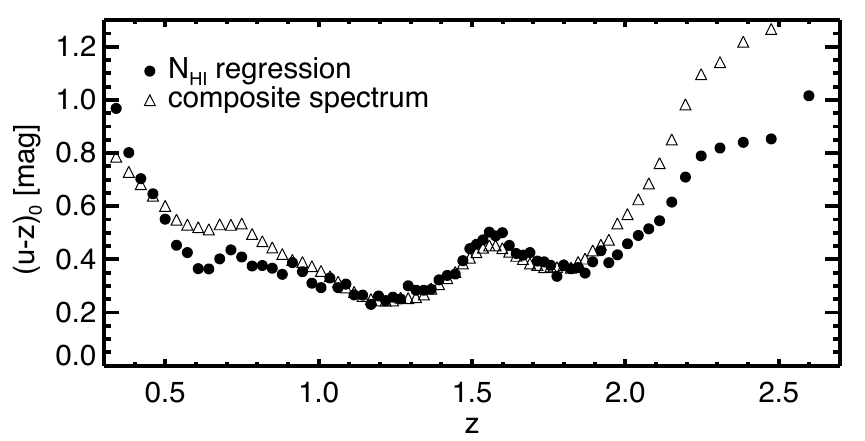}
\includegraphics[draft=false, angle=0]{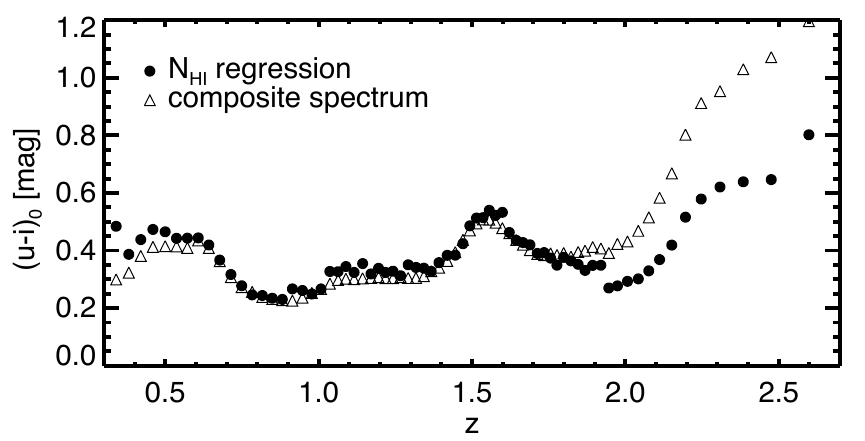}
\includegraphics[draft=false, angle=0]{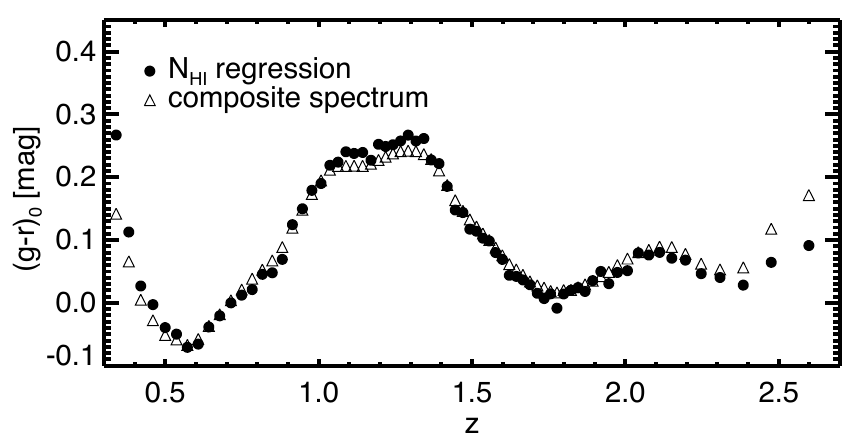}
\includegraphics[draft=false, angle=0]{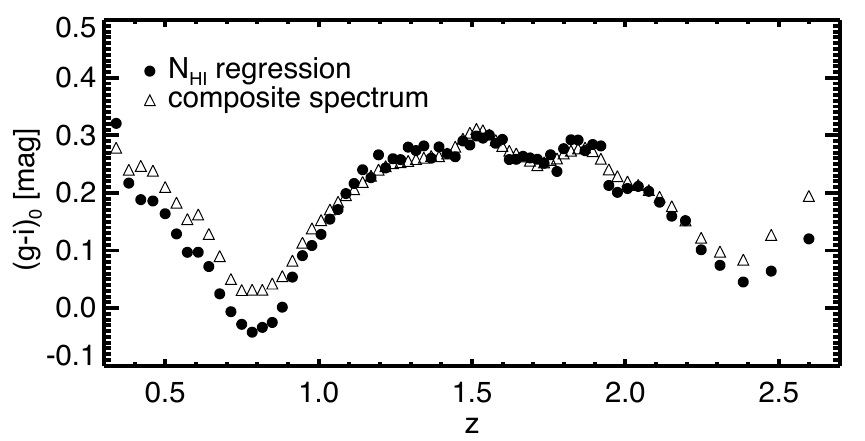}
\caption{\label{fig:qso_intr_color} Intrinsic colour of quasars as a function of redshift for $(u-z)$, $(u-i)$, $(g-r)$ and $(g-i)$, \textit{top} to \textit{bottom}. The black points give the intercept of the $(m_X-m_Y)$ vs.\ $N_\ion{H}{i}$ relation in each bin of redshift (see Eq.~\ref{eq:egr_vs_tau}). The triangles give the intrinsic colour of quasars computed with the composite spectrum (see Eq.~\ref{eq:int_col_template}).}
\end{figure}

From the SDSS quasar catalogue one can deduce directly the observed colours $(m_X - m_Y)$ but to estimate \Ebv\ using the formalism described in the previous section one also needs the source spectrum $f_\lambda$ and the intrinsic colour $(m_{X0} - m_{Y0})$. 
One of the advantages of using quasars is that they are known to have a fairly constant spectrum. Below we use the composite quasar spectrum of \citet{vanden_berk2001} for $f_\lambda$. The intrinsic colour $(m_{X0} - m_{Y0})$ is estimated by correlating $(m_X-m_Y)$ with a tracer of dust extinction. We use the \hi\ column density provided in the SDSS quasar catalogue as a proxy:\footnote{We preferred not to use the \Planck\ $\tau_{353}$ or the SFD map so as not to bias the analysis.}
\begin{equation}
\label{eq:egr_vs_tau}
(m_X-m_Y) = \psi \,N_\ion{H}{i} + (m_{X0}-m_{Y0})\,.
\end{equation}
The constant term of this linear regression is the average intrinsic colour of quasars but, as quasars are at different redshifts (up to $z \sim 5$), this intrinsic colour depends on $z$. Therefore we correlated $(m_X-m_Y)$ with $N_\ion{H}{i}$ for quasars in bins of $z$ where the width of each bin is set to have 1000 quasars per bin. Fig.~\ref{fig:qso_intr_color} shows the intrinsic colours found as a function of $z$ for the following four colours: $(m_u-m_z)$, $(m_u-m_i)$, $(m_g-m_r)$ and $(m_g-m_i)$. 

In order to validate the approach we also computed the intrinsic colour as a function of $z$ of the composite quasar spectrum:
\begin{equation}
\label{eq:int_col_template}
(m_{X0} - m_{Y0}) = -2.5 \log\left( \frac{F_{Y0}}{F_{X0}} \right)\,.
\end{equation}
The results for the same four colours are shown in red in Fig.~\ref{fig:qso_intr_color}. The good agreement between the two methods indicates that the composite spectrum is a good average representation. It also gives confidence in its use to estimate \Ebv\ using the formalism described earlier. 

The correspondence between the two methods is less good at high redshift. This can be attributed to the Lyman-$\alpha$ line that enters the shortest wavelength band. Indeed \citet{mortsell2013} mentioned that at $z>2.3$, Lyman-$\alpha$ introduces large variations in the intrinsic quasar colour computed with band $g$. A similar effect is seen at $z>1.7$ for colours using the $u$ band. We also noticed a larger dispersion in the residual ($(m_X-m_Y) - \psi \,N_\ion{H}{i}$) for bins with $z<0.7$. Based on this, we selected quasars in the redshift range $0.7 < z < 1.7$. The sample contains 53\,399 quasars.

For each quasar at a given redshift $z$, we computed the colour excesses $E(X-Y) = (m_X-m_Y)-(m_{X0}-m_{Y0})$ using the intrinsic colour estimated with the correlation with $N_\ion{H}{i}$ in bins of redshift (Eq.~\ref{eq:egr_vs_tau}). We then integrated numerically equations~\ref{eq:Fi} and \ref{eq:Fi0} using the composite quasar spectrum redshifted appropriately and solved for $E(B-V)$ using Eq.~\ref{eq:Exy_3}. Because we work with colours, the exact normalization of the quasar spectrum ($f_\lambda$) and of the SDSS transmissions ($T_i$) cancels out in Eq.~\ref{eq:Exy_3}. With this procedure we obtain an estimate of \Ebv\ for each quasar independently for each colour excess $E(X-Y)$. 

\begin{table}
\caption{\label{tab:colour} Value of $E(B-V)/\tau_{353}$ and $E(B-V)/{\cal{R}}$ obtained by correlation using \Ebv\ deduced for each quasar colour. }
\begin{center}
\begin{tabular}{ccc}\specialrule{\lightrulewidth}{0pt}{0pt} \specialrule{\lightrulewidth}{1.5pt}{\belowrulesep}
 & $E(B-V)/\tau_{353}$ & $E(B-V)/{\cal{R}}$ \\
Colour & & [m$^2$\,sr\,W$^{-1}$]\\ \midrule
$(g-r)$ & $(1.52 \pm 0.03) \times 10^4$ & ($5.53 \pm 0.10) \times 10^5$ \\
$(g-i)$ & $(1.47 \pm 0.03) \times 10^4$ & ($5.35 \pm 0.10) \times 10^5$ \\
$(u-z)$ & $(1.51 \pm 0.03) \times 10^4$ & ($5.46 \pm 0.10) \times 10^5$ \\
$(u-i)$ & $(1.46 \pm 0.03) \times 10^4$ & ($5.28 \pm 0.09) \times 10^5$ \\
All & $(1.49 \pm 0.03) \times 10^4$ & ($5.40 \pm 0.09) \times 10^5$ \\ \bottomrule[\lightrulewidth]
\end{tabular}
\end{center}
{\bf Note: } The last row is the value obtained by combining all colours (Fig.~\ref{fig:ebv_vs_tau}).
\end{table}

For each quasar position we extracted the values of $\tau_{353}$ and \radiance\, and then correlated globally with \Ebv\ for each colour $(X-Y)$. Table~\ref{tab:colour} gives the conversion factors $E(B-V)/\tau_{353}$ and $E(B-V)/{\cal{R}}$ estimated for each colour as well as the one obtained by combining all colours, weighted by their uncertainties. 

 Given the very low value of \Ebv, $<0.1$, the large number of measurements is key here to beat down the noise of the SDSS data but also the variations of the quasar spectra around the template spectrum. The intercepts of the regressions of \Ebv\ vs.\ $\tau_{353}$ and \Ebv\ vs.\ \radiance\ are small for each colour, indicating that there is a coherence between the zero levels of the \Planck\ and \IRAS\ maps used to build $\tau_{353}$ and the estimate of the intrinsic colours of quasars, both based on a correlation with $N_\ion{H}{i}$.


\section{Maps in the \Planck\ Legacy Archive}\label{revisedPLA}

Maps of the \Planck\ thermal dust emission model described in this paper can be obtained 
from the \Planck\ Legacy Archive (PLA).\footnote{\url{http://www.sciops.esa.int/index.php?project=planck&page=Planck_Legacy_Archive}}
The maps available give the three MBB parameters ($\tau_{353}$, $T_{\rm obs}$, and $\beta_{\rm obs}$) with their uncertainties ($\delta\tau_{353}$, $\delta T_{\rm obs}$, and $\delta\beta_{\rm obs}$) obtained using data from which point sources were not removed. In addition, the maps of \radiance\ (point sources in) and of $E(B-V)_{\rm xgal}$ are made available; the latter is \radiance\ obtained with data from which point sources were removed and scaled to \Ebv\ with the conversion factor computed using SDSS quasars.

A first release of the \Planck\ thermal dust model was made available on the PLA in March 2013, together with other 2013 \Planck\ products. This first model is superseded by the model presented in this paper.
The first model was made using \Planck\ data from which ZE was not removed, to be compatible with what was used for the cosmological analysis. The 3000\,GHz map used in this first edition was the \IRIS\ map. 
The second release described in this paper is based on \Planck\ data from which ZE was removed and on a combination of \IRIS\ and \citet{schlegel1998} for the 3000\,GHz map. 

The impact of this change is significant only in regions of the sky of low Galactic emission and low ecliptic latitude. It affects the parameters $T_{\rm obs}$ and $\beta_{\rm obs}$ only slightly, at the 2\,\% level. Considering the pixels in the low $N_\ion{H}{i}$ mask, the average $T_{\rm obs}$ went from 20.4 to 20.8\,K from the first to the second version, while the average $\beta_{\rm obs}$ went from 1.59 to 1.55. That means that the shape of the dust SED is not significantly different from one data set to the other. The main effect of the change of data set is on $\tau_{353}$ because the ZE removed data have less emission; from the first to the second version, $\langle\tau_{353}\rangle$ is reduced by about 25\,\% in the most diffuse areas of the sky, from $1.3\times 10^{-6}$ to $9.6\times 10^{-7}$. 

A significant difference between the two releases concerns the \Ebv\ map. In both releases the calibration was done the same way, using SDSS quasars, but in the first release the \Ebv\ map was based on $\tau_{353}$ instead of \radiance. Globally both \Ebv\ maps agree but not on small scales where the 353\,GHz the CIBA is much more present in $\tau_{353}$ than some spectrally-averaged CIBA is in \radiance. The CIBA in $\tau_{353}$ introduces an additional noise in \Ebv\ of the order of $0.003$ magnitude. In that respect the \Ebv\ map of the second version is a significant improvement.

\end{appendix}

\raggedright 

\end{document}